\documentclass[fleqn,usenatbib]{mnras}

\usepackage{newtxtext,newtxmath}
\usepackage{amsmath}

\usepackage[T1]{fontenc}

\DeclareRobustCommand{\VAN}[3]{#2}
\let\VANthebibliography\thebibliography
\def\thebibliography{\DeclareRobustCommand{\VAN}[3]{##3}\VANthebibliography}

\usepackage{graphicx}	
\usepackage{amsmath}	
\usepackage{makecell} 
\usepackage{adjustbox}

\usepackage{soul}
\usepackage{xcolor}

\def  \paul{\color{black}}

\newcommand{\teff}{T_{\rm eff}}
\newcommand{\logg}{\log g}
\newcommand{\mh}{\rm{[M/H]}}

\newcommand{\afe}{\rm{[\alpha/Fe]}}

\newcommand{\kms}{$\rm km\,s^{-1}$}

\newcommand{\vsini}{v\sin{i}}

\newcommand{\zrt}{\zeta_{\rm RT}}

\usepackage{makecell} 
\usepackage{adjustbox}

\title[The magnetic fields of slow rotators with \texttt{ZeeTurbo}]{Measuring small-scale magnetic fields of 44 M dwarfs from SPIRou spectra with \texttt{ZeeTurbo}}

\author[P. I. Cristofari et al.]{
P. I. Cristofari$^{1,2}$,\thanks{E-mail: paul.cristofari@cfa.harvard.edu (CfA)}
J.-F. Donati$^{1}$,
C. Moutou$^{1}$,
L. T. Lehmann$^{1}$,
P. Charpentier$^{1}$,
 P. Fouqu\'e$^{1}$,
 \newauthor
C. P. Folsom$^{3,4}$,
T. Masseron$^{5}$,
 A. Carmona$^{7}$,
 X. Delfosse$^{7}$,
 P. Petit$^{1}$,
 E. Artigau$^{8}$,
 N. J. Cook$^{8}$
 \newauthor
and the SLS consortium
\\
$^{1}$Univ. de Toulouse, CNRS, IRAP, 14 av. Belin, 31400 Toulouse, France\\
$^{2}$Center for Astrophysics | Harvard \& Smithsonian, 60 Garden street, Cambridge, MA 02138, United States \\
$^{3}$Tartu Observatory, University of Tartu, Observatooriumi 1, Tõravere, 61602 Tartumaa, Estonia\\
$^{4}$University of Western Ontario, Department of Physics \& Astronomy, London, ON, N6A 3K7, Canada \\
$^{5}$Instituto de Astrof\'isica de Canarias, 38205 La Laguna, Tenerife, Spain\\
$^{6}$Departamento de Astrofísica, Universidad de La Laguna, E-38206 La Laguna, Tenerife, Spain\\
$^{7}$Univ. Grenoble Alpes, CNRS, IPAG, 38000 Grenoble, France\\
$^{8}$Université de Montréal, Département de Physique, IREX, Montréal, QC, H3C 3J7, Canada
}

\date{Accepted XXX. Received YYY; in original form ZZZ}

\pubyear{2015}

\begin{document}
\label{firstpage}
\pagerange{\pageref{firstpage}--\pageref{lastpage}}
\maketitle

\begin{abstract}
We present the results of an analysis aimed at probing the small-scale magnetic fields of M dwarfs observed with SPIRou, the nIR high-resolution spectro-polarimeter installed at the Canada-France-Hawaii Telescope, in the context of the SPIRou Legacy Survey. Our analysis relies on high-resolution median spectra built from several tens of spectra recorded between 2019 and 2022, and on synthetic spectra computed with the \texttt{ZeeTurbo} code for various combination of atmospheric parameters and magnetic field strengths. We pursue the efforts undertaken in a previous study and focus on 44 weakly to moderately active M dwarfs. We derive average magnetic field strengths (<$B$>) ranging from 0.05 to 1.15\,kG, in good agreement with activity estimates and rotation periods. We found that including magnetic fields in our models has virtually no impact on our derived atmospheric parameters, and that a priori assumptions on the stellar surface gravity can affect our estimated <$B$>. 
Our results suggest that small-scale magnetic fields account for more than 70~\% of the overall average magnetic field for most targets whose large-scale fields were previously measured.
We derived low magnetic fluxes for several targets in our sample, and found no clear evidence that <$B$> decreases with increasing Rossby number in the unsaturated dynamo regime.
We even identified counterexamples (GJ~1289 and GJ~1286) where the small-scale field is unusually strong despite the long rotation period. Along with similar results on the large-scale fields, our findings further suggest that dynamo processes may operate in a non-conventional mode in these strongly magnetic, slowly-rotating stars.

\hphantom{\citep{donati-2023a}} 
\end{abstract}




\section{Introduction}

Magnetic fields are believed to play an essential role in stellar formation and evolution~\citep[e.g.,][]{donati-2009}. They can trigger star-planet interactions~\citep[e.g.,][]{vidotto-2018} and may also explain inflated radii of cools stars~\citep{feiden-2012, feiden-2014}. Magnetic fields are also responsible for activity phenomena hampering planet detection and characterization~\citep[e.g.,][]{hebrard-2016, dumusque-2021, bellotti-2022}. 
M dwarfs are none the less prime targets for finding out and characterizing nearby planetary systems, and in particular planets located in the habitable zone of their host star~\citep[e.g.,][]{bonfils-2013, dressing-2015, gaidos-2016}.
The study of the small-scale magnetic fields of M dwarfs thereby attracted increasing attention, not only for better understanding their origin~\citep[e.g., ][]{saar-1985, johns-krull-1996, shulyak-2014, kochukhov-2021, reiners-2022}, but also to investigate how they impact radial velocity (RV) variations~\citep[e.g.,][]{haywood-2022}

Among the most popular techniques to study magnetic fields, Zeeman-Doppler-Imaging~\citep{donati-1997, donati-2006} extracts information from polarized spectra to provide constraints on the large-scale magnetic field topologies of stars. Another approach consists in measuring the Zeeman broadening of line profiles in unpolarized spectra, allowing one to probe magnetic fields on smaller spatial scales at the surface of the star~\citep[e.g.,][]{kochukhov-2021}.

The presence of magnetic fields in M dwarfs can impact stellar atmospheric characterization, and conversely, erroneous estimates on the properties of M dwarfs may bias magnetic diagnostics. 
In recent years, studies aimed at modelling spectra of M dwarfs took advantage of the new generation of high-resolution near-infrared (nIR) spectrometers, such as CRIRES+~\citep{dorn-2023}, CARMENES~\citep{quirrenbach-2014} or SPIRou~\citep{donati-2020}, and yielded new constraints on the atmospheric properties of tens to hundreds of stars~\citep[e.g.,][]{rajpurohit-2018, passegger-2019, marfil-2021, sarmento-2021}. High-resolution spectroscopy also provides the means to refine magnetic diagnostics by accurately modelling the shape of spectral features affected by magnetic fields, including both Zeeman broadening and intensification of well selected lines~\citep[e.g.,][]{kochukhov-2020, petit-2021, hahlin-2023}. Because the splitting of the energy levels due to the Zeeman effect depends on the considered transition, this approach is particularly useful to disentangle spectral line broadening due to the presence of magnetic fields from rotation or macroturbulence. The detection of Zeeman signatures, however, remains difficult for weakly magnetized stars whose spectral lines may be dominantly broadened by rotation or macroturbulence.

Magnetic activity in low-mass stars is known to correlate with stellar rotation periods, and even better with the Rossby number Ro, defined as the rotation period normalized to the convective turnover time~\citep[e.g.,][]{mangeney-1984, noyes-1984, wright-2011, wright-2018, reiners-2014}.
 Measurements of small- and large-scale magnetic fields and their dependence with stellar parameters are therefore essential to further characterize such relations, and to refine models of the dynamo processes that generate such fields.
Both large- and small-scale magnetic fields strongly correlate with Ro, with field intensities increasing with decreasing Ro until they saturate for fast rotators with Ro<0.1~\citep[e.g.,][]{vidotto-2014, see-2016, reiners-2022}.
For the slowest rotators with Ro>1, however, uncertainties on fields measurements are larger, making it more difficult to assess whether this relation still holds.

With this paper, we pursue the work initiated in~\citet{cristofari-2023} that focused on rapidly rotating, extremely active targets. We now aim at characterizing the small-scale magnetic fields of weakly or moderately active M dwarfs from high-resolution nIR spectra collected with SPIRou in the framework of the SPIRou legacy Survey (SLS) at the Canada-France-Hawaii-Telescope (CFHT) where SPIRou is installed. We rely on models computed with our newly implemented code, \texttt{ZeeTurbo}, and on the process described in~\citet{cristofari-2023} to constrain atmospheric parameters and magnetic fields of 44 targets. This sample is the same as in~\citet{cristofari-2022b}, which relied on non-magnetic models to characterize the main atmospheric properties of all sample stars, i.e., the effective temperature ($\teff$), surface gravity ($\logg$), metallicity ($\mh$), and $\alpha$-enhancement ($\afe$), the latter scaling the abundances of O, Ne, Mg, Si, S, Ar, Ca, and Ti with respect to Fe.

We describe the data used for our analysis in Section~\ref{sec:observations}, and recall the main steps of our method in Section~\ref{sec:method}. We present our results in Section~\ref{sec:results} before assessing the impact of atmospheric characterization on magnetic field estimation in Sec~\ref{sec:impact_on_atmo}.  

\section{Observations and reduction}
\label{sec:observations}

The spectra analyzed in this paper were obtained in the context of the SLS between 2019  and 2022. We focus on 44 targets known to be no more than moderately active~\citep{fouque-2018, schofer-2019} and for which more than 25 visits were carried out. Data were processed with the SPIRou reduction pipeline, \texttt{APERO}~\citep[version 0.7.254, ][]{cook-2022}.
The telluric correction is performed by \texttt{APERO}, and takes advantage of the large number of observations obtained at various Barycentric Earth Radial Velocities~(BERV) and a large spectral library of observations of telluric standard stars observed since 2018. The telluric correction is performed through a hybrid method adjusting a simple telluric absorption model from TAPAS \citep{bertaux-2014} and with a higher-order correction from residuals measured in a sample of hot-star observations. The results are comparable in accuracy to those obtained with a PCA-based approach~\citep{artigau-2014} used in earlier versions of \texttt{APERO} but has the merit of being more robust. For each observation night, we average the telluric-corrected spectra to obtain a spectrum with higher signal-to-noise-ratio (SNR). We then compute the median of all spectra for each star in the barycentric frame to create a very high SNR spectrum for each target. We refer to this high-resolution and high-SNR median spectrum as `SPIRou template' in the rest of the paper. Our analysis is ultimately performed on the SPIRou templates, whose SNR per 2~\kms pixel can reach up to 2000 in the $H$ band. Table~\ref{tab:nbvisits} summarizes the typical SNR and number of visits for each of the 44 targets studied in this paper.

\begin{table}
	\caption{Number of visits, median SNR and SNR range for the targets studied in this paper.}
	\label{tab:nbvisits}
	\begin{tabular}{ccc}
		\hline
		Star & Nb. visits & Med. SNR [SNR Range] \\ 
		\hline
Gl\,338B & $57$ & $450$ [330--590] \\ 
Gl\,846 & $189$ & $310$ [120--560] \\ 
Gl\,410 & $128$ & $260$ [160--370] \\ 
Gl\,205 & $151$ & $600$ [250--680] \\ 
Gl\,514 & $161$ & $310$ [170--2710] \\ 
Gl\,880 & $164$ & $410$ [260--500] \\ 
Gl\,382 & $121$ & $310$ [160--440] \\ 
Gl\,412A & $177$ & $360$ [150--3040] \\ 
Gl\,15A & $187$ & $540$ [190--3980] \\ 
Gl\,411 & $174$ & $740$ [420--910] \\ 
Gl\,752A & $123$ & $350$ [120--510] \\ 
Gl\,48 & $184$ & $260$ [130--400] \\ 
Gl\,617B & $144$ & $240$ [50--300] \\ 
Gl\,436 & $80$ & $300$ [190--920] \\ 
Gl\,480 & $108$ & $220$ [170--260] \\ 
Gl\,849 & $189$ & $230$ [130--280] \\ 
Gl\,408 & $169$ & $280$ [80--340] \\ 
Gl\,687 & $208$ & $410$ [210--540] \\ 
Gl\,725A & $207$ & $440$ [230--510] \\ 
Gl\,317 & $74$ & $210$ [150--250] \\ 
Gl\,251 & $167$ & $280$ [110--330] \\ 
GJ\,4063 & $211$ & $200$ [140--330] \\ 
Gl\,581 & $28$ & $240$ [130--300] \\ 
PM~J09553$-$2715 & $76$ & $220$ [140--340] \\ 
GJ\,4333 & $180$ & $200$ [110--260] \\ 
GJ\,1012 & $145$ & $200$ [120--240] \\ 
Gl\,876 & $87$ & $310$ [280--550] \\ 
Gl\,725B & $205$ & $320$ [150--390] \\ 
GJ\,1148 & $102$ & $200$ [120--300] \\ 
PM~J08402$+$3127 & $138$ & $200$ [90--220] \\ 
Gl\,445 & $93$ & $220$ [70--300] \\ 
GJ\,3378 & $175$ & $210$ [60--320] \\ 
GJ\,1105 & $165$ & $200$ [80--300] \\ 
Gl\,169.1A & $168$ & $210$ [80--270] \\ 
Gl\,15B & $178$ & $200$ [110--240] \\ 
PM~J21463$+$3813 & $177$ & $200$ [70--240] \\ 
Gl\,699 & $240$ & $410$ [110--600] \\ 
GJ\,1289 & $201$ & $200$ [100--290] \\ 
Gl\,447 & $57$ & $260$ [130--330] \\ 
GJ\,1151 & $156$ & $200$ [120--230] \\ 
GJ\,1103 & $65$ & $200$ [90--230] \\ 
Gl\,905 & $213$ & $240$ [110--420] \\ 
GJ\,1286 & $113$ & $200$ [50--230] \\ 
GJ\,1002 & $145$ & $200$ [80--270] \\ 
		\hline
	\end{tabular}
\end{table}

In this work, we focus on the analysis of Stokes $I$ spectra, and do not use the polarized data also recorded  for our targets. The sample of stars is strictly the same as that of~\citet{cristofari-2022b}, and includes stars ranging from about 3000 to 4000\,K in $\teff$. Table~\ref{tab:first-table} lists luminosity and magnitude estimates for the 44 targets in our sample. 

\begin{table*}
	\caption{Reported properties for our sample of stars. Columns 2 to 4 list the spectral type, mass and radii from~\citet{cristofari-2022b}, column 5 lists $\tau$ estimates computed from mass with the relation of~\citet{wright-2018}. Rotation periods from~\citet{donati-2023b} and Rossby numbers are listed in columns 6 and 7. The rotation period of Gl\,447 was reported to be suspiciously short and therefore ignored in this paper~\citep{donati-2023b}.}
	\label{tab:first-table}
\begin{tabular}{ccccccc}
\hline
Star & Spectral Type & $M_\star/M_\odot$ & $R_\star/R_\odot$ & $\tau$ (d) & $P_{\rm rot}$ (d) & $\rm Ro$ \\ 
\hline
Gl\,338B & M0V & $0.58\pm 0.02$ & $0.609\pm 0.012$ & $37\pm 23$ & $42\pm4$ & $1.15\pm0.72$ \\ 
Gl\,846 & M0.5V & $0.57\pm 0.02$ & $0.568\pm 0.009$ & $38\pm 23$ & $22\pm0$ & $0.58\pm0.35$ \\ 
Gl\,410 & M1.0V & $0.55\pm 0.02$ & $0.543\pm 0.009$ & $40\pm 24$ & $14\pm0$ & $0.35\pm0.21$ \\ 
Gl\,205 & M1.5V & $0.58\pm 0.02$ & $0.588\pm 0.010$ & $37\pm 23$ & $35\pm0$ & $0.94\pm0.58$ \\ 
Gl\,514 & M1.0V & $0.50\pm 0.02$ & $0.497\pm 0.008$ & $45\pm 27$ & $30\pm0$ & $0.67\pm0.39$ \\ 
Gl\,880 & M1.5V & $0.55\pm 0.02$ & $0.563\pm 0.009$ & $40\pm 24$ & $37\pm0$ & $0.94\pm0.57$ \\ 
Gl\,382 & M2V & $0.51\pm 0.02$ & $0.511\pm 0.009$ & $44\pm 26$ & $22\pm0$ & $0.50\pm0.29$ \\ 
Gl\,412A & M1.0V & $0.39\pm 0.02$ & $0.391\pm 0.007$ & $62\pm 35$ & $37\pm2$ & $0.60\pm0.34$ \\ 
Gl\,15A & M2V & $0.39\pm 0.02$ & $0.345\pm 0.015$ & $62\pm 35$ & $43\pm0$ & $0.70\pm0.40$ \\ 
Gl\,411 & M2V & $0.39\pm 0.02$ & $0.383\pm 0.008$ & $62\pm 35$ & $427\pm34$ & $6.89\pm3.94$ \\ 
Gl\,752A & M3V & $0.47\pm 0.02$ & $0.469\pm 0.008$ & $49\pm 29$ & $45\pm4$ & $0.91\pm0.54$ \\ 
Gl\,48 & M3.0V & $0.46\pm 0.02$ & $0.469\pm 0.008$ & $51\pm 30$ & $52\pm2$ & $1.03\pm0.60$ \\ 
Gl\,617B & M3.0V & $0.45\pm 0.02$ & $0.460\pm 0.008$ & $52\pm 30$ & $43\pm3$ & $0.82\pm0.48$ \\ 
Gl\,436 & M3V & $0.42\pm 0.02$ & $0.425\pm 0.008$ & $57\pm 33$ & $48\pm13$ & $0.84\pm0.53$ \\ 
Gl\,480 & M3.5V & $0.45\pm 0.02$ & $0.449\pm 0.008$ & $52\pm 30$ & $25\pm0$ & $0.48\pm0.28$ \\ 
Gl\,849 & M3.5V & $0.46\pm 0.02$ & $0.458\pm 0.008$ & $51\pm 30$ & $42\pm1$ & $0.82\pm0.48$ \\ 
Gl\,408 & M4V & $0.38\pm 0.02$ & $0.390\pm 0.007$ & $64\pm 36$ & $172\pm7$ & $2.70\pm1.53$ \\ 
Gl\,687 & M3.0V & $0.39\pm 0.02$ & $0.414\pm 0.007$ & $62\pm 35$ & $57\pm1$ & $0.91\pm0.52$ \\ 
Gl\,725A & M3V & $0.33\pm 0.02$ & $0.345\pm 0.006$ & $74\pm 41$ & $102\pm4$ & $1.37\pm0.76$ \\ 
Gl\,317 & M3.5V & $0.42\pm 0.02$ & $0.423\pm 0.008$ & $57\pm 33$ & $39\pm4$ & $0.69\pm0.40$ \\ 
Gl\,251 & M3V & $0.35\pm 0.02$ & $0.365\pm 0.007$ & $70\pm 39$ & $93\pm7$ & $1.34\pm0.76$ \\ 
GJ\,4063 & M4V & $0.42\pm 0.02$ & $0.422\pm 0.008$ & $57\pm 33$ & $41\pm4$ & $0.72\pm0.41$ \\ 
Gl\,581 & M3V & $0.31\pm 0.02$ & $0.317\pm 0.006$ & $78\pm 43$ & ...~$\pm $~... & ...~$\pm $~... \\ 
PM~J09553$-$2715 & M3V & $0.29\pm 0.02$ & $0.302\pm 0.006$ & $83\pm 46$ & $73\pm4$ & $0.88\pm0.48$ \\ 
GJ\,1012 & M4.0V & $0.35\pm 0.02$ & $0.367\pm 0.007$ & $70\pm 39$ & ...~$\pm $~... & ...~$\pm $~... \\ 
GJ\,4333 & M3.5V & $0.37\pm 0.02$ & $0.386\pm 0.008$ & $66\pm 37$ & $71\pm2$ & $1.08\pm0.61$ \\ 
Gl\,725B & M3.5V & $0.25\pm 0.02$ & $0.280\pm 0.005$ & $94\pm 51$ & $135\pm15$ & $1.43\pm0.80$ \\ 
Gl\,876 & M3.5V & $0.33\pm 0.02$ & $0.333\pm 0.006$ & $74\pm 41$ & $84\pm3$ & $1.13\pm0.63$ \\ 
GJ\,1148 & M4.0V & $0.34\pm 0.02$ & $0.365\pm 0.007$ & $72\pm 40$ & ...~$\pm $~... & ...~$\pm $~... \\ 
PM~J08402$+$3127 & M3.5V & $0.28\pm 0.02$ & $0.299\pm 0.006$ & $86\pm 47$ & $90\pm8$ & $1.04\pm0.58$ \\ 
Gl\,445 & M4.0V & $0.24\pm 0.02$ & $0.266\pm 0.005$ & $97\pm 53$ & ...~$\pm $~... & ...~$\pm $~... \\ 
GJ\,3378 & M4.0V & $0.26\pm 0.02$ & $0.279\pm 0.005$ & $91\pm 50$ & $95\pm2$ & $1.04\pm0.57$ \\ 
GJ\,1105 & M3.5V & $0.27\pm 0.02$ & $0.283\pm 0.005$ & $89\pm 48$ & ...~$\pm $~... & ...~$\pm $~... \\ 
Gl\,169.1A & M4.0V & $0.28\pm 0.02$ & $0.292\pm 0.006$ & $86\pm 47$ & $92\pm4$ & $1.07\pm0.59$ \\ 
Gl\,15B & M3.5V & $0.16\pm 0.02$ & $0.182\pm 0.004$ & $125\pm 67$ & $113\pm4$ & $0.90\pm0.49$ \\ 
PM~J21463$+$3813 & M5V & $0.18\pm 0.02$ & $0.208\pm 0.004$ & $118\pm 63$ & $94\pm3$ & $0.80\pm0.43$ \\ 
Gl\,699 & M4V & $0.16\pm 0.02$ & $0.185\pm 0.004$ & $125\pm 67$ & $136\pm13$ & $1.09\pm0.59$ \\ 
GJ\,1289 & M4.5V & $0.21\pm 0.02$ & $0.233\pm 0.005$ & $107\pm 58$ & $74\pm1$ & $0.70\pm0.38$ \\ 
Gl\,447 & M4V & $0.18\pm 0.02$ & $0.201\pm 0.004$ & $118\pm 63$ & $^*24\pm4$ & $0.21\pm0.11$ \\ 
GJ\,1151 & M4.5V & $0.17\pm 0.02$ & $0.193\pm 0.004$ & $121\pm 65$ & $176\pm5$ & $1.45\pm0.78$ \\ 
GJ\,1103 & M4.5V & $0.19\pm 0.02$ & $0.224\pm 0.005$ & $114\pm 61$ & $143\pm10$ & $1.25\pm0.68$ \\ 
Gl\,905 & M5.0V & $0.15\pm 0.02$ & $0.165\pm 0.004$ & $129\pm 69$ & $114\pm3$ & $0.88\pm0.47$ \\ 
GJ\,1286 & M5.0V & $0.12\pm 0.02$ & $0.142\pm 0.004$ & $143\pm 76$ & $178\pm15$ & $1.25\pm0.67$ \\ 
GJ\,1002 & M5.5V & $0.12\pm 0.02$ & $0.139\pm 0.003$ & $143\pm 76$ & $90\pm3$ & $0.63\pm0.34$ \\ 
\hline
\end{tabular}
\end{table*}

\section{Zeeman broadening analysis}
\label{sec:method}

In this section, we briefly summarize the analysis described in~\citet{cristofari-2023}.
Our process relies on synthetic spectra computed from MARCS model atmospheres~\citep{gustafsson-2008} with \texttt{ZeeTurbo}~\citep{cristofari-2023}, a tool built from the \texttt{Turbospectrum}~\citep{alvarez-1998, plez-2012} and \texttt{Zeeman}~\citep{landstreet-1988, wade-2001, folsom-2016} codes. We compute models for various magnetic field strengths, assuming a radial magnetic field everywhere in the photosphere. Synthetic spectra were computed assuming a microturbulence of 1~\kms{} and local thermodynamic equilibrium (LTE), generally considered valid for M dwarfs~\citep[see e.g.,][]{hauschildt-1999, husser-2013, hahlin-2023}. All SPIRou templates are modeled with a linear combination of spectra computed for various magnetic field strengths, ranging from 0 to 10\,kG by steps of 2\,kG.
The model spectrum S can thus be written $S = \sum{f_iS_i}$, with $S_i$ denoting the spectrum associated with a magnetic strength of $i$\,kG and $f_i$ the filling factor associated with this component (all $f_i$s verifying $\sum f_i = 1$). This approach has already been used in several studies~\citep{shulyak-2010, shulyak-2014, kochukhov-2020, reiners-2022, cristofari-2023}. As in~\citet{cristofari-2023}, we rely on a Markov chain Monte Carlo (MCMC) process to estimate both the atmospheric parameters and the filling factors. We rely on a log-likelihood of the form, \[\ln{\mathcal{L}} = -\frac{1}{2}\Big(\sum_{i=1}^n\big(\frac{O_i-M_i}{\sigma_i}\big)^2\Big) - \frac{n}{2}\ln{(2\pi)} - \frac{1}{2}\sum_{i=1}^n\ln{\sigma_i^2},\] with $O_i$ the observed normalized flux, $M_i$ the synthetic spectrum, and $\sigma_i$ the uncertainty on the observed spectrum, for pixel $i$. The likelihood is closely related to the $\chi^2$, and we estimate the optimal atmospheric parameters and filling factors by averaging the walkers whose associated $\chi^2$ do not deviate by more than 1 from the minimum $\chi^2$. To estimate error bars, we look at the posterior distributions, and compare the $16^{\rm th}$ and $84^{\rm th}$ percentiles to the median of the distributions.
{\paul In order to account for some of the systematic uncertainties in the error bars, we run our process twice, enlarging the error bars on each pixel in the second run in order to ensure that the best fit leads to a reduced $\chi^2$ of 1~\citep{cristofari-2023}.}
The error bars derived from posterior distributions are referred to as `formal' error bars, and are typically 5--10\,K in $\teff$ and 0.005 to 0.020\,dex in $\logg$, $\mh$ and $\afe$. We found that formal error bars are significantly smaller than the dispersion due to modelling assumptions, and defined `empirical' error bars by quadratically adding 30\,K for $\teff$, 0.05\,dex for $\logg$, 0.10\,dex for $\mh$ and 0.04\,dex for $\afe$ to our formal error bars in order to account for some of this dispersion~\citep{cristofari-2022b}. In this work, we rely on these results, and compute empirical error bars for our derived atmospheric parameters.

Before computing the likelihood, synthetic spectra are convoled with a Gaussian profile of full width at half maximum (FWHM) of 4.3~\kms{} to account for the instrumental width of SPIRou, convolved with a rectangular function representing the 2.2~\kms{} wide pixels, and resampled on a reference SPIRou wavelength grid. The continua of both the SPIRou templates and synthetic spectra are then brought to the same level following the procedure described in~\citet{cristofari-2022b}. The comparison is performed on a limited number of spectral regions containing 30 identified atomic lines, 30 CO lines from one molecular band redward of 2293~nm, and 9 OH lines. This list is the same as that presented in~\citet{cristofari-2023}, and includes lines with Land\'e factors ranging from 0 to 2.5, allowing us to extract information from spectral lines that are either sensitive or insensitive to magnetic fields (see, e.g., Fig.~\ref{fig:fits}).
  
 Throughout this analysis, we neglected the effect of rotational broadening, and rather fit a radial-tangential macroturbulence ($\zrt$). This assumption is further motivated by the long rotation periods measured for most objects~\citep[][]{fouque-2023, donati-2023b}, yielding rotational velocities at the equator smaller than 2~\kms\ for all stars (median <0.2~\kms).

\section{Small-scale field measurements}
\label{sec:results}

We apply our process to the 44 targets in our sample, and report the retrieved atmospheric parameters and average magnetic field strengths (<$B$>) in Table~\ref{tab:res}.

\begin{table*}
	\caption{Derived stellar parameters, average magnetic field  and filling factors for the 44 targets in our sample.}
	\label{tab:res}
	\begin{adjustbox}{angle=90}   
		\resizebox{.97\textheight}{!}{
\begin{tabular}{cccccccc}
	\hline
Star & $\teff$ (K) & $\logg$ (dex) & $\mh$ (dex) & $\afe$ (dex) & $\zeta_{\rm RT}$ (\kms{}) & $<B>$ (kG) & \makecell{$f_0$, $f_2$, $f_4$, $f_6$, $f_8$, $f_{10}$} \\ 
\hline
Gl\,338B & $3944\pm 30$ & $4.77\pm 0.05$ & $-0.08\pm 0.10$ & $0.04\pm 0.10$ & $2.36\pm 0.11$ & $0.09\pm 0.03$ & \makecell{$0.955\pm 0.015$, $0.044\pm 0.015$, $0.000\pm 0.002$, $0.000\pm 0.001$, $0.000\pm 0.001$, $0.000\pm 0.001$} \\ 
Gl\,846 & $3826\pm 31$ & $4.71\pm 0.05$ & $0.05\pm 0.10$ & $-0.01\pm 0.10$ & $2.43\pm 0.11$ & $0.13\pm 0.04$ & \makecell{$0.936\pm 0.016$, $0.063\pm 0.017$, $0.000\pm 0.002$, $0.000\pm 0.002$, $0.000\pm 0.001$, $0.000\pm 0.001$} \\ 
Gl\,410 & $3818\pm 30$ & $4.79\pm 0.05$ & $0.01\pm 0.10$ & $0.03\pm 0.10$ & $3.08\pm 0.10$ & $0.71\pm 0.03$ & \makecell{$0.646\pm 0.015$, $0.352\pm 0.015$, $0.001\pm 0.002$, $0.001\pm 0.001$, $0.000\pm 0.001$, $0.000\pm 0.001$} \\ 
Gl\,205 & $3768\pm 31$ & $4.74\pm 0.05$ & $0.43\pm 0.10$ & $-0.07\pm 0.10$ & $2.29\pm 0.13$ & $0.07\pm 0.04$ & \makecell{$0.969\pm 0.017$, $0.029\pm 0.018$, $0.001\pm 0.002$, $0.000\pm 0.001$, $0.000\pm 0.001$, $0.000\pm 0.001$} \\ 
Gl\,514 & $3710\pm 30$ & $4.72\pm 0.05$ & $-0.12\pm 0.10$ & $0.04\pm 0.10$ & $2.15\pm 0.10$ & $0.02\pm 0.02$ & \makecell{$0.994\pm 0.007$, $0.004\pm 0.007$, $0.000\pm 0.002$, $0.001\pm 0.001$, $0.000\pm 0.001$, $0.000\pm 0.001$} \\ 
Gl\,880 & $3705\pm 30$ & $4.73\pm 0.05$ & $0.25\pm 0.10$ & $-0.05\pm 0.10$ & $2.18\pm 0.14$ & $0.12\pm 0.04$ & \makecell{$0.944\pm 0.018$, $0.054\pm 0.018$, $0.001\pm 0.002$, $0.000\pm 0.002$, $0.001\pm 0.001$, $0.000\pm 0.001$} \\ 
Gl\,382 & $3644\pm 31$ & $4.72\pm 0.05$ & $0.13\pm 0.10$ & $-0.02\pm 0.10$ & $2.54\pm 0.11$ & $0.21\pm 0.04$ & \makecell{$0.896\pm 0.017$, $0.103\pm 0.018$, $0.001\pm 0.002$, $0.000\pm 0.001$, $0.000\pm 0.001$, $0.000\pm 0.001$} \\ 
Gl\,412A & $3639\pm 31$ & $4.75\pm 0.05$ & $-0.43\pm 0.10$ & $0.10\pm 0.10$ & $2.42\pm 0.11$ & $0.29\pm 0.05$ & \makecell{$0.920\pm 0.017$, $0.058\pm 0.019$, $0.001\pm 0.007$, $0.002\pm 0.004$, $0.015\pm 0.007$, $0.004\pm 0.005$} \\ 
Gl\,15A & $3631\pm 31$ & $4.79\pm 0.05$ & $-0.33\pm 0.10$ & $0.07\pm 0.10$ & $2.43\pm 0.13$ & $0.05\pm 0.03$ & \makecell{$0.992\pm 0.008$, $0.002\pm 0.007$, $0.000\pm 0.003$, $0.004\pm 0.002$, $0.002\pm 0.003$, $0.000\pm 0.002$} \\ 
Gl\,411 & $3591\pm 31$ & $4.68\pm 0.05$ & $-0.37\pm 0.10$ & $0.16\pm 0.10$ & $2.44\pm 0.12$ & $0.49\pm 0.06$ & \makecell{$0.940\pm 0.011$, $0.004\pm 0.010$, $0.001\pm 0.005$, $0.001\pm 0.004$, $0.033\pm 0.014$, $0.020\pm 0.013$} \\ 
Gl\,752A & $3565\pm 31$ & $4.69\pm 0.05$ & $0.10\pm 0.10$ & $-0.01\pm 0.10$ & $2.23\pm 0.10$ & $0.06\pm 0.03$ & \makecell{$0.977\pm 0.014$, $0.020\pm 0.014$, $0.001\pm 0.003$, $0.000\pm 0.002$, $0.001\pm 0.001$, $0.000\pm 0.001$} \\ 
Gl\,48 & $3537\pm 31$ & $4.68\pm 0.05$ & $0.07\pm 0.10$ & $0.07\pm 0.10$ & $2.08\pm 0.12$ & $0.28\pm 0.05$ & \makecell{$0.912\pm 0.018$, $0.071\pm 0.019$, $0.001\pm 0.005$, $0.001\pm 0.003$, $0.010\pm 0.006$, $0.004\pm 0.004$} \\ 
Gl\,617B & $3532\pm 31$ & $4.78\pm 0.05$ & $0.14\pm 0.10$ & $-0.01\pm 0.10$ & $1.97\pm 0.14$ & $0.04\pm 0.03$ & \makecell{$0.982\pm 0.015$, $0.016\pm 0.016$, $0.001\pm 0.002$, $0.000\pm 0.001$, $0.000\pm 0.001$, $0.000\pm 0.001$} \\ 
Gl\,436 & $3521\pm 31$ & $4.73\pm 0.05$ & $-0.01\pm 0.10$ & $-0.00\pm 0.10$ & $2.26\pm 0.12$ & $0.02\pm 0.03$ & \makecell{$0.995\pm 0.008$, $0.002\pm 0.007$, $0.001\pm 0.002$, $0.000\pm 0.002$, $0.000\pm 0.002$, $0.001\pm 0.002$} \\ 
Gl\,480 & $3517\pm 31$ & $4.86\pm 0.05$ & $0.23\pm 0.10$ & $-0.01\pm 0.10$ & $2.19\pm 0.13$ & $0.05\pm 0.03$ & \makecell{$0.977\pm 0.016$, $0.021\pm 0.016$, $0.001\pm 0.002$, $0.000\pm 0.001$, $0.000\pm 0.001$, $0.000\pm 0.001$} \\ 
Gl\,849 & $3514\pm 31$ & $4.78\pm 0.05$ & $0.26\pm 0.10$ & $-0.04\pm 0.10$ & $2.12\pm 0.14$ & $0.01\pm 0.03$ & \makecell{$0.995\pm 0.009$, $0.003\pm 0.009$, $0.000\pm 0.002$, $0.000\pm 0.001$, $0.000\pm 0.001$, $0.000\pm 0.001$} \\ 
Gl\,408 & $3501\pm 30$ & $4.72\pm 0.05$ & $-0.16\pm 0.10$ & $0.01\pm 0.10$ & $2.39\pm 0.12$ & $0.20\pm 0.05$ & \makecell{$0.915\pm 0.019$, $0.079\pm 0.021$, $0.002\pm 0.005$, $0.001\pm 0.003$, $0.001\pm 0.003$, $0.001\pm 0.002$} \\ 
Gl\,687 & $3481\pm 31$ & $4.66\pm 0.05$ & $-0.01\pm 0.10$ & $0.05\pm 0.10$ & $2.30\pm 0.12$ & $0.32\pm 0.06$ & \makecell{$0.918\pm 0.018$, $0.055\pm 0.020$, $0.001\pm 0.006$, $0.001\pm 0.004$, $0.020\pm 0.008$, $0.004\pm 0.006$} \\ 
Gl\,725A & $3468\pm 31$ & $4.73\pm 0.05$ & $-0.25\pm 0.10$ & $0.14\pm 0.10$ & $2.51\pm 0.12$ & $0.49\pm 0.07$ & \makecell{$0.928\pm 0.016$, $0.016\pm 0.017$, $0.003\pm 0.007$, $0.002\pm 0.006$, $0.039\pm 0.014$, $0.013\pm 0.011$} \\ 
Gl\,317 & $3453\pm 31$ & $4.71\pm 0.05$ & $0.19\pm 0.10$ & $-0.04\pm 0.10$ & $2.22\pm 0.15$ & $0.08\pm 0.04$ & \makecell{$0.964\pm 0.019$, $0.034\pm 0.019$, $0.001\pm 0.004$, $0.000\pm 0.002$, $0.001\pm 0.002$, $0.000\pm 0.001$} \\ 
Gl\,251 & $3436\pm 31$ & $4.70\pm 0.05$ & $-0.06\pm 0.10$ & $-0.02\pm 0.10$ & $2.39\pm 0.13$ & $0.06\pm 0.04$ & \makecell{$0.984\pm 0.013$, $0.010\pm 0.012$, $0.003\pm 0.004$, $0.001\pm 0.003$, $0.002\pm 0.003$, $0.001\pm 0.003$} \\ 
GJ\,4063 & $3427\pm 31$ & $4.70\pm 0.06$ & $0.36\pm 0.10$ & $-0.08\pm 0.10$ & $2.15\pm 0.15$ & $0.02\pm 0.03$ & \makecell{$0.994\pm 0.011$, $0.004\pm 0.011$, $0.001\pm 0.003$, $0.000\pm 0.002$, $0.000\pm 0.001$, $0.000\pm 0.001$} \\ 
Gl\,581 & $3423\pm 31$ & $4.75\pm 0.05$ & $-0.14\pm 0.10$ & $-0.01\pm 0.10$ & $2.64\pm 0.12$ & $0.20\pm 0.06$ & \makecell{$0.939\pm 0.021$, $0.046\pm 0.022$, $0.003\pm 0.008$, $0.002\pm 0.004$, $0.008\pm 0.005$, $0.002\pm 0.004$} \\ 
PM~J09553$-$2715 & $3397\pm 31$ & $4.74\pm 0.05$ & $-0.10\pm 0.10$ & $-0.02\pm 0.10$ & $2.83\pm 0.18$ & $0.10\pm 0.05$ & \makecell{$0.961\pm 0.021$, $0.034\pm 0.022$, $0.001\pm 0.007$, $0.001\pm 0.003$, $0.001\pm 0.004$, $0.001\pm 0.003$} \\ 
GJ\,4333 & $3383\pm 31$ & $4.69\pm 0.06$ & $0.22\pm 0.10$ & $-0.03\pm 0.10$ & $2.35\pm 0.13$ & $0.19\pm 0.06$ & \makecell{$0.932\pm 0.022$, $0.058\pm 0.023$, $0.003\pm 0.006$, $0.001\pm 0.004$, $0.004\pm 0.004$, $0.002\pm 0.003$} \\ 
GJ\,1012 & $3382\pm 31$ & $4.62\pm 0.05$ & $0.02\pm 0.10$ & $0.01\pm 0.10$ & $2.48\pm 0.15$ & $0.33\pm 0.07$ & \makecell{$0.954\pm 0.014$, $0.008\pm 0.013$, $0.003\pm 0.006$, $0.001\pm 0.006$, $0.026\pm 0.011$, $0.008\pm 0.009$} \\ 
Gl\,876 & $3381\pm 31$ & $4.71\pm 0.06$ & $0.11\pm 0.10$ & $-0.05\pm 0.10$ & $2.66\pm 0.18$ & $0.05\pm 0.04$ & \makecell{$0.983\pm 0.014$, $0.013\pm 0.014$, $0.002\pm 0.004$, $0.001\pm 0.003$, $0.000\pm 0.002$, $0.000\pm 0.002$} \\ 
Gl\,725B & $3381\pm 31$ & $4.77\pm 0.06$ & $-0.27\pm 0.10$ & $0.13\pm 0.10$ & $2.82\pm 0.15$ & $0.44\pm 0.08$ & \makecell{$0.927\pm 0.020$, $0.022\pm 0.021$, $0.006\pm 0.009$, $0.002\pm 0.006$, $0.033\pm 0.013$, $0.010\pm 0.011$} \\ 
GJ\,1148 & $3374\pm 31$ & $4.67\pm 0.05$ & $0.07\pm 0.10$ & $-0.00\pm 0.10$ & $2.49\pm 0.13$ & $0.24\pm 0.07$ & \makecell{$0.964\pm 0.015$, $0.009\pm 0.014$, $0.002\pm 0.006$, $0.001\pm 0.005$, $0.020\pm 0.008$, $0.004\pm 0.006$} \\ 
PM~J08402$+$3127 & $3371\pm 31$ & $4.73\pm 0.05$ & $-0.15\pm 0.10$ & $0.00\pm 0.10$ & $2.57\pm 0.19$ & $0.21\pm 0.06$ & \makecell{$0.916\pm 0.025$, $0.077\pm 0.027$, $0.001\pm 0.007$, $0.001\pm 0.004$, $0.003\pm 0.005$, $0.002\pm 0.004$} \\ 
Gl\,445 & $3356\pm 31$ & $4.78\pm 0.06$ & $-0.24\pm 0.10$ & $0.13\pm 0.10$ & $3.04\pm 0.14$ & $0.41\pm 0.08$ & \makecell{$0.944\pm 0.016$, $0.008\pm 0.014$, $0.003\pm 0.006$, $0.002\pm 0.005$, $0.035\pm 0.013$, $0.009\pm 0.012$} \\ 
GJ\,3378 & $3342\pm 31$ & $4.78\pm 0.05$ & $-0.10\pm 0.10$ & $-0.01\pm 0.10$ & $2.77\pm 0.10$ & $0.15\pm 0.05$ & \makecell{$0.937\pm 0.020$, $0.057\pm 0.024$, $0.003\pm 0.010$, $0.001\pm 0.003$, $0.002\pm 0.002$, $0.001\pm 0.002$} \\ 
GJ\,1105 & $3335\pm 31$ & $4.68\pm 0.06$ & $-0.04\pm 0.10$ & $-0.05\pm 0.10$ & $2.72\pm 0.17$ & $0.08\pm 0.05$ & \makecell{$0.986\pm 0.013$, $0.004\pm 0.012$, $0.004\pm 0.006$, $0.003\pm 0.004$, $0.002\pm 0.004$, $0.002\pm 0.003$} \\ 
Gl\,169.1A & $3311\pm 31$ & $4.69\pm 0.06$ & $0.12\pm 0.10$ & $-0.07\pm 0.10$ & $2.69\pm 0.15$ & $0.23\pm 0.05$ & \makecell{$0.931\pm 0.023$, $0.049\pm 0.026$, $0.012\pm 0.010$, $0.004\pm 0.004$, $0.003\pm 0.003$, $0.002\pm 0.002$} \\ 
Gl\,15B & $3287\pm 31$ & $4.83\pm 0.06$ & $-0.43\pm 0.10$ & $0.01\pm 0.10$ & $3.27\pm 0.18$ & $0.10\pm 0.07$ & \makecell{$0.984\pm 0.019$, $0.005\pm 0.019$, $0.002\pm 0.007$, $0.002\pm 0.005$, $0.004\pm 0.006$, $0.003\pm 0.005$} \\ 
PM~J21463$+$3813 & $3281\pm 31$ & $4.84\pm 0.06$ & $-0.41\pm 0.10$ & $0.21\pm 0.10$ & $3.46\pm 0.20$ & $0.54\pm 0.11$ & \makecell{$0.930\pm 0.021$, $0.007\pm 0.018$, $0.003\pm 0.009$, $0.002\pm 0.008$, $0.039\pm 0.019$, $0.019\pm 0.016$} \\ 
Gl\,699 & $3269\pm 31$ & $4.83\pm 0.06$ & $-0.46\pm 0.10$ & $0.12\pm 0.10$ & $3.44\pm 0.18$ & $0.50\pm 0.10$ & \makecell{$0.911\pm 0.027$, $0.037\pm 0.029$, $0.005\pm 0.011$, $0.002\pm 0.008$, $0.029\pm 0.014$, $0.016\pm 0.013$} \\ 
GJ\,1289 & $3242\pm 31$ & $4.75\pm 0.06$ & $-0.11\pm 0.10$ & $-0.03\pm 0.10$ & $2.86\pm 0.23$ & $1.01\pm 0.08$ & \makecell{$0.545\pm 0.032$, $0.409\pm 0.038$, $0.042\pm 0.017$, $0.002\pm 0.007$, $0.001\pm 0.006$, $0.001\pm 0.004$} \\ 
Gl\,447 & $3207\pm 31$ & $4.70\pm 0.06$ & $-0.21\pm 0.10$ & $-0.02\pm 0.10$ & $3.06\pm 0.19$ & $0.32\pm 0.09$ & \makecell{$0.912\pm 0.028$, $0.061\pm 0.030$, $0.005\pm 0.012$, $0.002\pm 0.008$, $0.017\pm 0.009$, $0.003\pm 0.007$} \\ 
GJ\,1151 & $3191\pm 31$ & $4.72\pm 0.06$ & $-0.16\pm 0.10$ & $-0.03\pm 0.10$ & $3.23\pm 0.15$ & $0.44\pm 0.10$ & \makecell{$0.851\pm 0.039$, $0.124\pm 0.042$, $0.005\pm 0.013$, $0.002\pm 0.007$, $0.013\pm 0.008$, $0.006\pm 0.006$} \\ 
GJ\,1103 & $3184\pm 31$ & $4.68\pm 0.06$ & $-0.04\pm 0.10$ & $-0.00\pm 0.10$ & $3.16\pm 0.19$ & $0.29\pm 0.09$ & \makecell{$0.958\pm 0.019$, $0.009\pm 0.017$, $0.002\pm 0.008$, $0.001\pm 0.006$, $0.017\pm 0.011$, $0.012\pm 0.010$} \\ 
Gl\,905 & $3074\pm 32$ & $4.68\pm 0.06$ & $0.05\pm 0.10$ & $-0.06\pm 0.10$ & $3.49\pm 0.19$ & $0.42\pm 0.12$ & \makecell{$0.817\pm 0.049$, $0.167\pm 0.054$, $0.009\pm 0.015$, $0.005\pm 0.007$, $0.002\pm 0.006$, $0.001\pm 0.005$} \\ 
GJ\,1286 & $2967\pm 32$ & $4.58\pm 0.06$ & $-0.20\pm 0.10$ & $-0.03\pm 0.10$ & $4.28\pm 0.20$ & $1.13\pm 0.17$ & \makecell{$0.610\pm 0.064$, $0.332\pm 0.074$, $0.010\pm 0.026$, $0.005\pm 0.013$, $0.018\pm 0.017$, $0.025\pm 0.014$} \\ 
GJ\,1002 & $2961\pm 32$ & $4.63\pm 0.06$ & $-0.26\pm 0.10$ & $-0.00\pm 0.10$ & $5.02\pm 0.24$ & $0.78\pm 0.17$ & \makecell{$0.837\pm 0.055$, $0.094\pm 0.058$, $0.006\pm 0.019$, $0.004\pm 0.012$, $0.027\pm 0.020$, $0.033\pm 0.019$} \\ 
\hline
\end{tabular}
		}
	\end{adjustbox}
\end{table*}

\subsection{Deriving an average magnetic field strength}

\begin{figure}
	\includegraphics[width=\columnwidth]{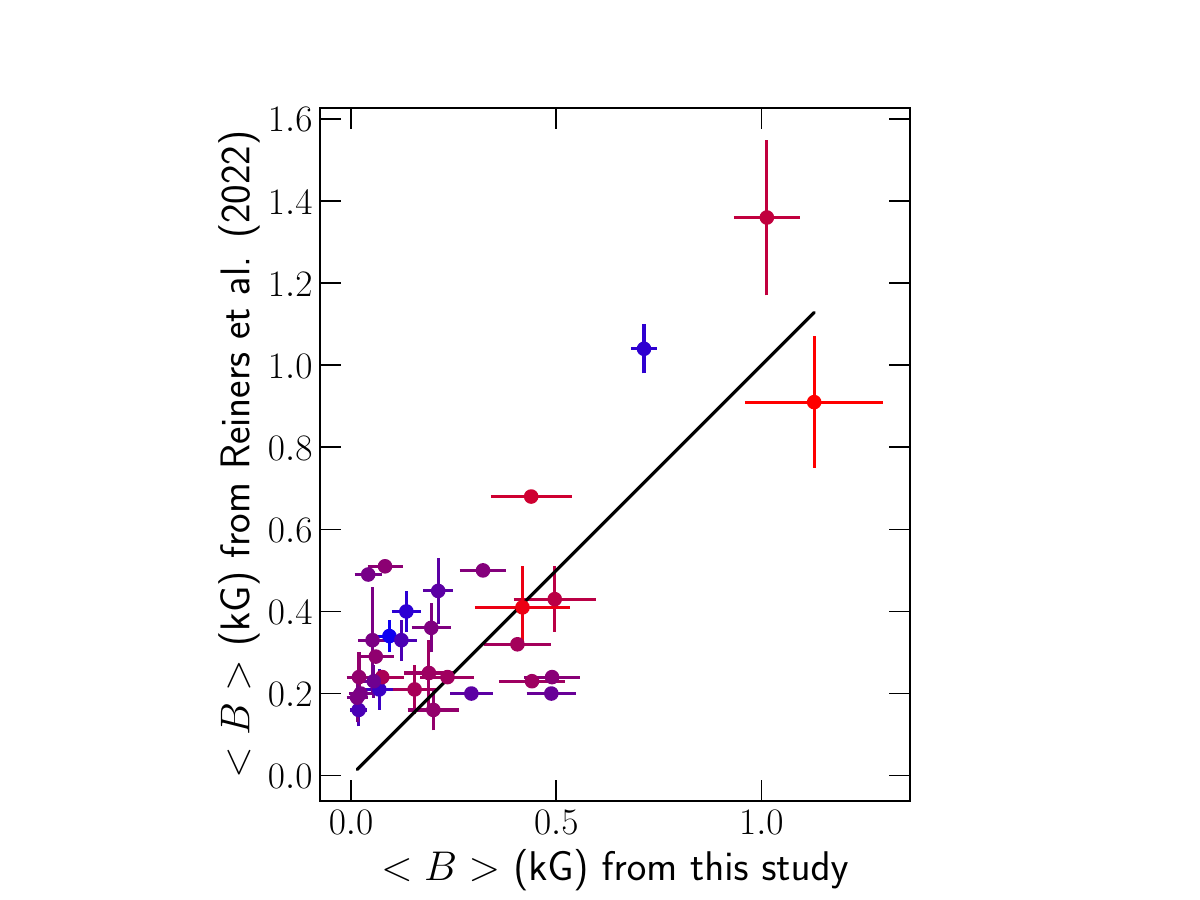}
	\caption{Comparison between our retrieved <$B$> and those of~\citet{reiners-2022}. The color gradient illustrates the effective temperature from cold (red) to hot (blue). The black line marks the equality.
	}
	\label{fig:bf-reiners}
\end{figure}

~\citet{reiners-2022} reported average magnetic fields for 33 of our 44 targets. Figure~\ref{fig:bf-reiners}~\&~\ref{fig:bf-reiners-label} presents a comparison between their estimates and ours. We find that the two sets of values are consistent with one another, with differences up to 0.4\,kG. Such differences result from several modelling steps, such as the choice of model used, the line list the analyses rely on, the fitting procedure and the choice of fundamental parameters for analysed targets (see Sec.~\ref{sec:impact-logg}). 
We also note that both our study and~\citet{reiners-2022} found Gl\,410, GJ\,1289 and GJ\,1286 to be the most magnetic stars among the 33 included in both works. This is also consistent with other activity diagnostics, such as the work of~\citet{schofer-2019}, who measured the H$\alpha$ equivalent widths of 30 targets included in our sample, and also found Gl\,410, GJ\,1289 and GJ\,1286 to be the most active.

\begin{figure}
	\includegraphics[width=\linewidth]{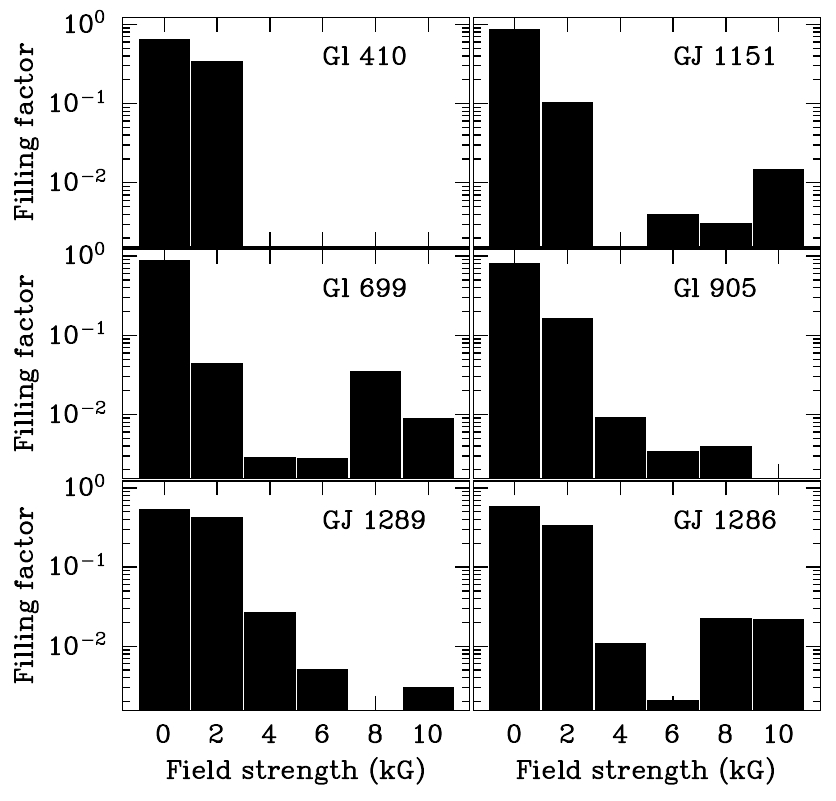}
	\caption{Distribution of the filling factors for 6 targets in our sample.}
	\label{fig:distribs}
\end{figure}

\begin{figure}
	\includegraphics[width=\columnwidth]{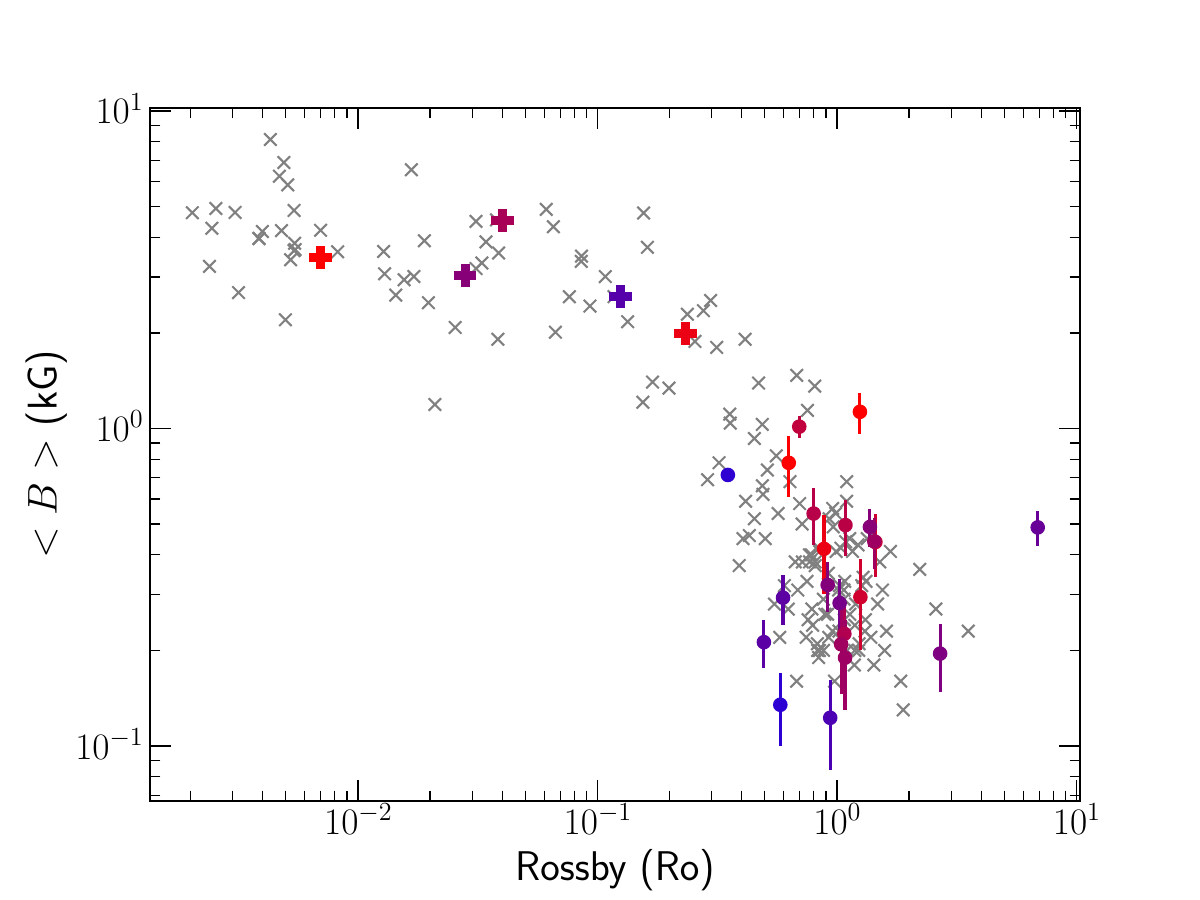}
	\caption{<$B$>--Ro diagram for {\paul the stars in our sample} relying on the rotation periods of Donati et al. (2023).  Only the stars for which a 3~$\sigma$ detection of the magnetic field was achieved are shown on the figure. The gray `x' markers show the results of~\citet{reiners-2022} for several stars not included in our sample. The plus symbols show the results of~\citet{cristofari-2023} for a few strongly magnetic targets.}
	\label{fig:rossby-donati}
\end{figure}

For most targets, the second magnetic component (2\,kG) accounts for most of the average magnetic field (see e.g., Fig.~\ref{fig:distribs})
Our estimated magnetic fields go as low as 0.06\,kG, reaching values below those reported by~\citet{reiners-2022} for their sample of stars (see Figs.~\ref{fig:bf-reiners}~\&~\ref{fig:rossby-donati}). Figures~\ref{fig:corner-gj1289},~\ref{fig:corner-gl411} and~\ref{fig:corner-gl699} present examples of posterior distributions obtained for all fitted parameters.

Two recent studies,~\citet{fouque-2023} and~\citet{donati-2023b}, have provided constraints on the rotation periods for 27 and 38 stars of our sample, respectively, relying on the detection of large-scale magnetic fields with SPIRou. {\paul Figures~\ref{fig:rossby-donati},~\ref{fig:rossby-donati-zoom}~\&~\ref{fig:rossby-donati-full} 
present} the 38 stars for which~\citet{donati-2023b} reports rotation periods in a <$B$>--Rossby number (Ro) diagram.

The relatively long rotation periods of the stars of our sample~\citep[][]{fouque-2023, donati-2023b}, as well as previous activity estimates~\citep{fouque-2018, schofer-2019}  and our retrieved average magnetic fields are all fully consistent with our targets falling in the unsaturated dynamo regime but do not clearly follow any trend with Ro.
The stars in our sample are found to have Rossby numbers below 2, with the exception of Gl\,411, whose reported rotation period is longer than any other star in this sample~\citep[$P_{\rm rot}=471\pm41$~d; ][]{fouque-2023}.

Through re-analyses performed by ignoring some of the magnetic components, we found that while the contributions of the 4, 6, 8 and 10\,kG fields are small, the first three are found to often significantly contribute to the fits to our spectra and to the overall derived magnetic field, especially in the case of the stars with strongest magnetic fields. Similar tests carried out with finer steps in magnetic field strength yield similar results, and we, therefore, chose to stick with steps of 2\,kG.

\section{Assessing the impact of magnetic fields on stellar characterization}
\label{sec:impact_on_atmo}

The analysis of magnetic stars with non-magnetic models can bias the estimation of atmospheric parameters~\citep{lopez-valdivia-2021,cristofari-2023}. In this section, we assess the extent to which magnetic fields impact the stellar characterization of the quiet targets included in our sample and vice-versa.

\subsection{Effect on derived atmospheric parameters}

\begin{figure}
	\includegraphics[width=\linewidth]{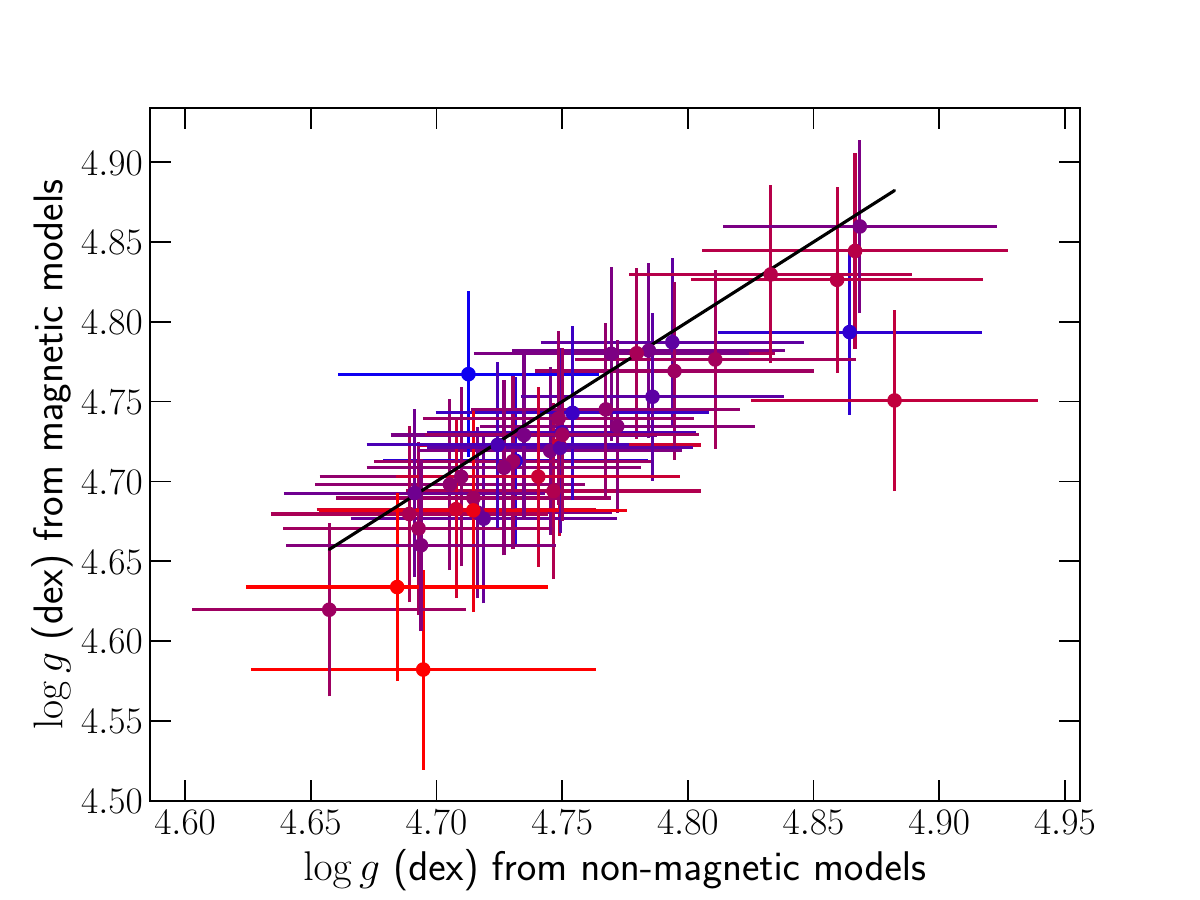}
	\caption{Comparison  between our $\logg$ estimates obtained with and without magnetic models.}
	\label{fig:logg-logg}
\end{figure}

We compare the results of our analysis relying on 6 magnetic components (0 to 10\,kG in steps of 2\,kG) to those obtained when relying only on non-magnetic models.
 Including magnetic fields in the models has virtually no impact on the estimated $\teff$, $\mh$ and $\afe$, with differences comparable to our formal error bars. We find slightly larger differences for $\logg$, with values on average 0.03\,dex lower when using magnetic models than without (see Fig.~\ref{fig:logg-logg}). Both surface gravity and magnetic fields are known to impact the width of spectral lines, which can partly account for this correlation. The observed discrepancies remain lower than our empirical error bar on $\logg$, estimated to be about 0.05\,dex. The largest discrepancies are observed for Gl\,410, GJ\,1286 and GJ\,1289, also found to be the most magnetic stars in our sample. We also retrieve similar macroturbulence estimates with non-magnetic and magnetic models, with the notable exception of Gl\,410, for which $\zrt$ is 0.5~\kms{} lower with magnetic models.

\begin{figure*}
	\includegraphics[width=\linewidth]{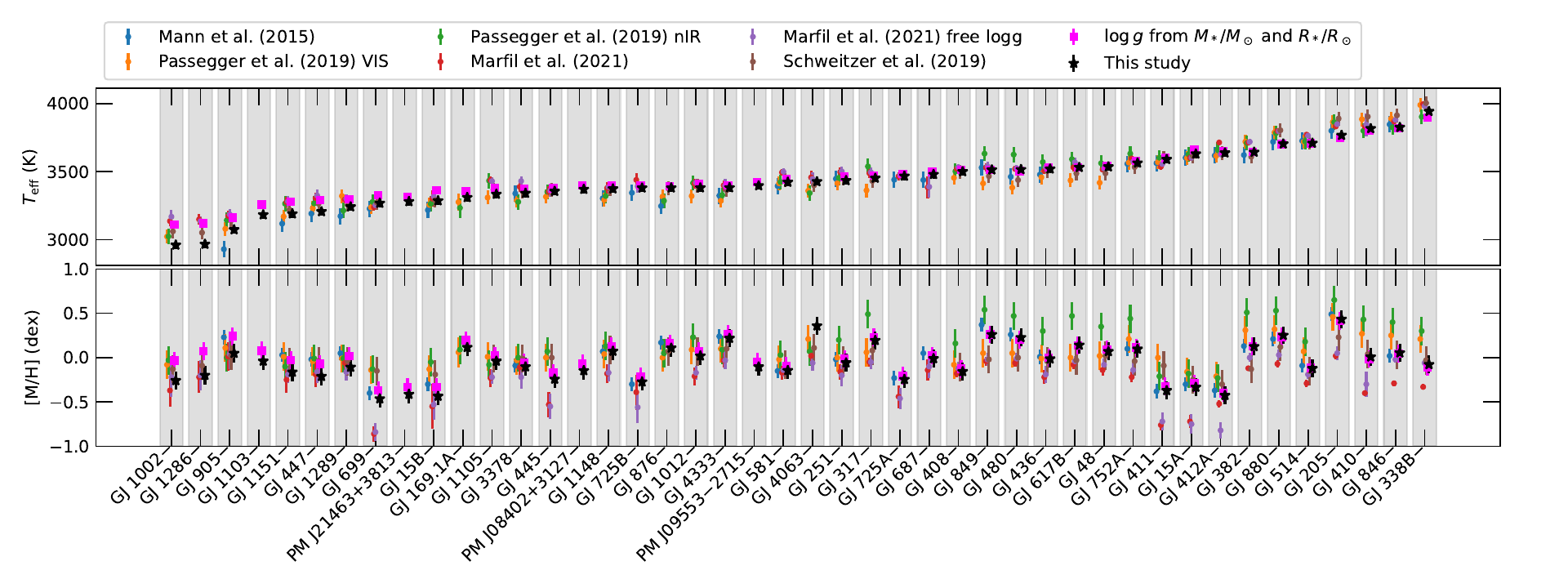}
	\caption{$\teff$ and $\mh$ estimates reported by several studies.}
	\label{fig:all-literature}
\end{figure*}

Like in our previous study~\citep{cristofari-2022b}, our derived atmospheric parameters are also found to be in good agreement with those of~\citet[][see Figs.~\ref{fig:res-teff},~\ref{fig:res-logg}~\&~\ref{fig:res-mh}]{mann-2015}. Here again we find $\logg$ to be among the most difficult parameter to constrain. We also find that our retrieved $\teff$ and $\mh$ are consistent with a large number of reported estimates (see Fig.~\ref{fig:all-literature}). Because $\logg$ is particularly challenging to constrain, several studies fixed its value from other quantities or imposed stringent priors in their analyses~\citep[e.g.,][]{passegger-2019, marfil-2021}.

\subsection{The impact of \texorpdfstring{$\logg$}{} on magnetic field estimates}
\label{sec:impact-logg}

\begin{figure}
	\includegraphics[width=\linewidth]{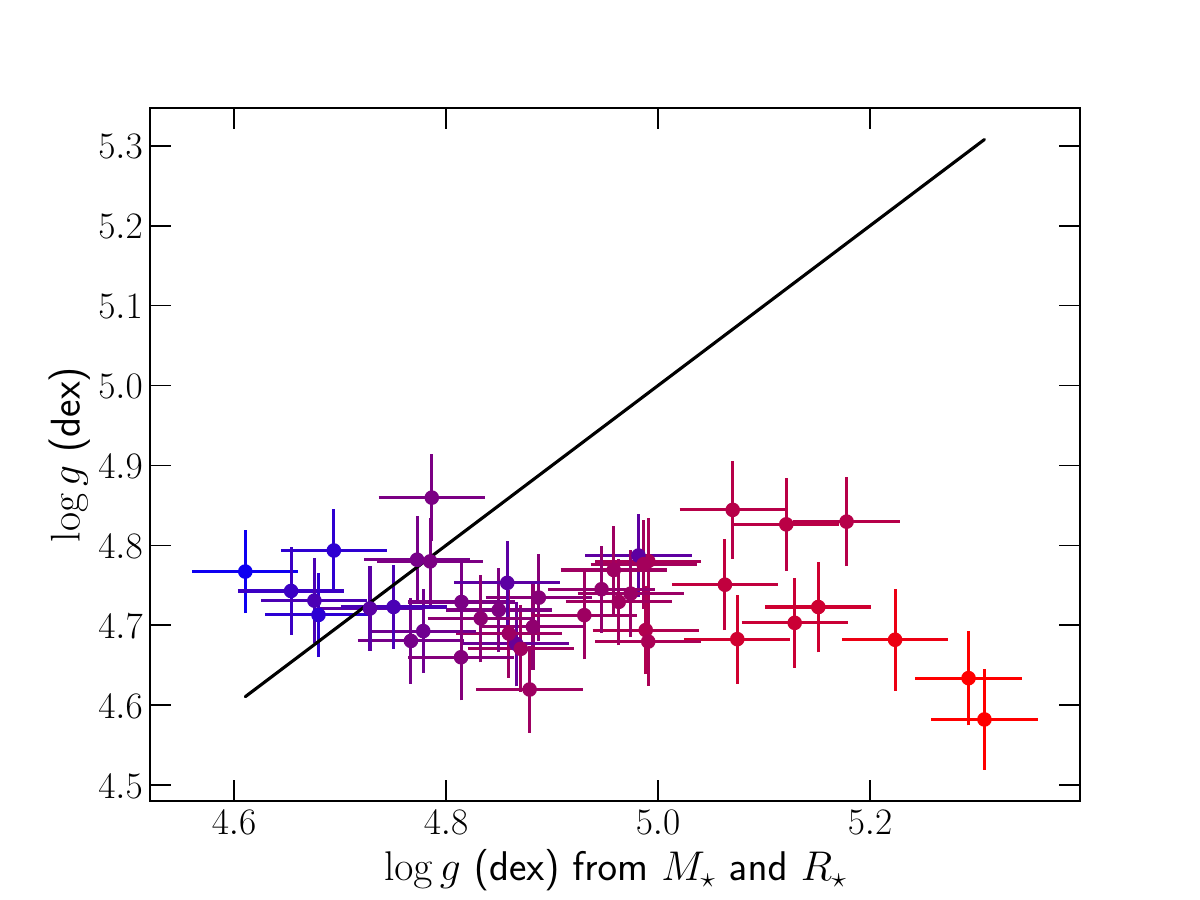}
	\caption{Comparison between our fitted $\logg$ and those derived from $M_\star$ and $R_\star$.}
	\label{fig:logg-autologg}
\end{figure}

Because the inclusion of magnetic fields impacts our $\logg$ estimates, we performed another analysis, fixing $\log{g}$ for each star. 
Rather than fixing the value of $\log{g}$ a priori,  we computed the radius of each star from effective temperature and luminosity relying on the Stefan-Boltzmann law, at each step of the MCMC analysis. We derived a mass for each target relying on the mass-magnitude-metallicity relation of~\citet{mann-2019} and computed $\logg$ from mass and radius. This approach allows us to ensure that $\logg$ remains consistent with $L_\star/L_\odot$ and $\teff$ throughout the analysis. Figure~\ref{fig:logg-autologg} 
presents a comparison between the $\logg$ obtained from $M_\star$ and $R_\star$ and those derived in our initial analysis. The full results recovered with the additional constraint on $\logg$ are presented in Table~\ref{tab:res-autologg}.

\begin{figure*}
	\includegraphics[width=\linewidth]{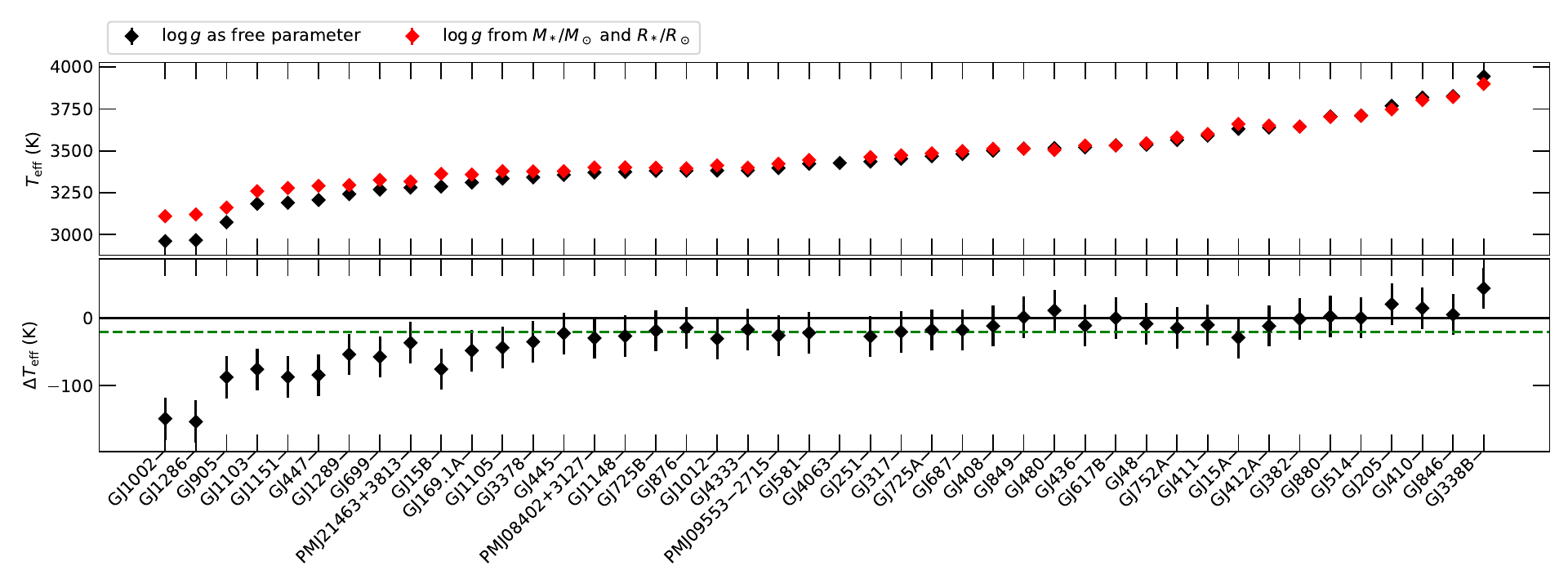}
	\caption{The top panel presents the comparison between our retrieved $T_{\rm eff}$ with and without fitting $\logg$ as a free parameter (black  and red, respectively). The bottom panel shows the residuals, with a median of $-22$\,K (green dashed line), and the zero difference line (black solid line).}
	\label{fig:compare-teff-autologg}
\end{figure*}

We compare the results obtained while fixing $\logg$ from $M_\star$ and $R_\star$ to those derived while fitting $\logg$ as a free parameter. The additional constraints on $\logg$ leads to larger $\teff$ values for the coolest stars  in our sample (see Fig.~\ref{fig:compare-teff-autologg}), with differences reaching up to 150\,K. Fixing the value of $\logg$ also impacts the estimation of $\mh$, with a RMS on the residuals of 0.07\,dex, lower than our empirical error bars on this parameter, but larger than our formal error bars. $\afe$ estimates are also found to be about 0.03\,dex larger with the additional constraint on $\logg$. Similar effects were reported in~\citet{cristofari-2022b} when fixing the $\logg$ for our targets with non-magnetic models.

\begin{figure*}
	\includegraphics[width=\linewidth]{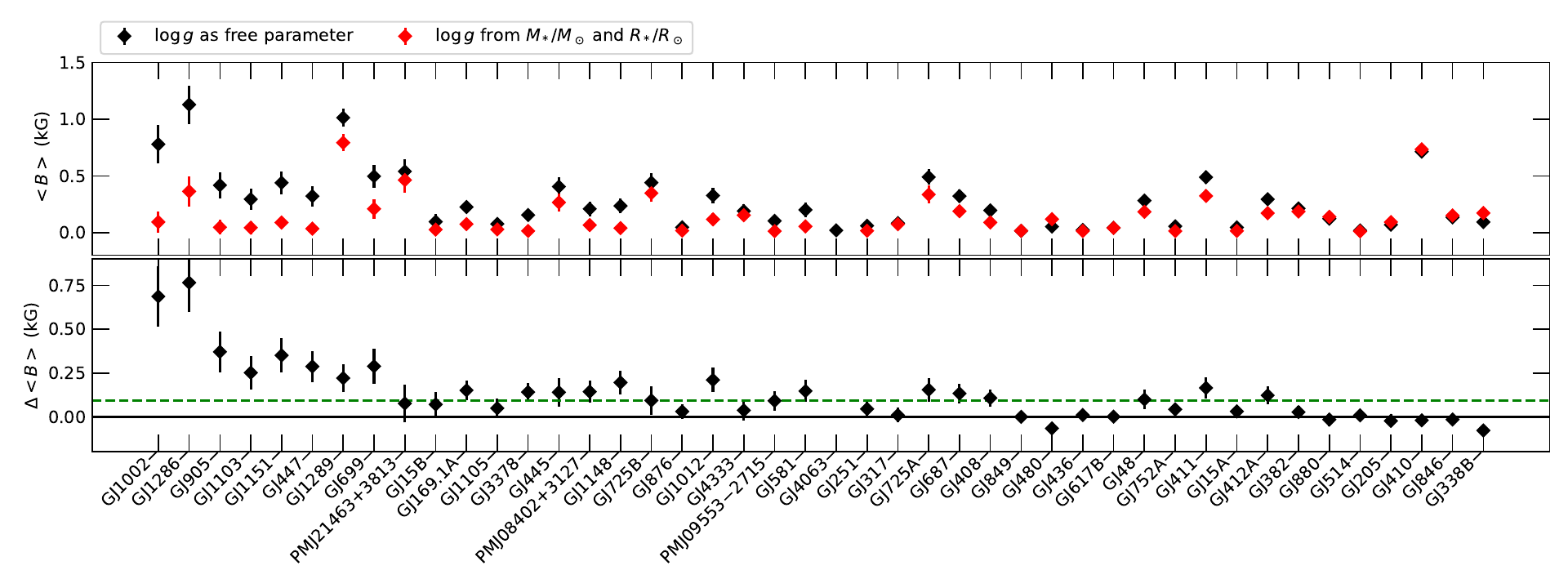}
	\caption{Same as Fig.~\ref{fig:compare-teff-autologg} for <$B$>. On the bottom panel, the dashed green line marks the median residual (of 0.1\,kG).}
	\label{fig:compare-bf-autologg}
\end{figure*}

\begin{figure*}
	\includegraphics[width=\linewidth]{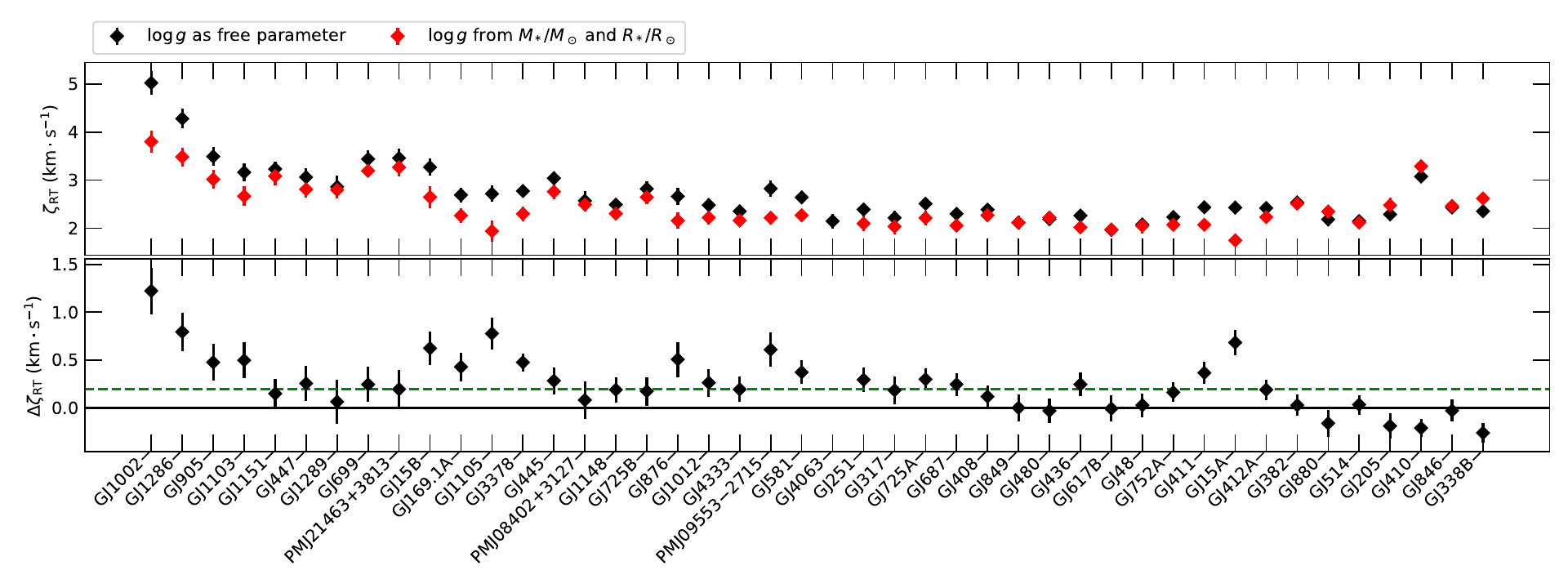}
	\caption{Same as Fig.~\ref{fig:compare-zrt-autologg} for $\zrt$. On the bottom panel, the dashed green line marks the median residual (of 0.2~\kms{}).}
	\label{fig:compare-zrt-autologg}
\end{figure*}

Fixing $\logg$ also leads to lower estimates of the average magnetic field, <$B$> (see Fig.~\ref{fig:compare-bf-autologg}) with estimates lower than those reported by~\citet{reiners-2022} for most stars common to both samples (see Fig.~\ref{fig:bf-reiners-autologg}). These discrepancies are most visible for some of the most magnetic targets of our sample, reaching up to 0.75\,kG, and illustrate how assumptions on atmospheric parameters can impact magnetic diagnostics. 
Fixing $\logg$ from $M_\star$ and $R_\star$ also leads to significantly smaller derived $\zrt$, particularly for the coolest stars of our sample (see Fig.~\ref{fig:compare-zrt-autologg}).

\begin{table*}
	\caption{Same as Table~\ref{tab:res} but with $\logg$ estimated from $\teff$ and $L_\star/L_\odot$ at each step of the MCMC process, for all stars in our sample but GJ\,4063.}
	\label{tab:res-autologg}
	\begin{adjustbox}{angle=90}   
		\resizebox{0.97\textheight}{!}{
	\begin{tabular}{cccccccc}
		\hline
		Star & $\teff$ (K) & $\logg$ (dex) & $\mh$ (dex) & $\afe$ (dex) & $\zeta_{\rm RT}$ (\kms{}) & <$B$> (kG) & \makecell{$f_0$, $f_2$, $f_4$, $f_6$, $f_8$, $f_{10}$} \\ 
		\hline
Gl\,338B & $3899\pm 30$ & $4.61\pm 0.05$ & $-0.11\pm 0.10$ & $-0.00\pm 0.10$ & $0.00\pm 0.10$ & $0.17\pm 0.03$ & \makecell{$0.919\pm 0.014$, $0.079\pm 0.015$, $0.000\pm 0.003$, $0.001\pm 0.002$, $0.001\pm 0.001$, $0.000\pm 0.001$} \\ 
Gl\,846 & $3821\pm 30$ & $4.68\pm 0.05$ & $0.05\pm 0.10$ & $-0.02\pm 0.10$ & $-0.02\pm 0.11$ & $0.15\pm 0.03$ & \makecell{$0.928\pm 0.016$, $0.070\pm 0.017$, $0.000\pm 0.003$, $0.001\pm 0.002$, $0.000\pm 0.001$, $0.000\pm 0.001$} \\ 
Gl\,410 & $3803\pm 30$ & $4.69\pm 0.05$ & $-0.02\pm 0.10$ & $0.01\pm 0.10$ & $0.02\pm 0.11$ & $0.73\pm 0.03$ & \makecell{$0.638\pm 0.014$, $0.359\pm 0.015$, $0.001\pm 0.004$, $0.001\pm 0.002$, $0.000\pm 0.001$, $0.000\pm 0.001$} \\ 
Gl\,205 & $3747\pm 31$ & $4.65\pm 0.05$ & $0.41\pm 0.10$ & $-0.08\pm 0.10$ & $0.03\pm 0.15$ & $0.09\pm 0.04$ & \makecell{$0.958\pm 0.019$, $0.041\pm 0.019$, $0.001\pm 0.003$, $0.000\pm 0.002$, $0.000\pm 0.001$, $0.001\pm 0.001$} \\ 
Gl\,514 & $3710\pm 30$ & $4.75\pm 0.05$ & $-0.12\pm 0.10$ & $0.04\pm 0.10$ & $0.00\pm 0.10$ & $0.01\pm 0.02$ & \makecell{$0.998\pm 0.006$, $0.001\pm 0.006$, $0.000\pm 0.002$, $0.000\pm 0.001$, $0.000\pm 0.001$, $0.000\pm 0.001$} \\ 
Gl\,880 & $3702\pm 30$ & $4.68\pm 0.05$ & $0.24\pm 0.10$ & $-0.05\pm 0.10$ & $0.03\pm 0.15$ & $0.14\pm 0.04$ & \makecell{$0.936\pm 0.017$, $0.061\pm 0.018$, $0.001\pm 0.003$, $0.001\pm 0.002$, $0.001\pm 0.001$, $0.000\pm 0.001$} \\ 
Gl\,382 & $3645\pm 31$ & $4.73\pm 0.05$ & $0.13\pm 0.10$ & $-0.02\pm 0.10$ & $0.01\pm 0.11$ & $0.19\pm 0.03$ & \makecell{$0.909\pm 0.016$, $0.090\pm 0.016$, $0.000\pm 0.002$, $0.000\pm 0.001$, $0.000\pm 0.001$, $0.000\pm 0.001$} \\ 
Gl\,412A & $3650\pm 30$ & $4.86\pm 0.05$ & $-0.40\pm 0.10$ & $0.12\pm 0.10$ & $-0.02\pm 0.10$ & $0.17\pm 0.04$ & \makecell{$0.948\pm 0.015$, $0.040\pm 0.017$, $0.002\pm 0.005$, $0.001\pm 0.003$, $0.006\pm 0.005$, $0.003\pm 0.003$} \\ 
Gl\,15A & $3660\pm 31$ & $4.98\pm 0.05$ & $-0.29\pm 0.10$ & $0.11\pm 0.10$ & $-0.10\pm 0.13$ & $0.01\pm 0.02$ & \makecell{$0.998\pm 0.004$, $0.000\pm 0.003$, $0.000\pm 0.001$, $0.001\pm 0.001$, $0.000\pm 0.001$, $0.001\pm 0.001$} \\ 
Gl\,411 & $3601\pm 30$ & $4.87\pm 0.05$ & $-0.32\pm 0.10$ & $0.20\pm 0.10$ & $-0.01\pm 0.13$ & $0.32\pm 0.05$ & \makecell{$0.964\pm 0.007$, $0.001\pm 0.005$, $0.002\pm 0.003$, $0.000\pm 0.003$, $0.007\pm 0.009$, $0.025\pm 0.009$} \\ 
Gl\,752A & $3579\pm 31$ & $4.78\pm 0.05$ & $0.11\pm 0.10$ & $-0.00\pm 0.10$ & $0.00\pm 0.11$ & $0.01\pm 0.02$ & \makecell{$0.996\pm 0.009$, $0.002\pm 0.009$, $0.000\pm 0.002$, $0.001\pm 0.001$, $0.000\pm 0.001$, $0.000\pm 0.001$} \\ 
Gl\,48 & $3545\pm 31$ & $4.77\pm 0.05$ & $0.09\pm 0.10$ & $0.08\pm 0.10$ & $0.10\pm 0.16$ & $0.18\pm 0.05$ & \makecell{$0.936\pm 0.017$, $0.056\pm 0.017$, $0.000\pm 0.005$, $0.001\pm 0.003$, $0.006\pm 0.003$, $0.002\pm 0.003$} \\ 
Gl\,617B & $3532\pm 31$ & $4.77\pm 0.05$ & $0.14\pm 0.10$ & $-0.01\pm 0.10$ & $-0.02\pm 0.11$ & $0.04\pm 0.03$ & \makecell{$0.982\pm 0.014$, $0.017\pm 0.014$, $0.000\pm 0.002$, $0.000\pm 0.001$, $0.000\pm 0.001$, $0.000\pm 0.001$} \\ 
Gl\,436 & $3531\pm 30$ & $4.81\pm 0.05$ & $-0.00\pm 0.10$ & $0.00\pm 0.10$ & $-0.06\pm 0.11$ & $0.01\pm 0.02$ & \makecell{$0.996\pm 0.005$, $0.002\pm 0.004$, $0.001\pm 0.002$, $0.000\pm 0.001$, $0.000\pm 0.001$, $0.000\pm 0.001$} \\ 
Gl\,480 & $3505\pm 30$ & $4.79\pm 0.05$ & $0.20\pm 0.10$ & $-0.02\pm 0.10$ & $-0.05\pm 0.13$ & $0.12\pm 0.04$ & \makecell{$0.942\pm 0.020$, $0.057\pm 0.020$, $0.001\pm 0.002$, $0.000\pm 0.001$, $0.000\pm 0.001$, $0.000\pm 0.001$} \\ 
Gl\,849 & $3512\pm 31$ & $4.79\pm 0.05$ & $0.26\pm 0.10$ & $-0.04\pm 0.10$ & $0.04\pm 0.13$ & $0.02\pm 0.02$ & \makecell{$0.995\pm 0.009$, $0.004\pm 0.008$, $0.000\pm 0.002$, $0.000\pm 0.001$, $0.000\pm 0.001$, $0.000\pm 0.001$} \\ 
Gl\,408 & $3512\pm 31$ & $4.85\pm 0.05$ & $-0.13\pm 0.10$ & $0.04\pm 0.10$ & $0.01\pm 0.12$ & $0.09\pm 0.04$ & \makecell{$0.961\pm 0.017$, $0.036\pm 0.018$, $0.002\pm 0.004$, $0.000\pm 0.002$, $0.000\pm 0.002$, $0.001\pm 0.002$} \\ 
Gl\,687 & $3498\pm 30$ & $4.81\pm 0.05$ & $0.02\pm 0.10$ & $0.08\pm 0.10$ & $-0.03\pm 0.11$ & $0.19\pm 0.05$ & \makecell{$0.965\pm 0.013$, $0.016\pm 0.014$, $0.001\pm 0.004$, $0.001\pm 0.003$, $0.011\pm 0.006$, $0.006\pm 0.005$} \\ 
Gl\,725A & $3485\pm 31$ & $4.89\pm 0.05$ & $-0.21\pm 0.10$ & $0.17\pm 0.10$ & $-0.10\pm 0.15$ & $0.34\pm 0.08$ & \makecell{$0.947\pm 0.012$, $0.017\pm 0.010$, $0.002\pm 0.005$, $0.001\pm 0.004$, $0.024\pm 0.012$, $0.010\pm 0.013$} \\ 
Gl\,317 & $3473\pm 31$ & $4.83\pm 0.05$ & $0.23\pm 0.10$ & $-0.03\pm 0.10$ & $2.04\pm 0.16$ & $0.07\pm 0.03$ & \makecell{$0.983\pm 0.011$, $0.011\pm 0.011$, $0.002\pm 0.002$, $0.001\pm 0.001$, $0.001\pm 0.001$, $0.001\pm 0.001$} \\ 
Gl\,251 & $3463\pm 31$ & $4.88\pm 0.05$ & $-0.02\pm 0.10$ & $-0.00\pm 0.10$ & $0.06\pm 0.16$ & $0.02\pm 0.03$ & \makecell{$0.997\pm 0.006$, $0.001\pm 0.005$, $0.001\pm 0.002$, $0.001\pm 0.002$, $0.000\pm 0.002$, $0.000\pm 0.002$} \\ 
Gl\,581 & $3445\pm 31$ & $4.95\pm 0.05$ & $-0.09\pm 0.10$ & $0.02\pm 0.10$ & $-0.03\pm 0.12$ & $0.05\pm 0.04$ & \makecell{$0.982\pm 0.010$, $0.016\pm 0.009$, $0.000\pm 0.003$, $0.000\pm 0.002$, $0.002\pm 0.003$, $0.001\pm 0.003$} \\ 
PM~J09553$-$2715 & $3423\pm 30$ & $4.97\pm 0.05$ & $-0.05\pm 0.10$ & $0.00\pm 0.10$ & $-0.10\pm 0.12$ & $0.01\pm 0.03$ & \makecell{$0.997\pm 0.008$, $0.002\pm 0.007$, $0.001\pm 0.002$, $0.000\pm 0.002$, $0.000\pm 0.002$, $0.000\pm 0.002$} \\ 
GJ\,4333 & $3400\pm 30$ & $4.86\pm 0.05$ & $0.26\pm 0.10$ & $-0.01\pm 0.10$ & $0.09\pm 0.11$ & $0.15\pm 0.03$ & \makecell{$0.938\pm 0.015$, $0.055\pm 0.018$, $0.002\pm 0.007$, $0.003\pm 0.002$, $0.001\pm 0.002$, $0.001\pm 0.002$} \\ 
GJ\,1012 & $3412\pm 31$ & $4.88\pm 0.05$ & $0.08\pm 0.10$ & $0.06\pm 0.10$ & $0.07\pm 0.15$ & $0.12\pm 0.05$ & \makecell{$0.985\pm 0.008$, $0.003\pm 0.005$, $0.001\pm 0.003$, $0.000\pm 0.003$, $0.003\pm 0.004$, $0.009\pm 0.005$} \\ 
Gl\,876 & $3395\pm 30$ & $4.93\pm 0.05$ & $0.16\pm 0.10$ & $-0.03\pm 0.10$ & $-0.01\pm 0.17$ & $0.02\pm 0.02$ & \makecell{$0.996\pm 0.006$, $0.002\pm 0.006$, $0.000\pm 0.002$, $0.001\pm 0.001$, $0.000\pm 0.001$, $0.000\pm 0.001$} \\ 
Gl\,725B & $3399\pm 30$ & $4.96\pm 0.05$ & $-0.21\pm 0.10$ & $0.17\pm 0.10$ & $0.04\pm 0.15$ & $0.35\pm 0.08$ & \makecell{$0.962\pm 0.013$, $0.003\pm 0.011$, $0.001\pm 0.005$, $0.001\pm 0.004$, $0.002\pm 0.011$, $0.032\pm 0.011$} \\ 
GJ\,1148 & $3401\pm 30$ & $4.87\pm 0.05$ & $0.12\pm 0.10$ & $0.03\pm 0.10$ & $0.01\pm 0.11$ & $0.04\pm 0.05$ & \makecell{$0.992\pm 0.009$, $0.004\pm 0.007$, $0.000\pm 0.003$, $0.000\pm 0.002$, $0.001\pm 0.004$, $0.002\pm 0.004$} \\ 
PM~J08402$+$3127 & $3400\pm 30$ & $4.96\pm 0.05$ & $-0.07\pm 0.10$ & $0.05\pm 0.10$ & $0.12\pm 0.15$ & $0.07\pm 0.04$ & \makecell{$0.987\pm 0.012$, $0.004\pm 0.012$, $0.003\pm 0.004$, $0.001\pm 0.002$, $0.002\pm 0.003$, $0.002\pm 0.003$} \\ 
Gl\,445 & $3379\pm 31$ & $4.99\pm 0.05$ & $-0.17\pm 0.10$ & $0.18\pm 0.10$ & $0.02\pm 0.16$ & $0.27\pm 0.08$ & \makecell{$0.970\pm 0.012$, $0.002\pm 0.008$, $0.002\pm 0.004$, $0.001\pm 0.004$, $0.004\pm 0.009$, $0.022\pm 0.009$} \\ 
GJ\,3378 & $3377\pm 30$ & $4.99\pm 0.05$ & $-0.05\pm 0.10$ & $0.02\pm 0.10$ & $-0.03\pm 0.15$ & $0.01\pm 0.04$ & \makecell{$0.996\pm 0.011$, $0.002\pm 0.010$, $0.000\pm 0.003$, $0.000\pm 0.002$, $0.001\pm 0.002$, $0.000\pm 0.002$} \\ 
GJ\,1105 & $3378\pm 30$ & $4.99\pm 0.05$ & $0.02\pm 0.10$ & $-0.01\pm 0.10$ & $-0.11\pm 0.22$ & $0.03\pm 0.03$ & \makecell{$0.995\pm 0.006$, $0.001\pm 0.005$, $0.000\pm 0.002$, $0.001\pm 0.002$, $0.001\pm 0.002$, $0.001\pm 0.002$} \\ 
Gl\,169.1A & $3359\pm 31$ & $4.99\pm 0.05$ & $0.20\pm 0.10$ & $-0.04\pm 0.10$ & $2.27\pm 0.15$ & $0.07\pm 0.03$ & \makecell{$0.985\pm 0.009$, $0.008\pm 0.008$, $0.003\pm 0.003$, $0.002\pm 0.002$, $0.001\pm 0.001$, $0.001\pm 0.001$} \\ 
Gl\,15B & $3362\pm 31$ & $5.18\pm 0.05$ & $-0.33\pm 0.10$ & $0.06\pm 0.10$ & $-0.11\pm 0.23$ & $0.03\pm 0.03$ & \makecell{$0.992\pm 0.009$, $0.006\pm 0.008$, $0.001\pm 0.003$, $0.001\pm 0.002$, $0.001\pm 0.002$, $0.000\pm 0.002$} \\ 
PM~J21463$+$3813 & $3317\pm 31$ & $5.07\pm 0.05$ & $-0.33\pm 0.10$ & $0.25\pm 0.10$ & $0.08\pm 0.20$ & $0.46\pm 0.11$ & \makecell{$0.950\pm 0.016$, $0.001\pm 0.011$, $0.002\pm 0.007$, $0.001\pm 0.006$, $0.003\pm 0.013$, $0.042\pm 0.015$} \\ 
Gl\,699 & $3326\pm 31$ & $5.12\pm 0.05$ & $-0.37\pm 0.10$ & $0.17\pm 0.10$ & $-0.02\pm 0.13$ & $0.21\pm 0.08$ & \makecell{$0.969\pm 0.018$, $0.008\pm 0.016$, $0.002\pm 0.006$, $0.002\pm 0.005$, $0.006\pm 0.007$, $0.013\pm 0.008$} \\ 
GJ\,1289 & $3296\pm 30$ & $5.06\pm 0.05$ & $0.02\pm 0.10$ & $0.00\pm 0.10$ & $0.02\pm 0.17$ & $0.79\pm 0.08$ & \makecell{$0.616\pm 0.035$, $0.379\pm 0.037$, $0.002\pm 0.009$, $0.001\pm 0.004$, $0.000\pm 0.004$, $0.002\pm 0.003$} \\ 
Gl\,447 & $3291\pm 30$ & $5.13\pm 0.05$ & $-0.07\pm 0.10$ & $0.03\pm 0.10$ & $0.08\pm 0.17$ & $0.03\pm 0.05$ & \makecell{$0.991\pm 0.011$, $0.006\pm 0.010$, $0.001\pm 0.004$, $0.000\pm 0.003$, $0.001\pm 0.003$, $0.001\pm 0.003$} \\ 
GJ\,1151 & $3278\pm 31$ & $5.15\pm 0.05$ & $-0.03\pm 0.10$ & $0.01\pm 0.10$ & $0.40\pm 0.20$ & $0.09\pm 0.05$ & \makecell{$0.972\pm 0.018$, $0.021\pm 0.018$, $0.002\pm 0.005$, $0.001\pm 0.003$, $0.001\pm 0.003$, $0.002\pm 0.003$} \\ 
GJ\,1103 & $3259\pm 31$ & $5.07\pm 0.05$ & $0.08\pm 0.10$ & $0.07\pm 0.10$ & $0.05\pm 0.20$ & $0.04\pm 0.06$ & \makecell{$0.993\pm 0.011$, $0.002\pm 0.008$, $0.000\pm 0.003$, $0.000\pm 0.003$, $0.000\pm 0.004$, $0.003\pm 0.004$} \\ 
Gl\,905 & $3161\pm 31$ & $5.22\pm 0.05$ & $0.24\pm 0.10$ & $-0.01\pm 0.10$ & $-0.02\pm 0.19$ & $0.05\pm 0.06$ & \makecell{$0.987\pm 0.024$, $0.007\pm 0.025$, $0.003\pm 0.006$, $0.000\pm 0.003$, $0.001\pm 0.003$, $0.001\pm 0.003$} \\ 
GJ\,1286 & $3121\pm 31$ & $5.31\pm 0.05$ & $0.08\pm 0.10$ & $0.04\pm 0.10$ & $-0.08\pm 0.20$ & $0.36\pm 0.13$ & \makecell{$0.841\pm 0.060$, $0.149\pm 0.064$, $0.002\pm 0.012$, $0.006\pm 0.006$, $0.001\pm 0.005$, $0.001\pm 0.005$} \\ 
GJ\,1002 & $3110\pm 31$ & $5.29\pm 0.05$ & $-0.03\pm 0.10$ & $0.09\pm 0.10$ & $0.07\pm 0.23$ & $0.09\pm 0.09$ & \makecell{$0.979\pm 0.029$, $0.008\pm 0.028$, $0.006\pm 0.010$, $0.005\pm 0.006$, $0.002\pm 0.006$, $0.001\pm 0.005$} \\ 
\hline
	\end{tabular}
}
\end{adjustbox}
\end{table*}

\subsection{The \texorpdfstring{$\zrt$--$\teff$}{} relation}

\begin{figure}
	\includegraphics[width=\columnwidth]{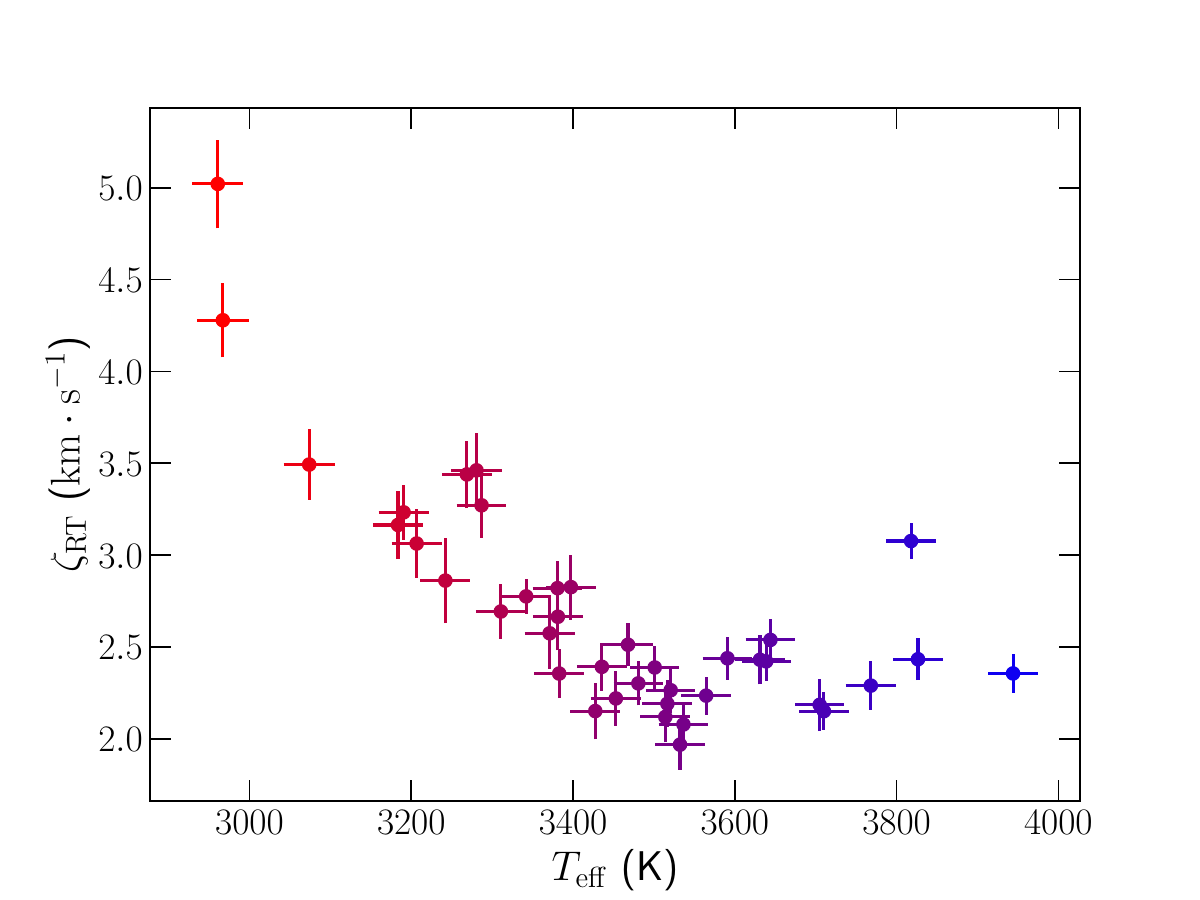}
	\caption{Retrieved $\zrt$ as a function of $\teff$ for the 44 targets of our sample. The color of the points vary with $\teff$ from coldest (red) to hottest (blue).}
	\label{fig:macro-teff}
\end{figure}

Throughout our analysis, we chose to neglect rotational broadening, and only include a radial-tangential macroturbulence to fit our models to SPIRou templates. Figures~\ref{fig:macro-teff} and~\ref{fig:macro-teff-label} present our retrieved $\zrt$ as a function of $\teff$. Our estimates of $\zrt$ tend to decrease with increasing $\teff$. An opposite trend is typically reported by previous studies~\citep{doyle-2014, brewer-2016} although reliable constraints are difficult to obtain for cool M dwarfs. We reprocessed our data fixing $\zrt=2.5$~\kms{}, and found the impact on our derived atmospheric parameters to be quite small, except for the two coolest stars in our sample, GJ\,1286 and GJ\,1002, for which $\teff$ increased by 50\,K and $\logg$ by 0.1\,dex. The impact on <$B$> was negligible, with differences of less than 0.05\,kG. Thus, the choice of $\zrt$ does not appear to impact or magnetic field estimates, although the $\zrt$ estimates should be taken with caution, in particular for the coolest stars in our sample, as systematics in the models could bias its derivation.

\section{Discussion and conclusions}
\label{sec:conclusions}

In this paper, we describe a revised analysis of targets previously studied in~\citet{cristofari-2022b}, with the tools introduced in~\citet{cristofari-2023}. Relying on spectra computed with \texttt{ZeeTurbo} from MARCS model atmospheres we investigate the small-scale magnetic fields of 44 moderately to weakly active M dwarfs monitored in the context of the SPIRou Legacy Survey.

Our approach consists in inferring $\teff$, $\logg$, $\mh$, $\afe$ and the magnetic filling factors at once, allowing us to improve our fits to the data, to derive the magnetic parameters of our sample stars, and to assess the impact of magnetic fields on stellar characterization. We rely on a line list containing about 30 atomic lines, 10 OH lines and 30 CO lines. This list contains transitions with different Land\'e factors ranging from 0 to 2.5. This allows us to disentangle the effect of magnetic fields from those of atmospheric parameter and macroturbulence, by matching the shapes of lines affected in different ways by Zeeman broadening and intensification.
Our tools and method were successfully applied to a few strongly magnetic stars observed with SPIRou~\citep{cristofari-2023}.

We assessed the impact of several modeling assumptions on our results, including the effect of varying atmospheric parameters. In particular, we found our $\logg$ estimates are lower than those expected from $\teff$ and $L_\star/L_\odot$ using empirical relations, and show that fixing the value of $\logg$ leads to significant differences in the retrieved <$B$>, reaching up to 0.75\,kG for GJ\,1286. Fixing $\logg$ also leads to increased $\teff$  and $\mh$ for the coolest stars in our sample, such as GJ\,1286 and GJ\,1002. 
In this case, the average magnetic field of GJ\,1286 is no longer among the highest in our sample, and drops below that of Gl\,725B and Gl\,752A.
Fixing $\logg$ also impact our <$B$> estimates for most stars also, that move further down in the <$B$>--Ro diagram (see Fig.~\ref{fig:rossby-donati-autologg}), below the estimates of~\citet{reiners-2022}.
We note that GJ\,1286 was reported to be among the most active~\citep{fouque-2018, schofer-2019} and most magnetic~\citep{reiners-2022} stars in our sample, but its rotation period was estimated to be among the longest~(203$\pm$21 and 178$\pm15$~days from~\citealt{fouque-2023} and~\citealt{donati-2023b}, respectively). Further improvements in the models of M dwarfs spectra are needed in order to improve the magnetic characterization of such stars.
Deriving accurate constraints on atmospheric parameters remains a challenge, particularly for cool stars, with larger dispersions observed in literature estimates~(see Fig.~\ref{fig:all-literature}). 
The surface gravity is known to be particularly challenging to constrain from high-resolution spectra~\citep{cristofari-2022, cristofari-2022b}, and several studies tend to fix its value from prior knowledge~\citep{passegger-2019, marfil-2021}. 
None the less, we derived atmospheric parameters consistent with previous studies, and found that our initial process with unconstrained $\logg$ would rather lead to over-estimates rather than under-estimates on <$B$>.

We further assessed the impact of $\zrt$ on our results, and found that the best fits to our data were obtained for decreasing $\zrt$ with increasing $\teff$. Fixing $\zrt$ had little impact on our derived <$B$>, further demonstrating that the two effects can be disentangled from high-resolution spectra.
Our approach consisted in neglecting rotation, motivated by the long rotation periods implying equatorial rotational velocities lower than $1.5$~\kms{} for most stars. For the fastest rotator in our sample, Gl\,410,~\citet{cristofari-2023} derived $\zrt=2.7\pm 0.1$~\kms{} for Gl\,410, assuming $\vsini=1.5$~\kms{}, while we obtained $3.1\pm0.1$~\kms{} in the present work, indicating that the effect is indeed small.

Our analysis provides average magnetic fields estimates that are consistent with previous measurements by~\citet{reiners-2022}
for the 33 stars common to both samples, with a RMS difference of 0.2\,kG.
The differences between the two studies can be due to several effects, such as temporal variations of magnetic fields, differences in spectral synthesis, line selection, continuum adjustment or assumed atmospheric parameters.
For the majority of our targets, the 2\,kG component accounts for most of the average magnetic field strength, suggesting that these stars do not host stronger large-scale magnetic fields. The impact of the stronger field components is not negligible, however, and relying solely on the 2\,kG component leads to lower magnetic field estimates.

Placing our targets in a <$B$>--Ro diagram, we found that our estimated magnetic fields and the reported rotation periods~\citep[][]{fouque-2023, donati-2023b} are consistent with our targets falling in the unsaturated dynamo regime (see Figs.~\ref{fig:rossby-donati},~\ref{fig:rossby-donati-zoom}~\&~\ref{fig:rossby-donati-full}), although no clear trend with Ro is observed. 
{\paul 
We find that the stars whose magnetic fields were detected at a 3$\sigma$ level, apart for a few outliers hosting particularly strong fields (e.g. GJ1286, Lehmann et al., submitted), show small-scale field that somewhat match the \text{$<\!B\!>$~--~Ro} relationship found in~\citep{reiners-2022},
 the dispersion of our $<\!B\!>$ estimates with respect to this relation being similar in both studies (see Fig.~\ref{fig:rossby-donati}).
	We note that several stars with Rossby numbers ranging between 0.3 to 1.0 show low average magnetic fields, with values consistent with 0\,kG within 3\,$\sigma$ (see Fig.~\ref{fig:rossby-donati-full}). The spectra of those targets are well reproduced by non-magnetic models (Fig.~\ref{fig:fits-2}), despite their rotation periods being similar to those of more magnetic stars. }

{\paul We find that our $\logg$ estimates are similar for almost all stars in our sample, although empirically calibrated relations~\citep[e.g.,][]{mann-2019}, photometric measurements, evolutionary models~\citep{feiden-2012, baraffe-2015} and interferometric measurements~\citep{boyajian-2012} suggest that $\logg$ should increase with decreasing $\teff$. Nonetheless, setting priors on $\log{g}$ does not provide $\teff$ or $\mh$ estimates more consistent with the literature, and leads to significant increases in $\teff$ for the coolest stars in our sample (Fig.~\ref{fig:compare-teff-autologg}), suggesting that the effect could arise from systematic differences between models and observations.
	 We find that}
fixing $\logg$ from $\teff$ and $L_\star/L_\odot$ leads to a larger dispersion in the <$B$> estimates for stars with similar Ro{\paul , with the magnetic field of several targets passing below the 3\,$\sigma$ detection threshold (see Fig.~\ref{fig:rossby-donati-autologg}).} 

\begin{figure}
	\includegraphics[width=\linewidth]{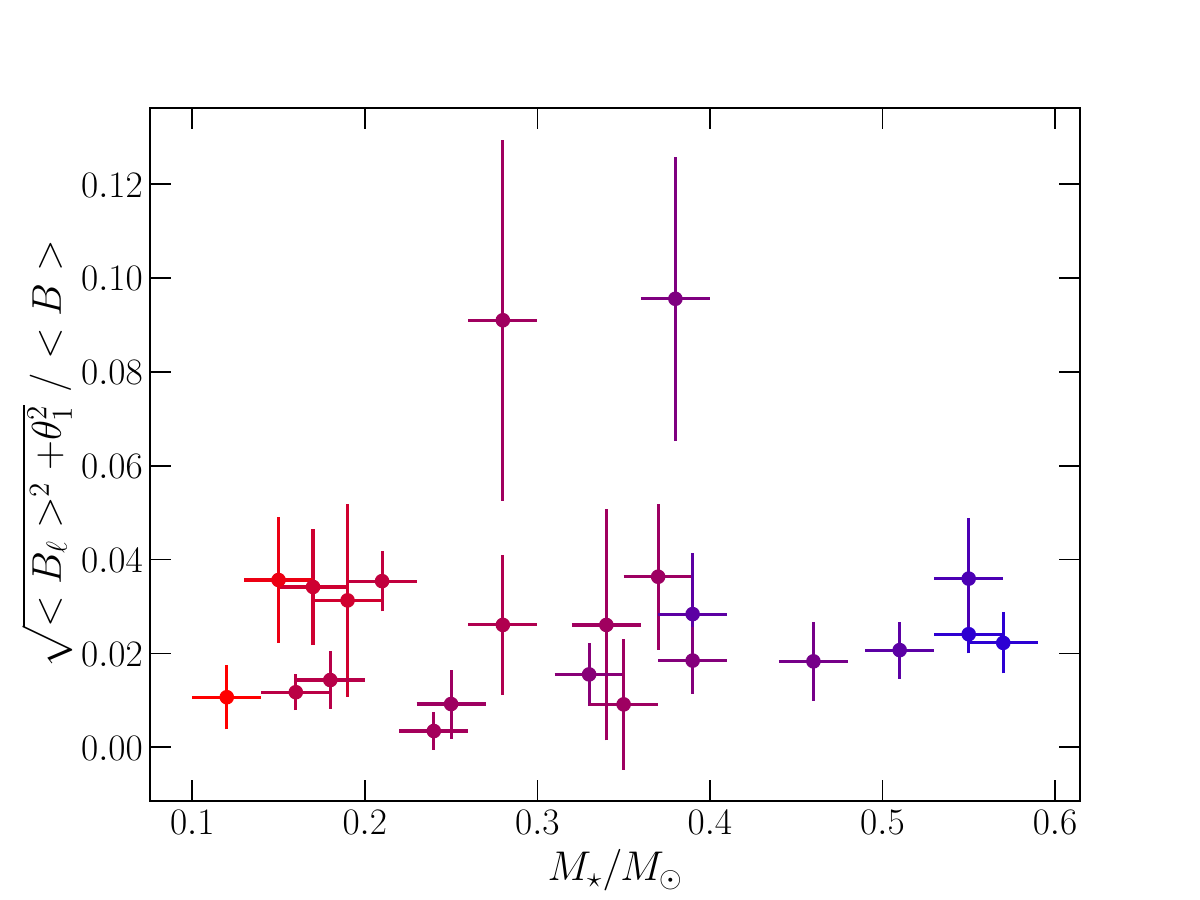}
	\caption{Ratio between the large scale magnetic field proxy ($\sqrt{{\rm <}B_\ell{\rm >}^2 + \theta_1^2}$ where $B_\ell$\ is the average longitudinal field and $\theta_1$ the semi-amplitude of its modulation, see~\citealt{donati-2023b}) and our <$B$> estimates as a function of $M_\star / M_\odot$. Only stars for which a 3~$\sigma$ detection of the magnetic field was achieved are shown on this figure.}
	\label{fig:bl-mass}
\end{figure}

	We compared our average magnetic fields estimates to the large-scale field measurements reported by~\citet{donati-2023b}. We choose to compare our values to the square root of the quadratic sum of the average longitudinal field (<$B_\ell$>) and the amplitude of the large scale field modulation with time~($\theta_1$,~\citealt{donati-2023b}, see Figs.~\ref{fig:bl-mass}~\&~\ref{fig:bl-mass-label}). {\paul We find that the longitudinal field accounts for up to 10\% of the total magnetic field,  consistent with previous results for low mass stars with non-axisymmetric magnetic fields~\citep{morin-2010, kochukhov-2021}. Given that the longitudinal field is typically about 3 or 4 times smaller than the large-scale magnetic field as a result of projection effects~\citep[e.g.,][]{preston-1967, kochukhov-2021, donati-2023b}, those results indicate that the large-scale magnetic field of the stars in our sample amounts to less than 30~--~40\% of <$B$>.}

Our analysis will benefit from updates of line lists and model atmospheres that recent and ongoing works aim at improving,  in particular for cool stars~\citep[e.g.,][]{valyavin-2004, stift-2016, jarvinen-2020, olander-2021, gerber-2023}. Adding lines to our analysis, including magnetically-sensitive molecular lines, will help us further improve constraints on magnetic field measurements~(Crozet et al. in prep), and on $\log{g}$.
Our analysis was performed on template spectra built from several observations. A detailed analysis relying on high-resolution data acquired for each night will allow us to study the evolution of small-scale magnetic fields over time, and in particular to search for rotational modulation of small-scale magnetic fields of M dwarfs, similarly to what was done for AU~Mic~\citep{donati-2023a}. Our tools will also allow us to study pre-main-sequence stars, whose characterization will require further implementation, but is of great interest to the study of planet formation~\citep{flores-2020, lopez-valdivia-2021, lopez-valdivia-2023}.

\section*{Acknowledgements}

This project received funding from the European Research Council (ERC, grant 740651, NewWorlds) under the innovation research and innovation program H2020. 
We also acknowledge funding from the French National Research Agency (ANR, grant ANR18CE310019 / SPlaSH) and from the « Origin of Life » project of the Grenoble-Alpes University (grant ANR-15-IDEX-02).
TM acknowledges financial support from the Spanish Ministry of Science and Innovation (MICINN) through the Spanish State Research Agency, under the Severo Ochoa Programme 2020–2023 (CEX2019-000920-S) as well as support from the ACIISI, Consejer\'ia de Econom\'iıa, Conocimiento y Empleo del Gobiernode Canarias, and the European Regional Development Fund (ERDF) under grant with reference PROID2021010128. 

This work is based on observations obtained at the Canada-France-Hawaii Telescope (CFHT), operated by the National Re- search Council (NRC) of Canada, the Institut National des Sciences de l’Univers of the Centre National de la Recherche Scientifique (CNRS) of France, and the University of Hawaii. The observations at the CFHT were performed with care and respect from the summit of Mauna Kea, which is a significant cultural and historic site.

\section*{Data Availability}

The data used in this work were recorded in the context of the SLS, and will be available to the public at the Canadian Astronomy Data Center one year after completion of the program.



\bibliographystyle{mnras}
\bibliography{BibPaper3} 




\appendix

\section{Additional figures for \texorpdfstring{<$B$>}{} }

\begin{figure}
	\includegraphics[width=\columnwidth]{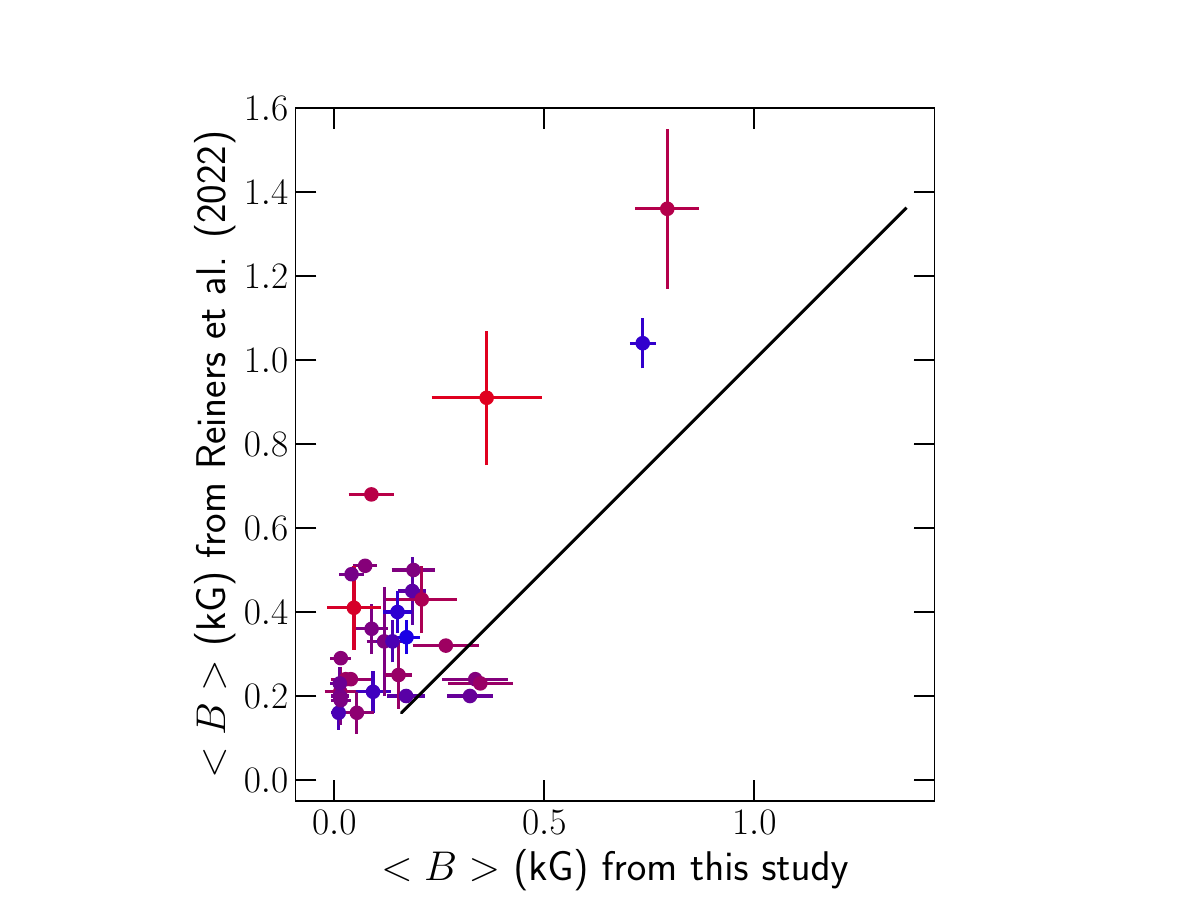}
	\caption{Same as Fig.~\ref{fig:bf-reiners} when fixing $\logg$ from $M_\star$ and $R_\star$. 
	}
	\label{fig:bf-reiners-autologg}
\end{figure}

\begin{figure}
	\includegraphics[width=\columnwidth]{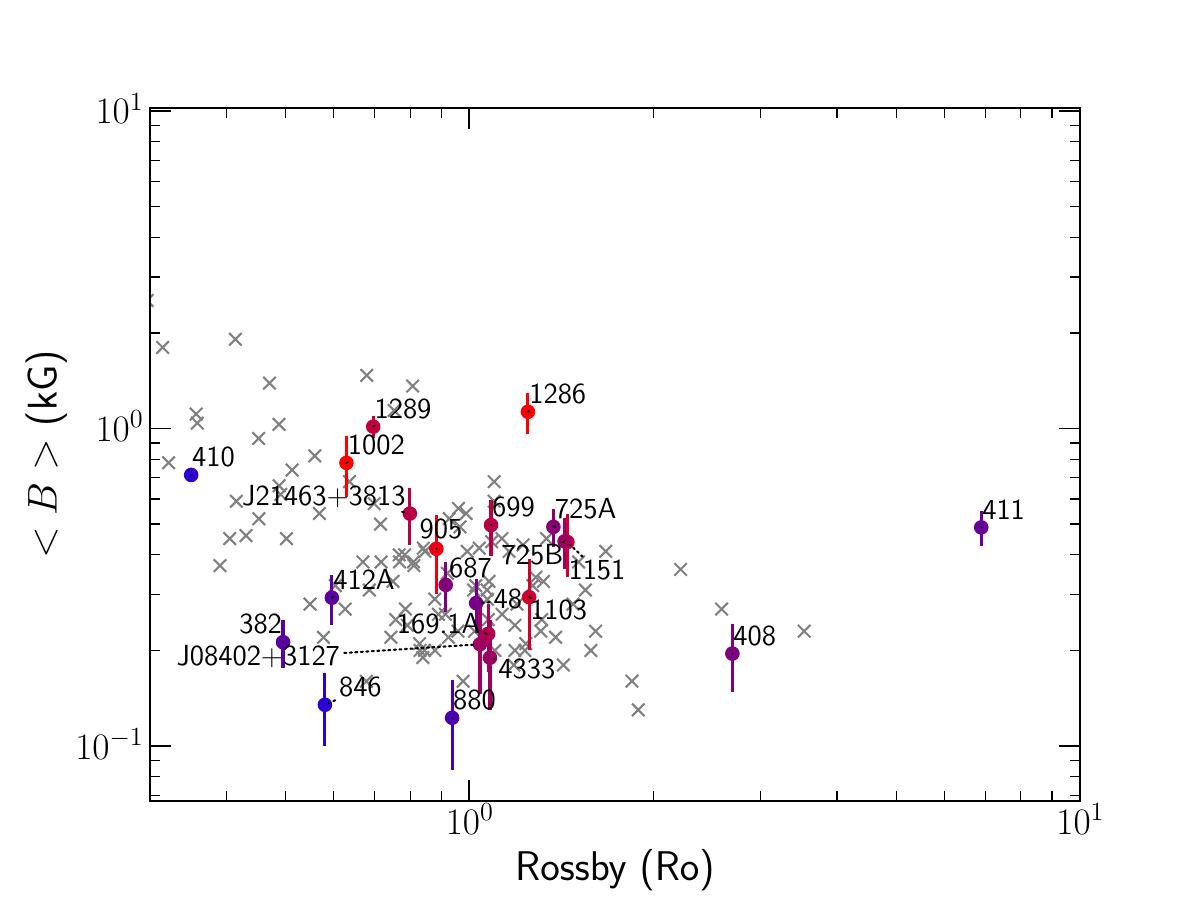}
	\caption{Same as Fig.~\ref{fig:rossby-donati} but showing only the targets with $Ro>0.3$.}
	\label{fig:rossby-donati-zoom}
\end{figure}

\begin{figure}
	\includegraphics[width=\columnwidth]{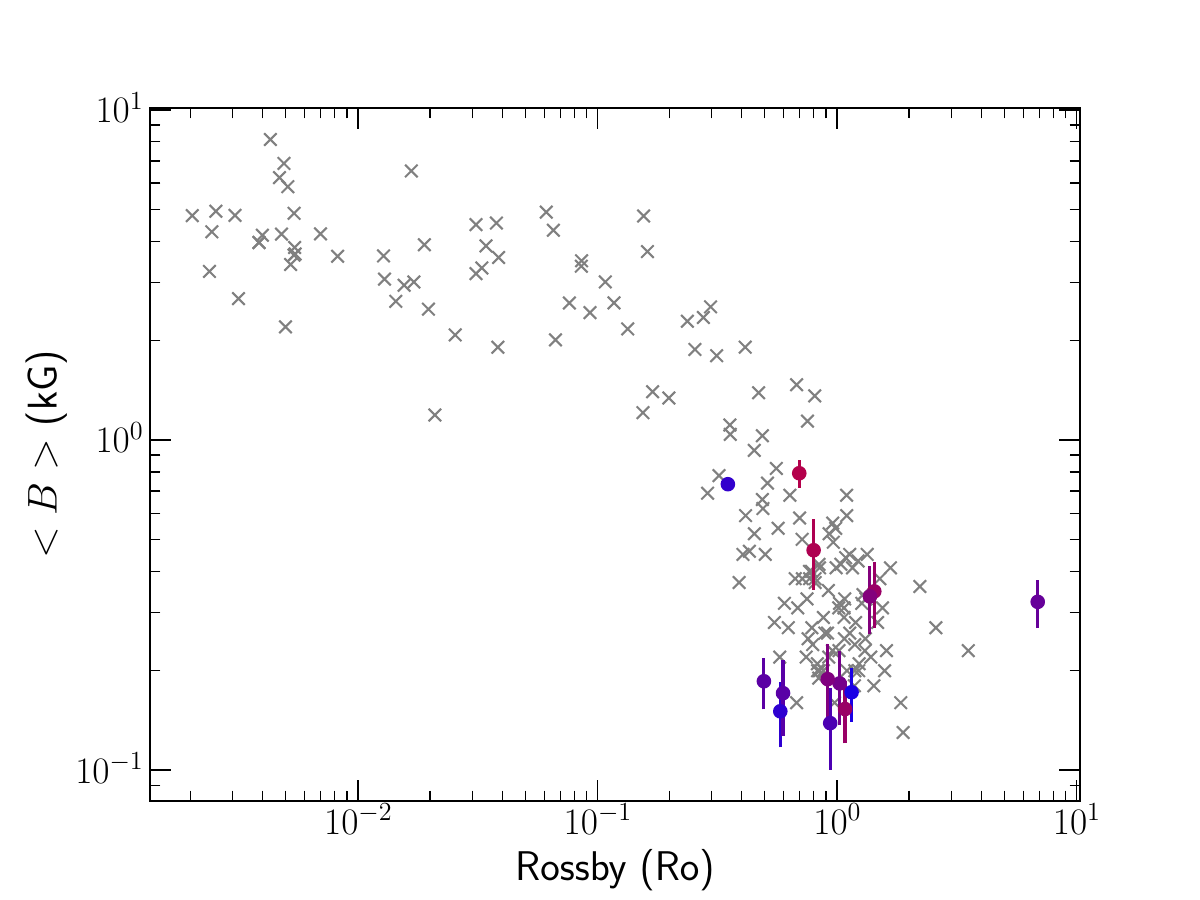}
	\caption{Same as Fig.~\ref{fig:rossby-donati} with results obtained while fixing $\logg$ from $\teff$ and $L_\star/L_\odot$ }
	\label{fig:rossby-donati-autologg}
\end{figure}

Figure~\ref{fig:bf-reiners-autologg} presents an additional comparison between our results and those of~\citet{reiners-2022}. Figures~\ref{fig:rossby-donati-zoom}~\ref{fig:rossby-donati-autologg} present additional Ro-<$B$> diagrams.
	
\begin{figure}
	\includegraphics[width=\columnwidth]{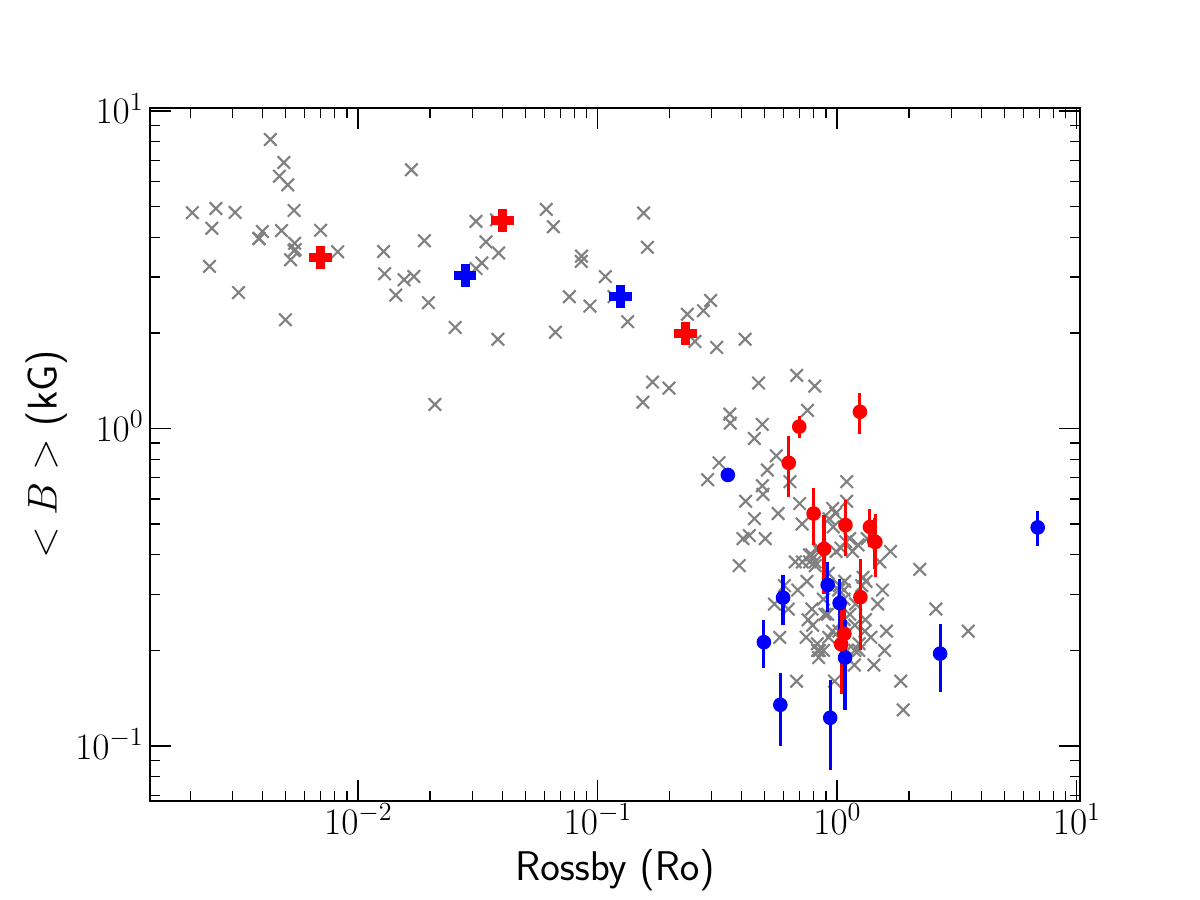}
	\caption{Same as Fig.~\ref{fig:rossby-donati} but with red and blue indicating stars with masses smaller and larger than 0.35~$M_\odot$, respectively.}
	\label{fig:rossby-donati-convective}
\end{figure}

\begin{figure}
	\includegraphics[width=\columnwidth]{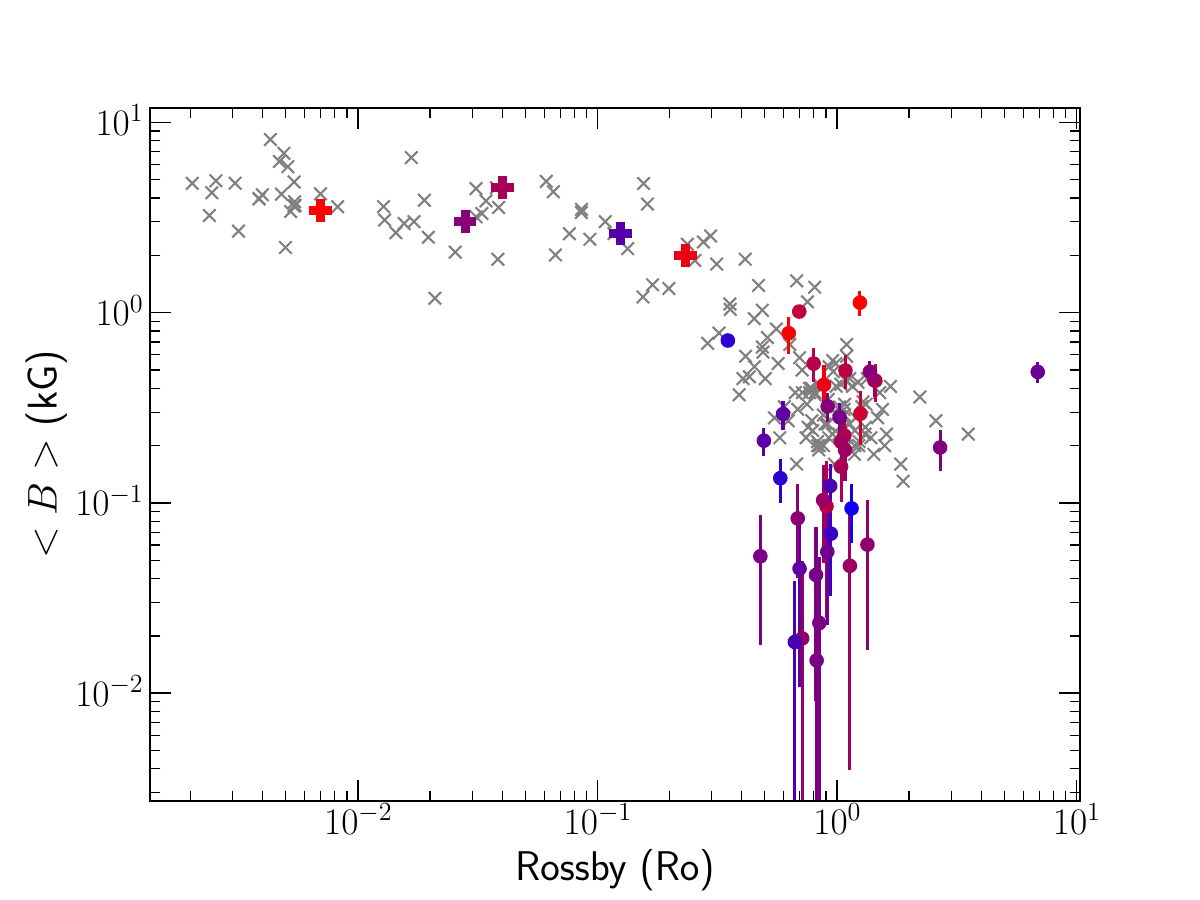}
	\caption{{\paul Same as Fig.~\ref{fig:rossby-donati} showing the results obtained for the 38 stars whose rotation period were derived by~\citet{donati-2023b}}}
	\label{fig:rossby-donati-full}
\end{figure}

\section{Fits}
Figure~\ref{fig:fits} present a comparison between our best fits and the SPIRou templates of 6 targets.

\begin{figure*}
	\includegraphics[width=.85\linewidth, trim={0cm 0cm 0 0cm},clip]{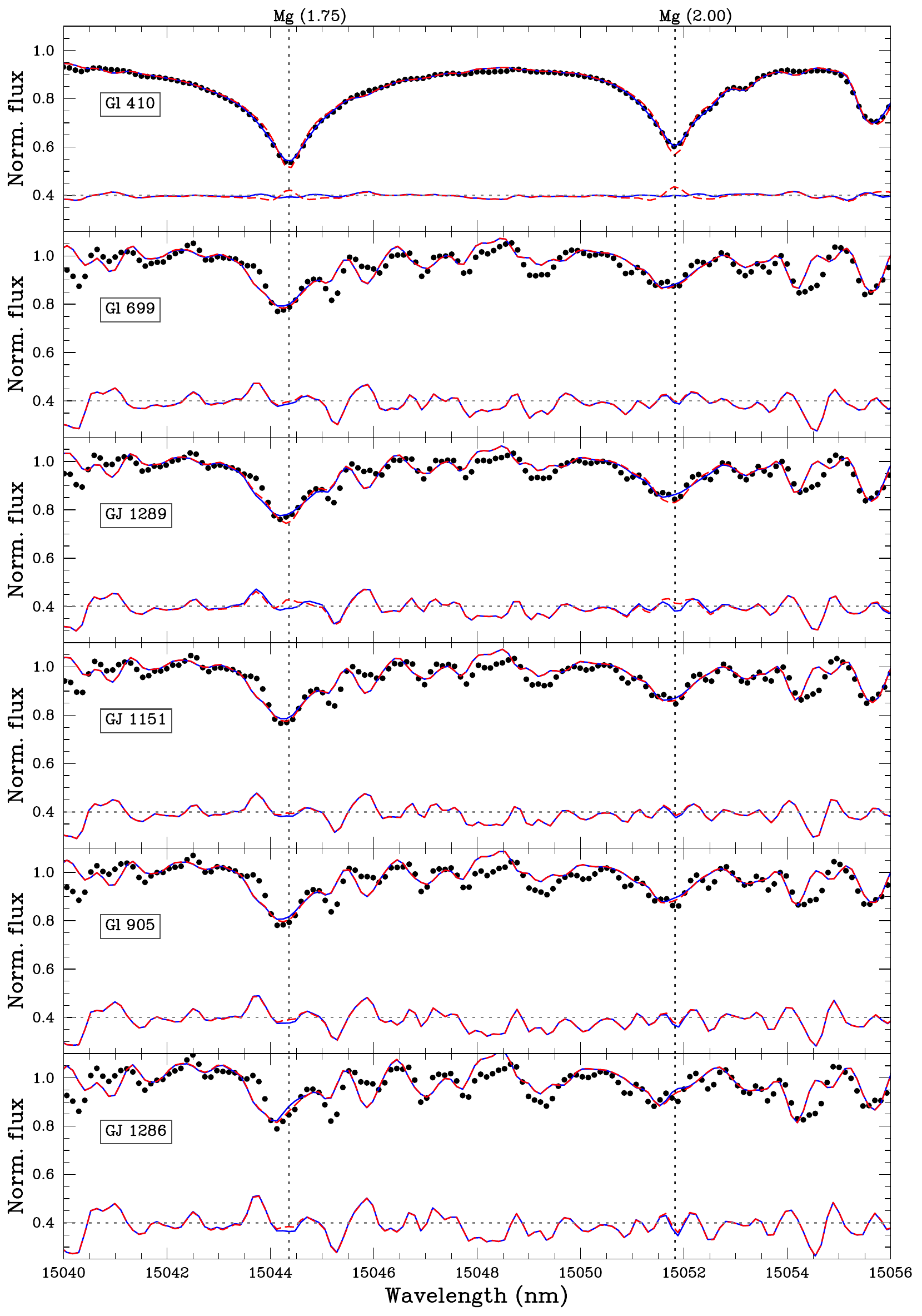}
	\caption{Comparison between the SPIRou template (black points) and the best fit (blue line) for 6 targets in our sample. The red dashed lines show the non-magnetic model. The residuals obtained with the magnetic and non-magnetic models (blue line and red dashed line respectively) are presented for each star, and centred on 0.4 for better readability. The Land\'e factor of the atomic lines is specified in parentheses.}
	\label{fig:fits}
\end{figure*}

\begin{figure*}
	\includegraphics[width=.85\linewidth, trim={0cm 0cm 0 0cm},clip]{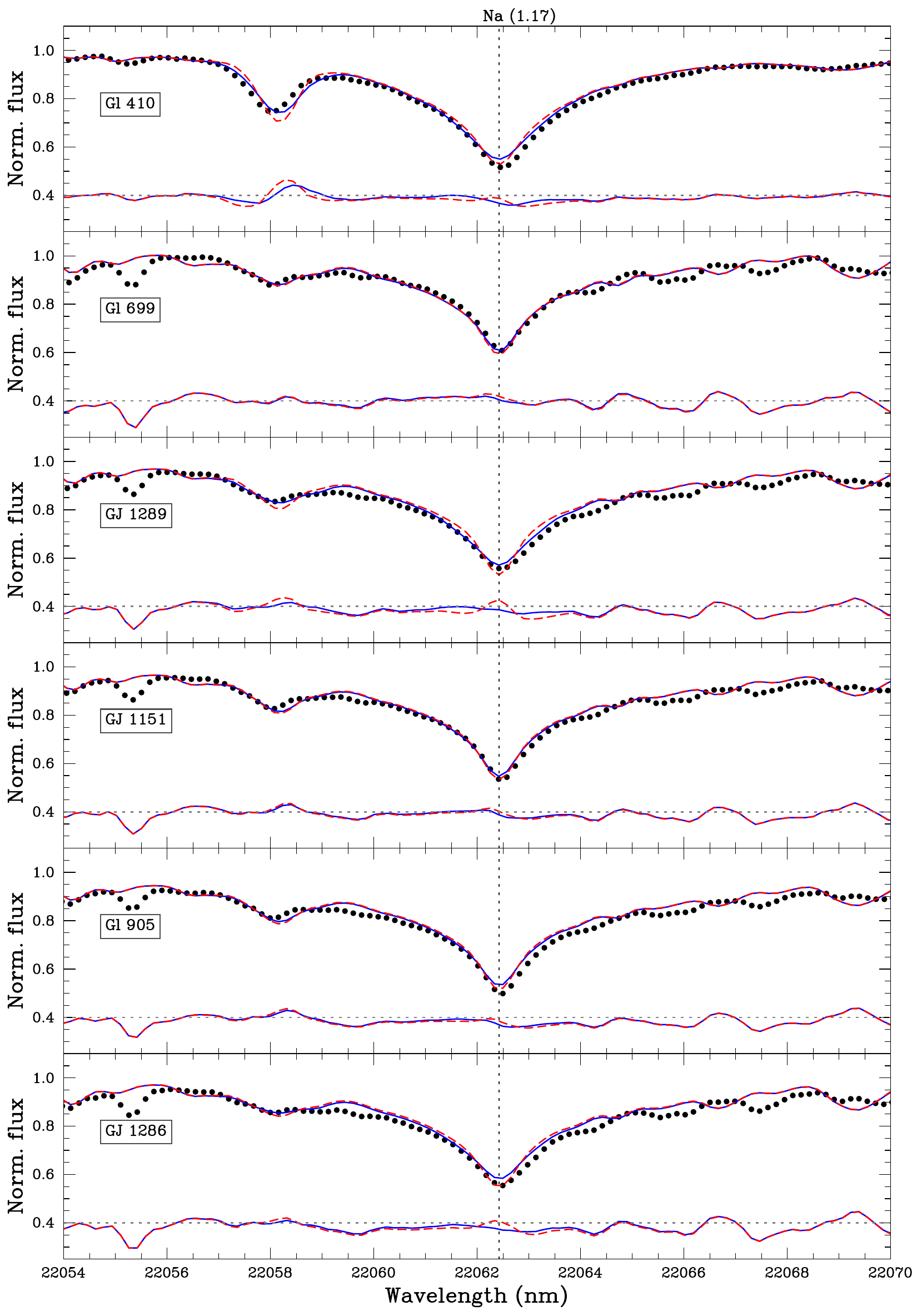}
	\contcaption{}
\end{figure*}

\begin{figure*}
	\includegraphics[width=.85\linewidth, trim={0cm 0cm 0 0cm},clip]{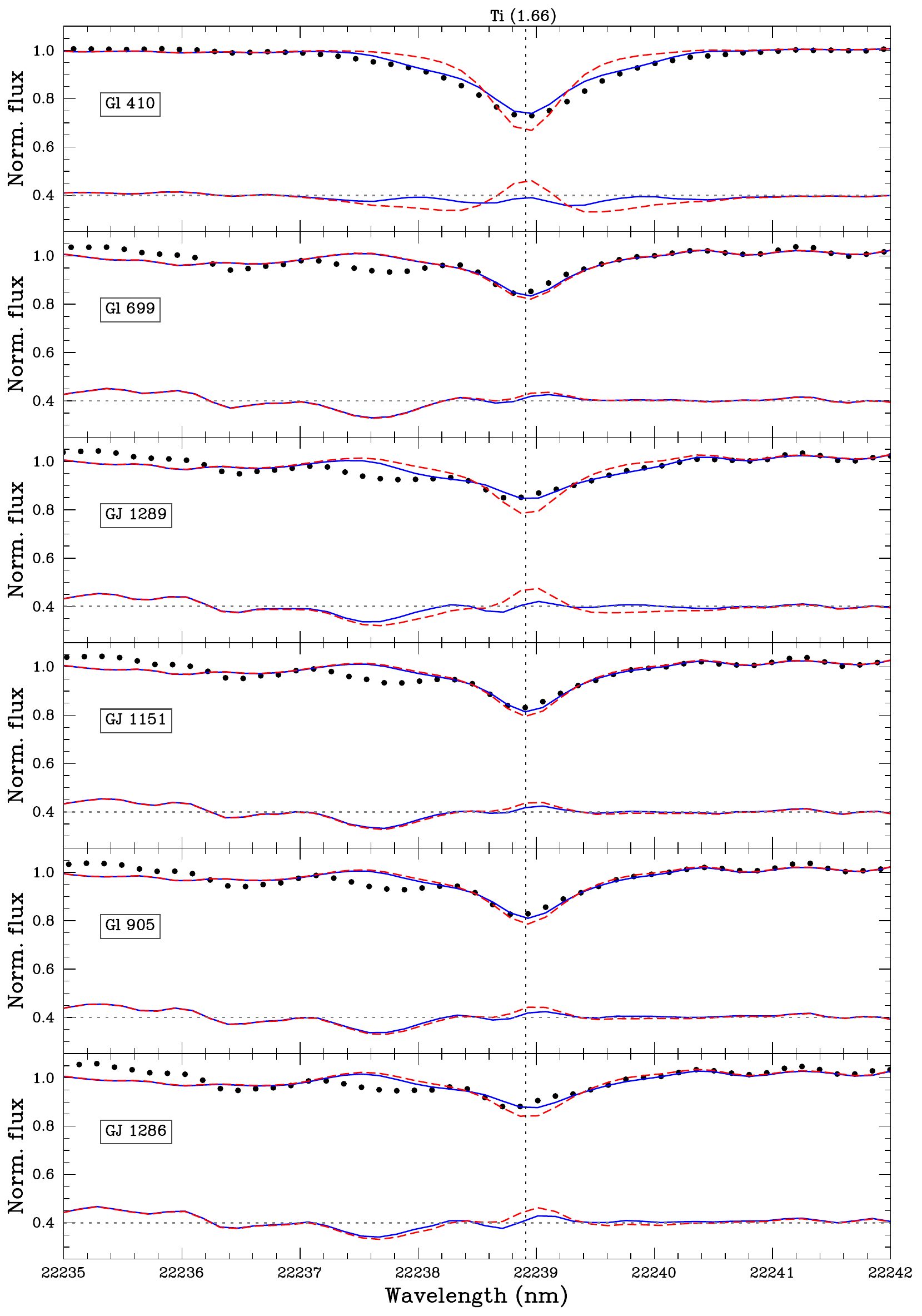}
	\contcaption{}
\end{figure*}

\begin{figure*}
	\includegraphics[width=.85\linewidth, trim={0cm 0cm 0 0cm},clip]{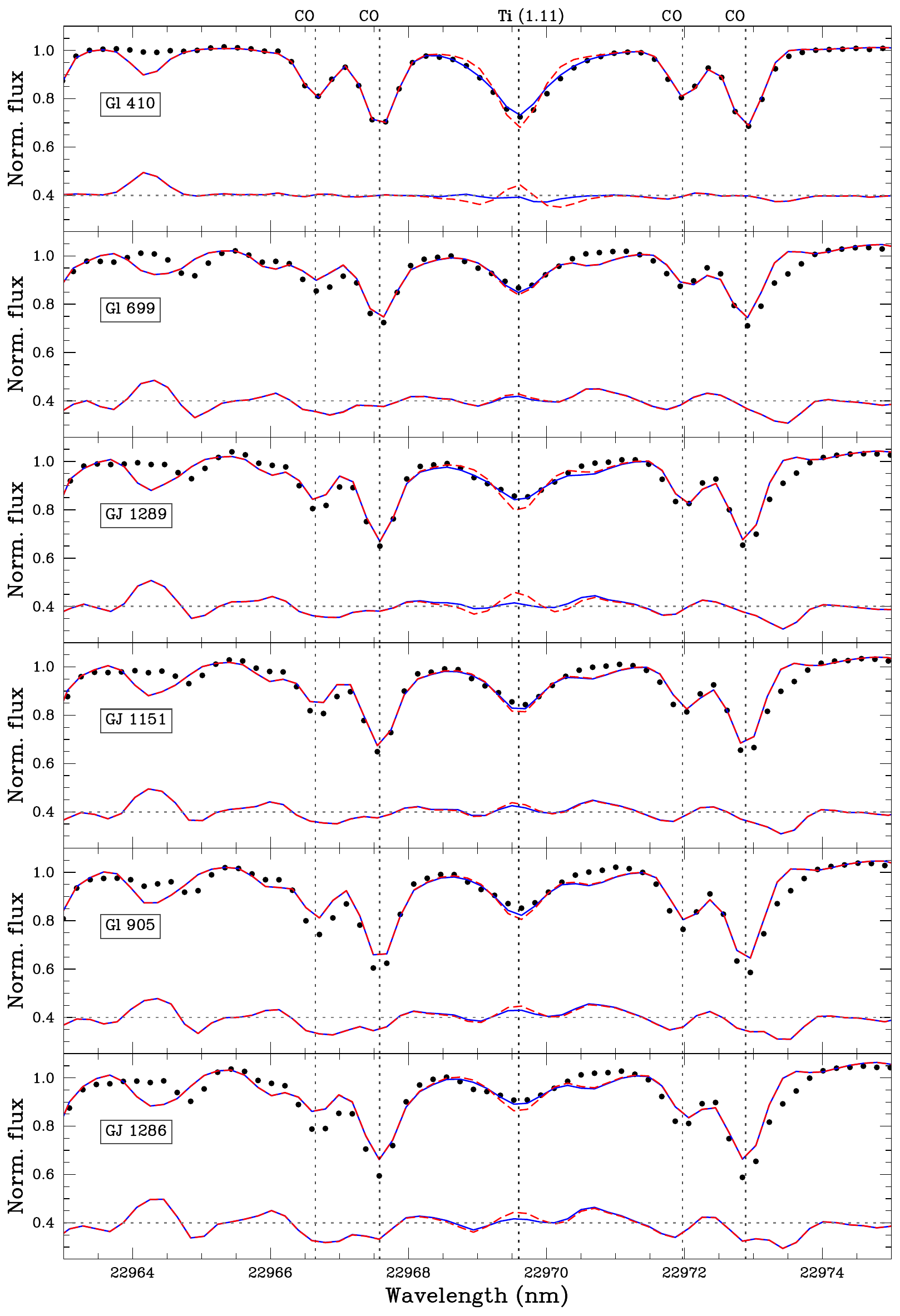}
	\contcaption{}
\end{figure*}

\begin{figure*}
	\includegraphics[width=.85\linewidth, trim={0cm 0cm 0 0cm},clip]{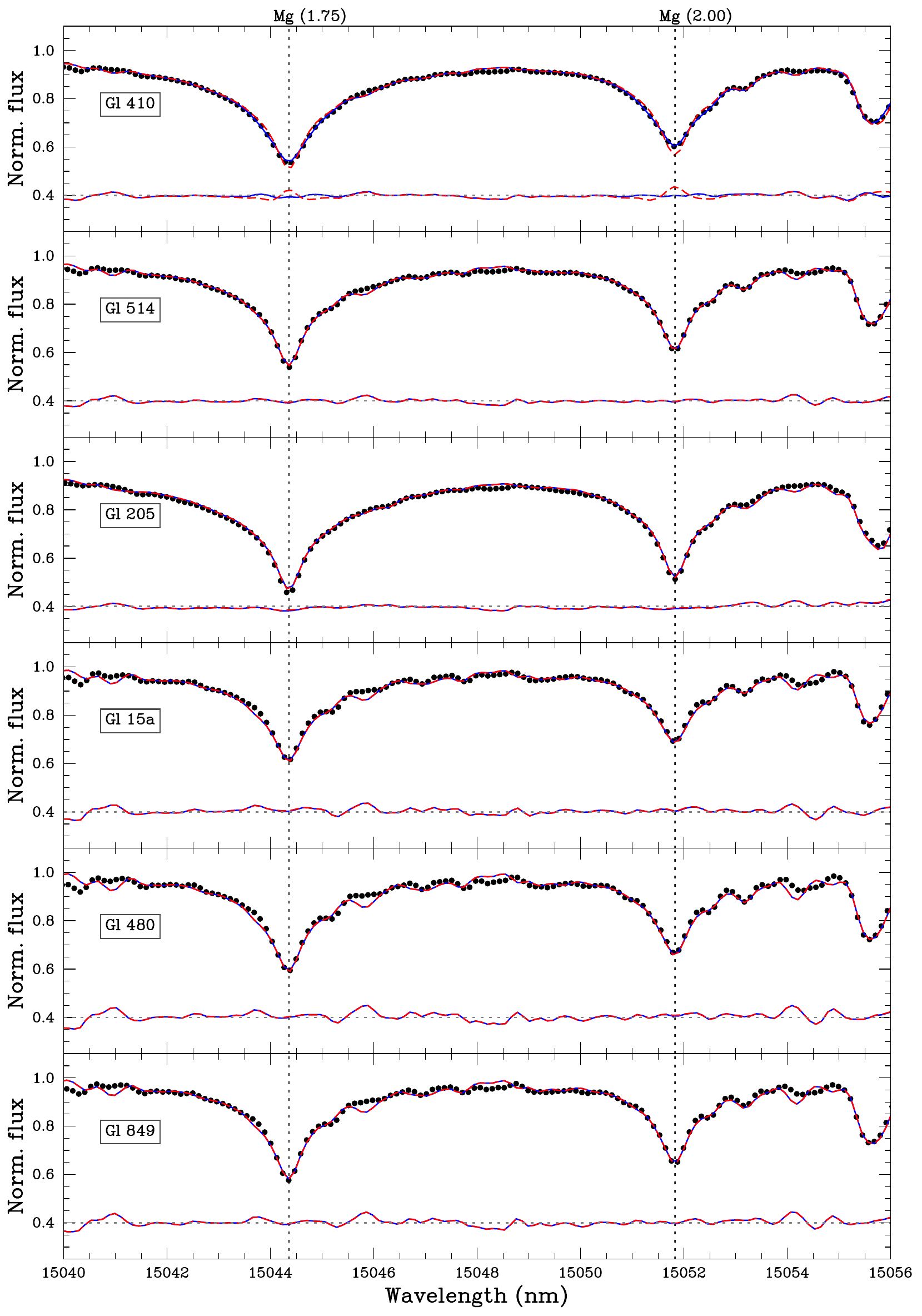}
	\caption{{\paul Same as~\ref{fig:fits} comparing the spectra of Gl\,410 to additional targets.}}
	\label{fig:fits-2}
\end{figure*}

\begin{figure*}
	\includegraphics[width=.85\linewidth, trim={0cm 0cm 0 0cm},clip]{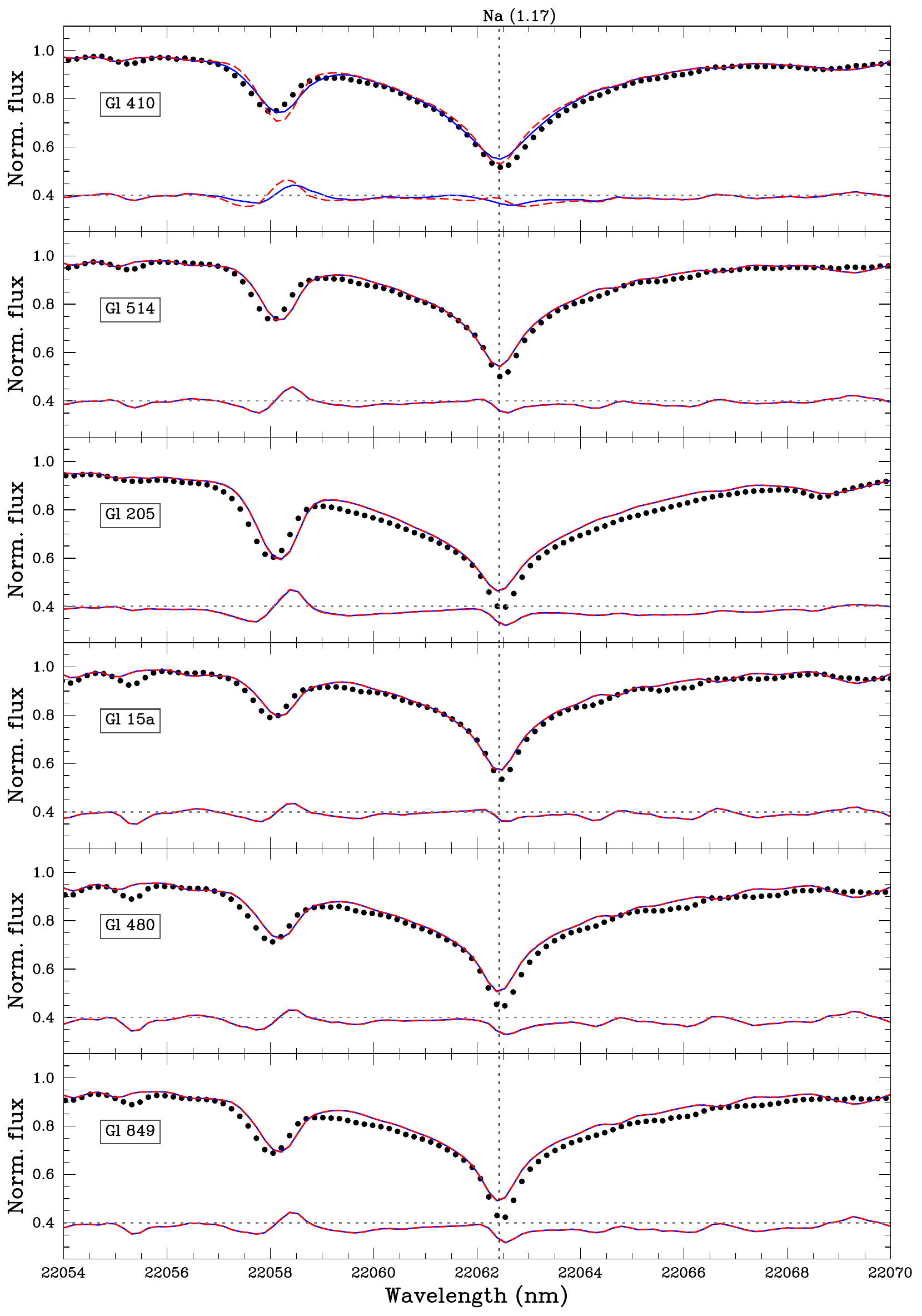}
	\contcaption{}
\end{figure*}

\begin{figure*}
	\includegraphics[width=.85\linewidth, trim={0cm 0cm 0 0cm},clip]{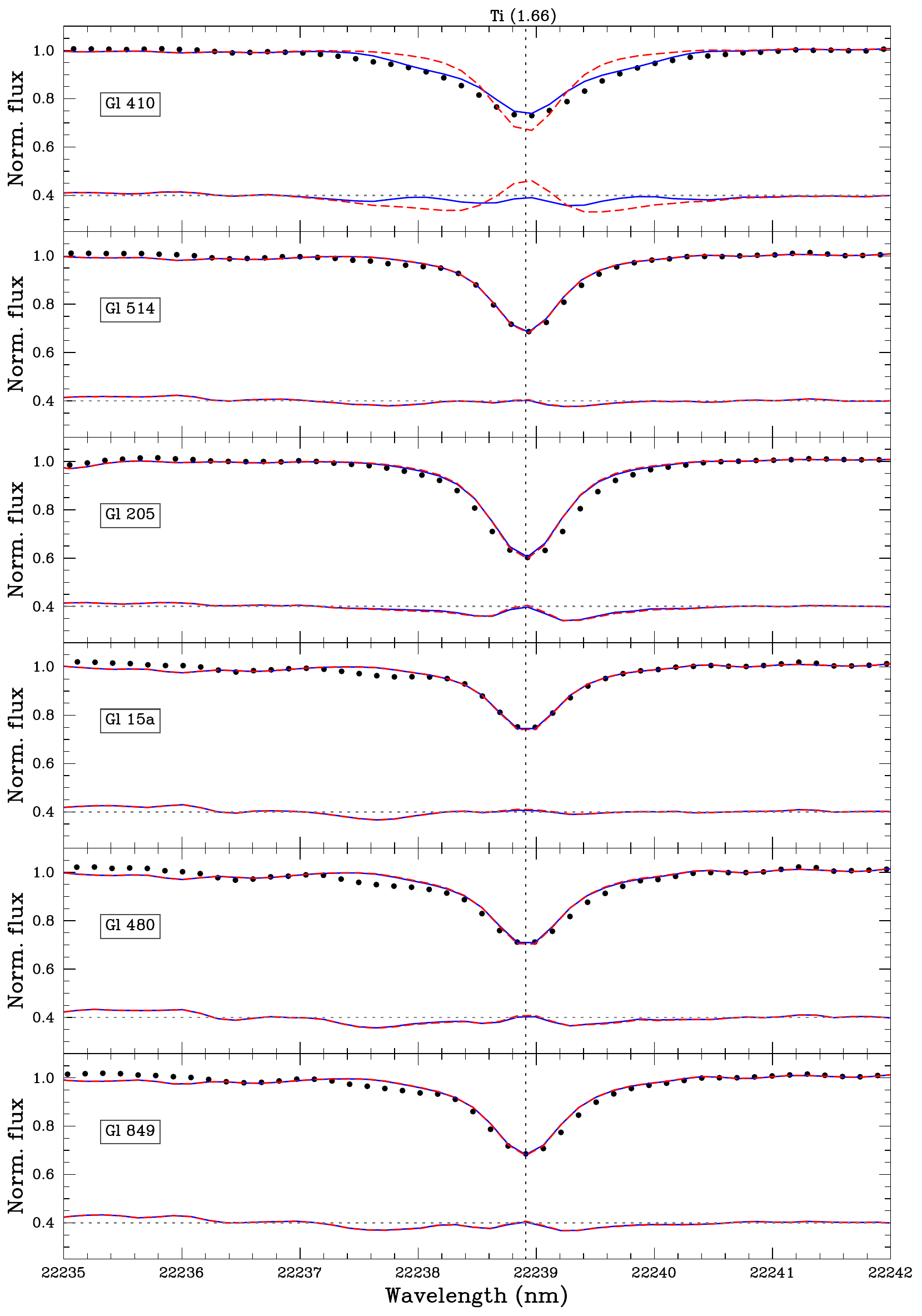}
	\contcaption{}
\end{figure*}

\begin{figure*}
	\includegraphics[width=.85\linewidth, trim={0cm 0cm 0 0cm},clip]{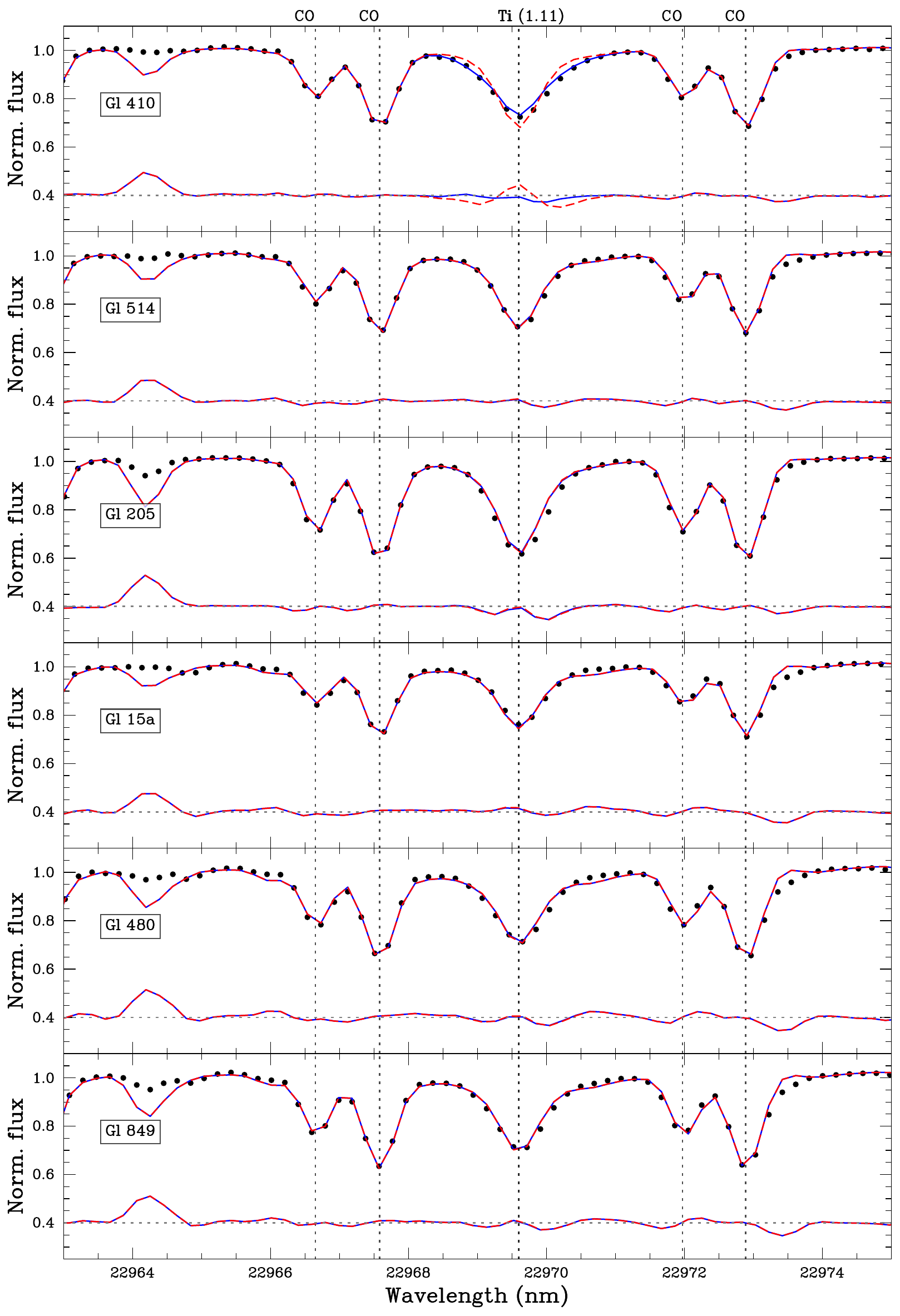}
	\contcaption{}
\end{figure*}

\section{Corner plots and filling factors}
\begin{figure*}
	\includegraphics[width=\columnwidth]{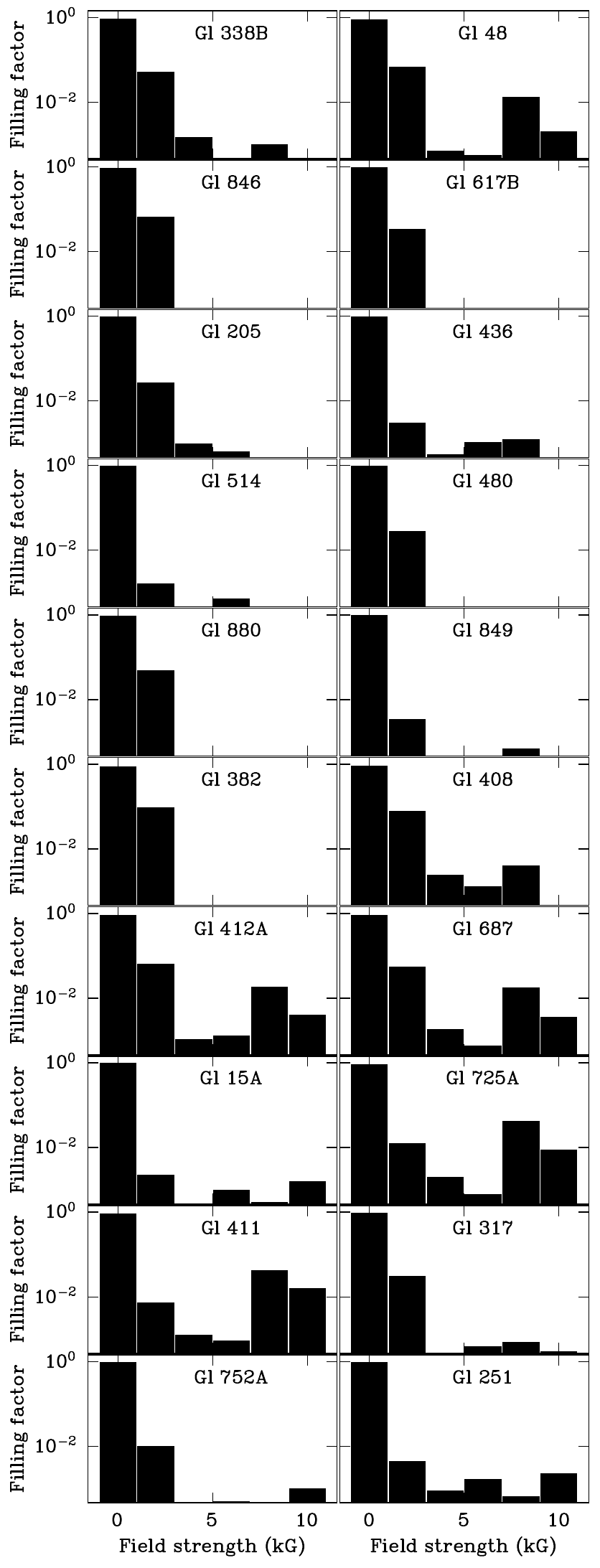}\hspace{0.5cm}\includegraphics[width=\columnwidth]{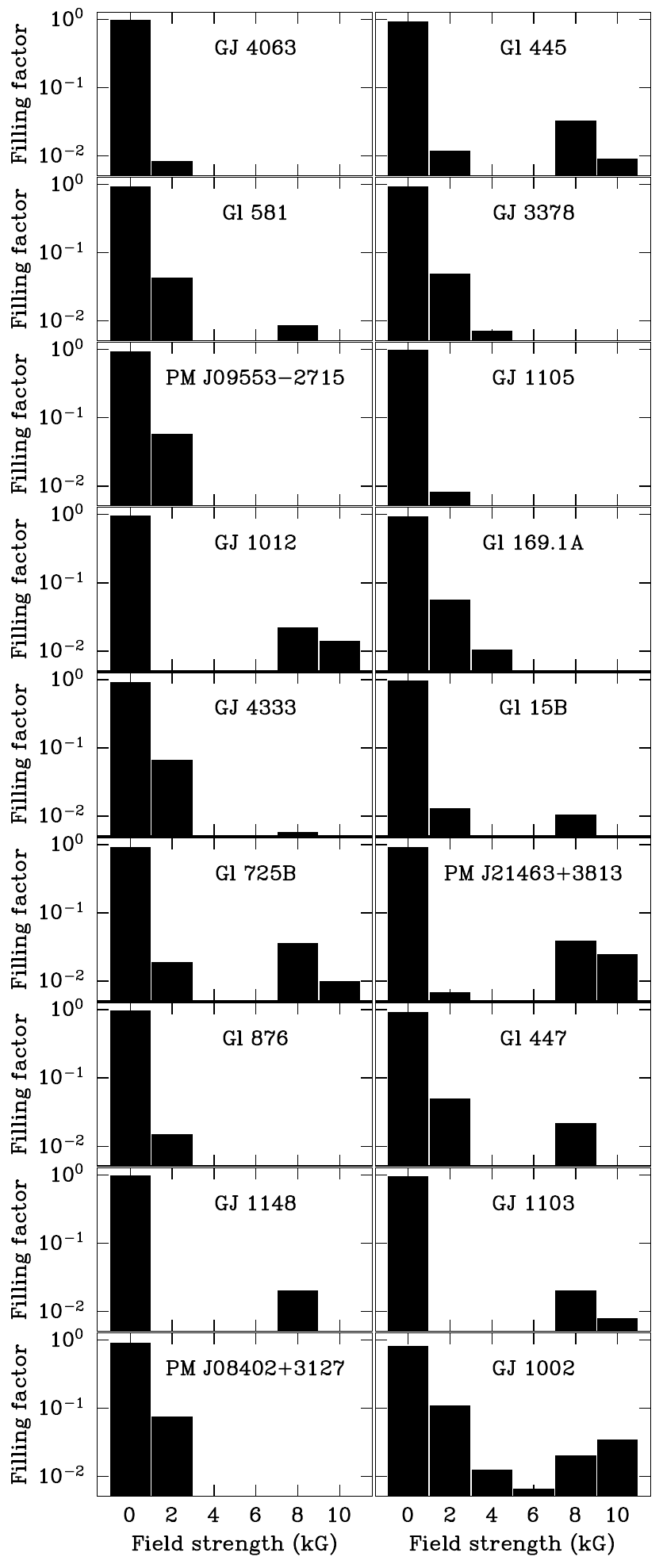}
	
	\caption{Same as Fig.~\ref{fig:distribs} for the other stars in our sample.}
	\label{fig:distrib-other}
\end{figure*}

\begin{figure*}
	\includegraphics[width=\linewidth]{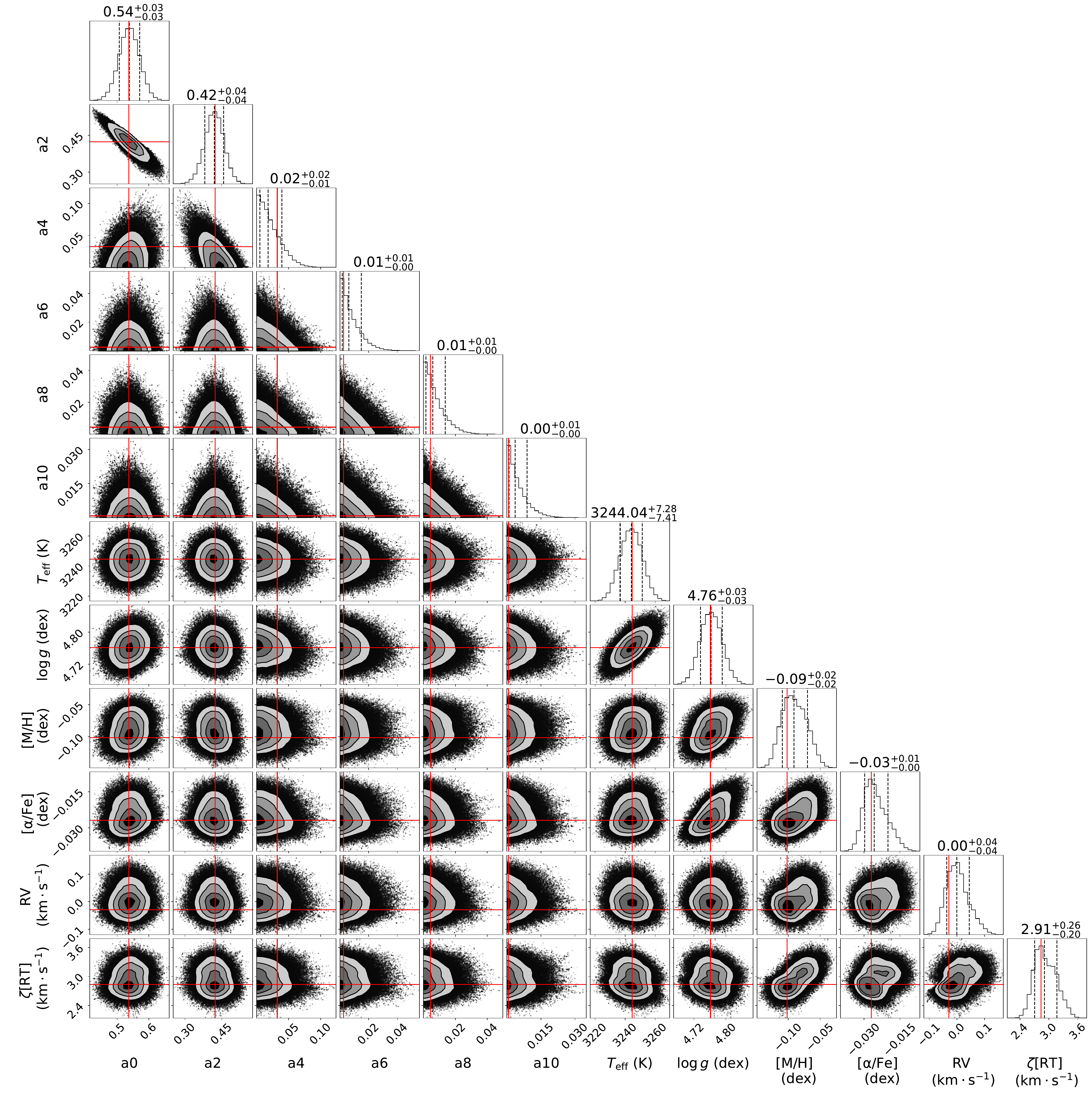}
	\caption{Posterior distributions for the atmospheric parameters and filling factors of GJ\,1289. The red lines mark the average position of the walkers whose associated $\chi^2$ do not deviate by more than 1 from the minimum $\chi^2$.}
	\label{fig:corner-gj1289}
\end{figure*}

\begin{figure*}
	\includegraphics[width=\linewidth]{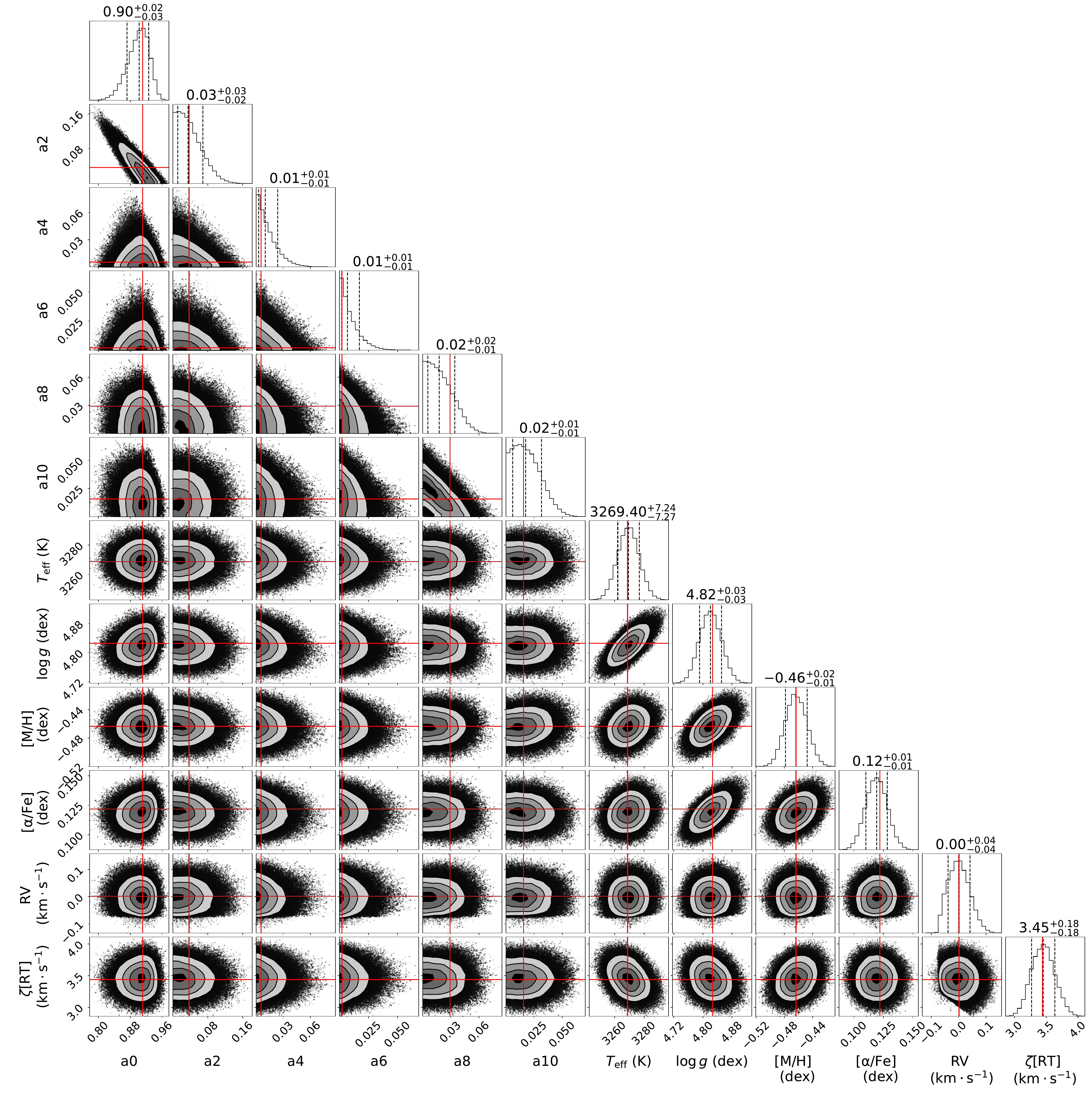}
	\caption{Same as Fig.~\ref{fig:corner-gj1289} for Gl\,699.}
	\label{fig:corner-gl699}
\end{figure*}

\begin{figure*}
	\includegraphics[width=\linewidth]{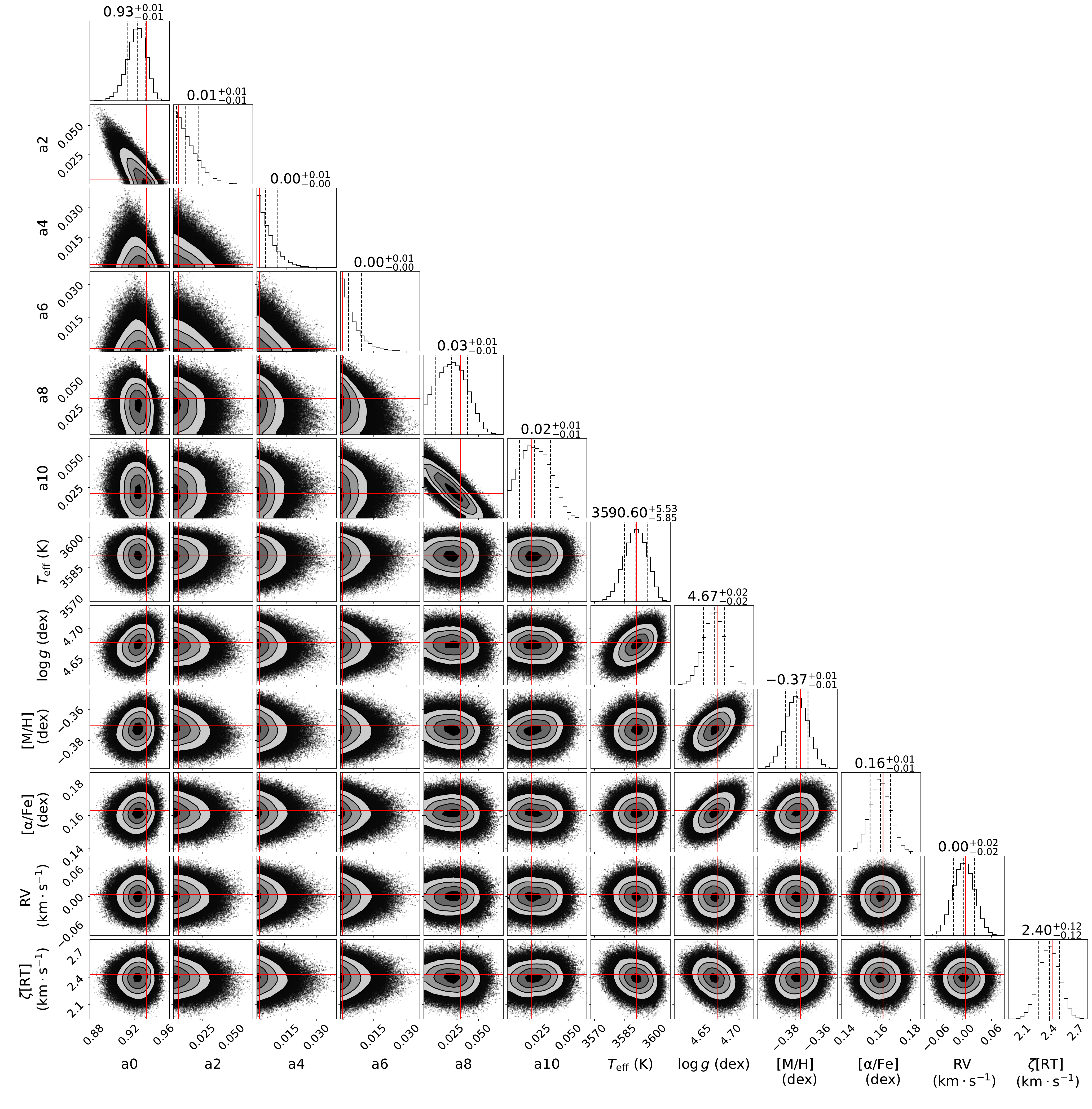}
	\caption{Same as Fig.~\ref{fig:corner-gj1289} for Gl\,411.}
	\label{fig:corner-gl411}
\end{figure*}

Figure~\ref{fig:distrib-other} presents the distribution of filling factors recovered for 38 targets in our sample (see Fig.~\ref{fig:distribs} for the other 6 targets). 
Figures~\ref{fig:corner-gj1289}, ~\ref{fig:corner-gl699} and~\ref{fig:corner-gl411} present the posterior distribution obtained for all the fitted parameters for GJ\,1289, Gl\,699 and Gl\,411, respectively.

\section{Additional comparisons to literature estimates}

\begin{figure}
	\includegraphics[width=\columnwidth]{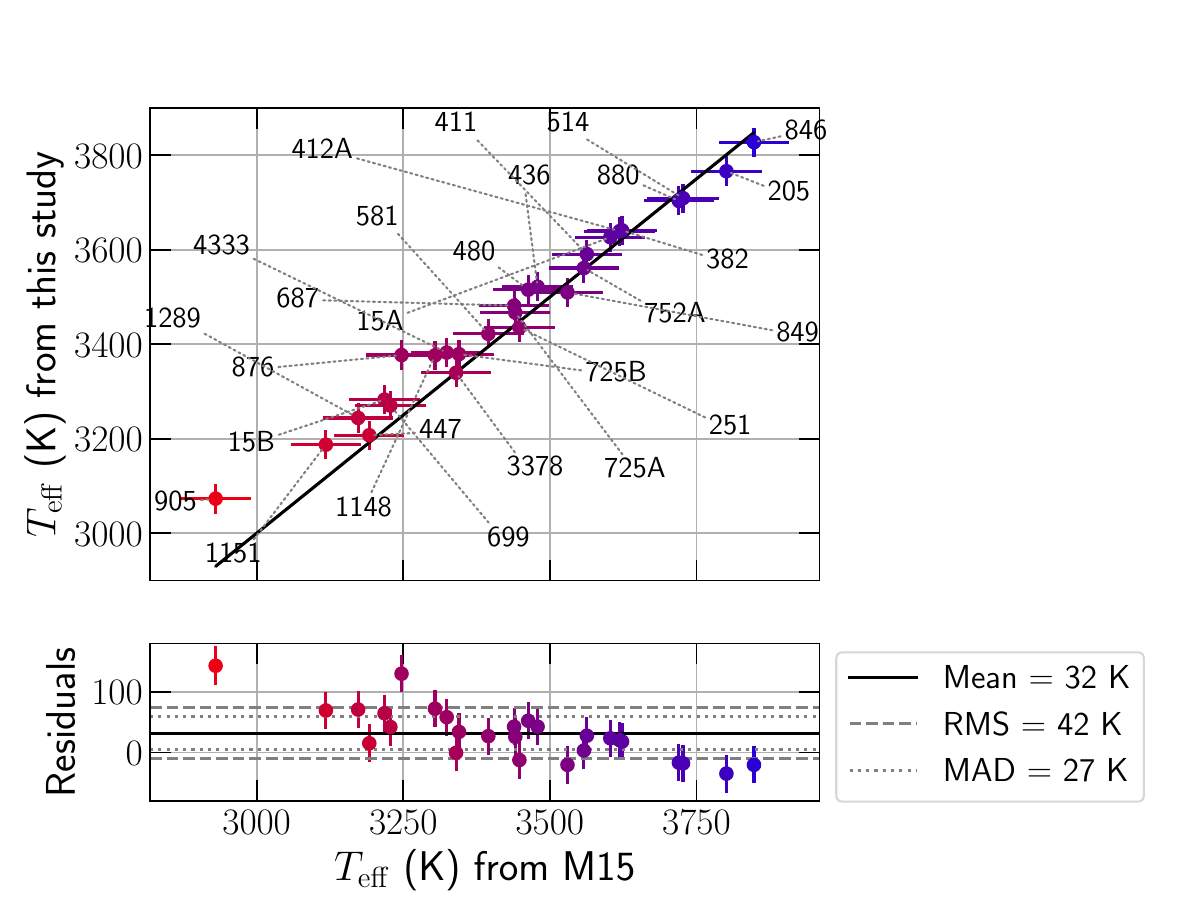}
	\caption{Comparison between our retrieved $\teff$ and those of~\citet[][M15]{mann-2015}.}
	\label{fig:res-teff}
\end{figure}

\begin{figure}
	\includegraphics[width=\columnwidth]{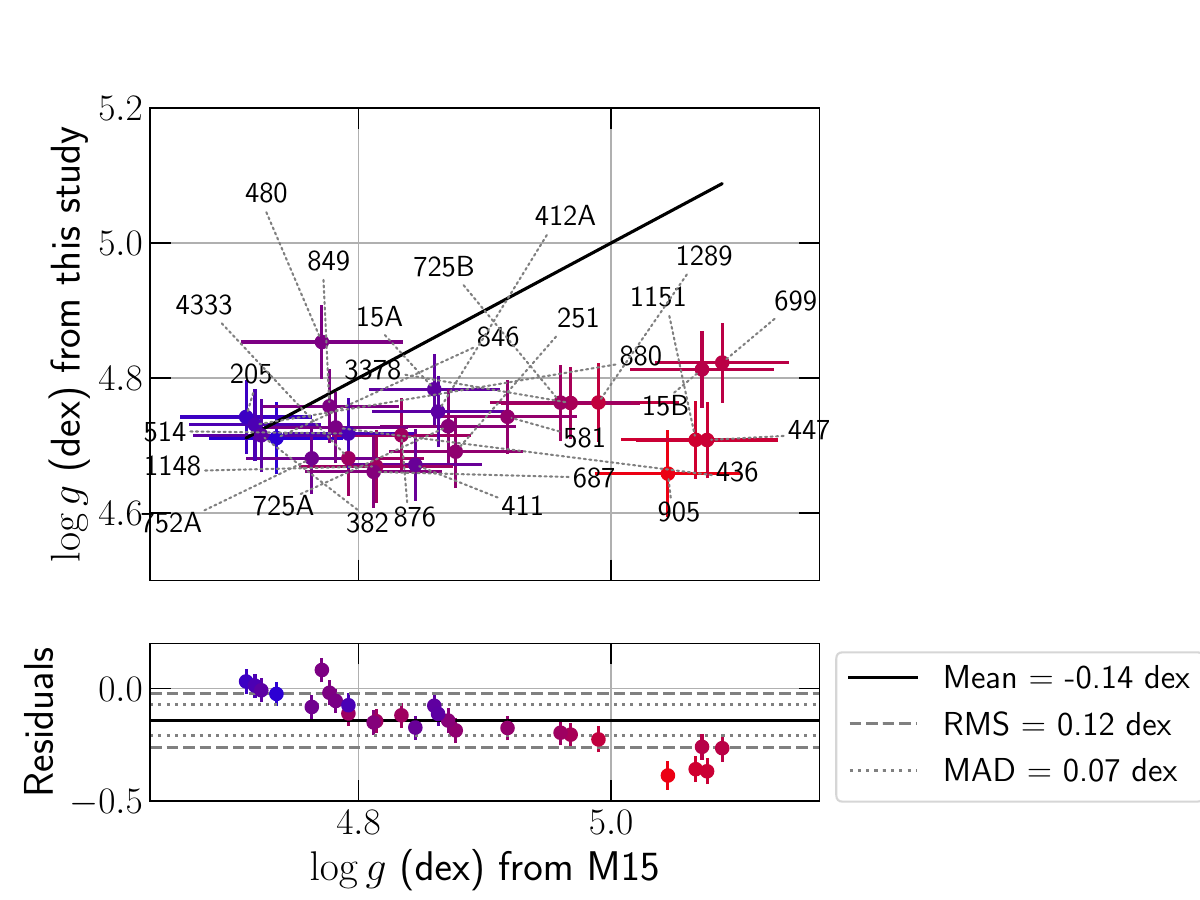}
	\caption{Same as Fig.~\ref{fig:res-teff} for $\logg$.}
	\label{fig:res-logg}
\end{figure}

\begin{figure}
	\includegraphics[width=\columnwidth]{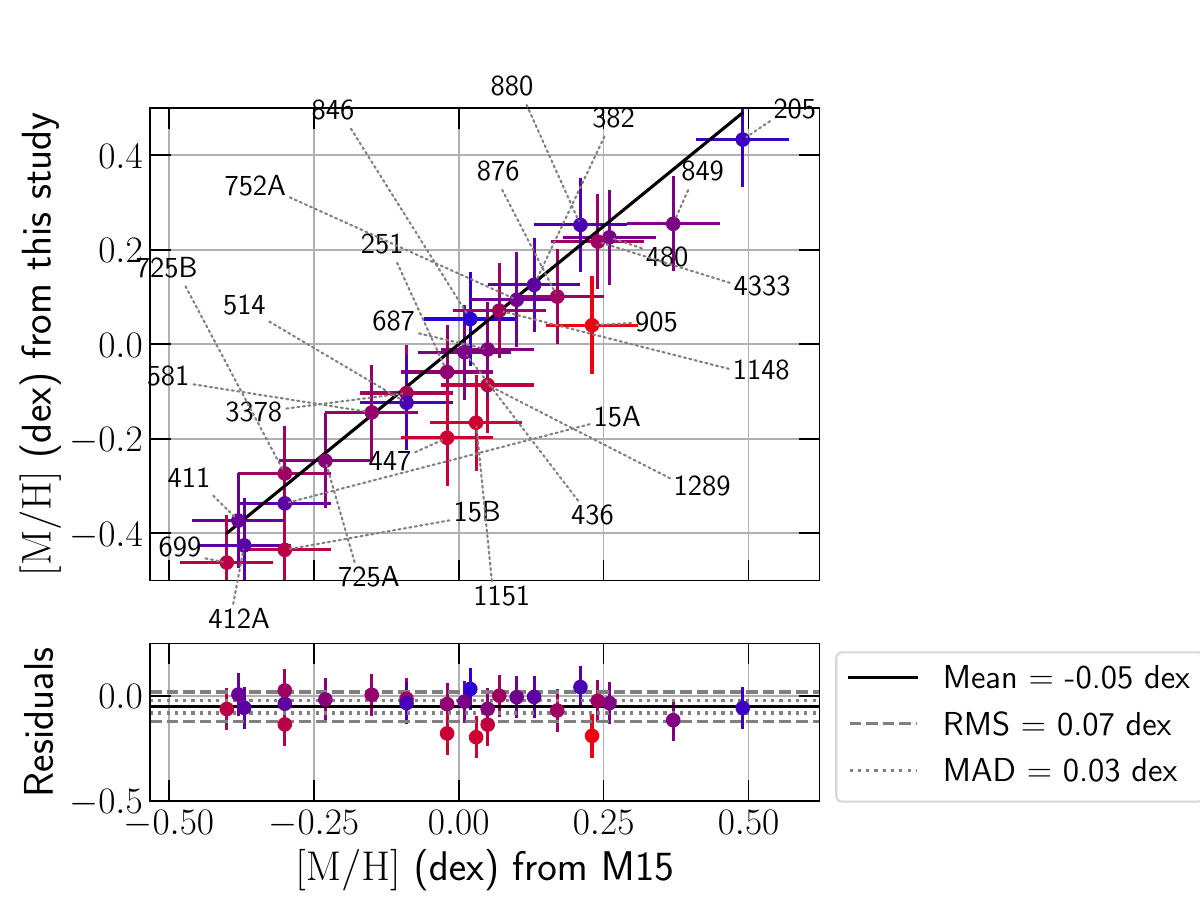}
	\caption{Same as Fig.~\ref{fig:res-teff} for $\mh$.}
	\label{fig:res-mh}
\end{figure}

Figures~\ref{fig:res-teff},~\ref{fig:res-logg} and~\ref{fig:res-mh} present a comparison between our derived $\teff$, $\logg$ and $\mh$ to those of~\citet{mann-2015}, respectively.

\section{Figures with labels}
\begin{figure}
	\includegraphics[width=\linewidth]{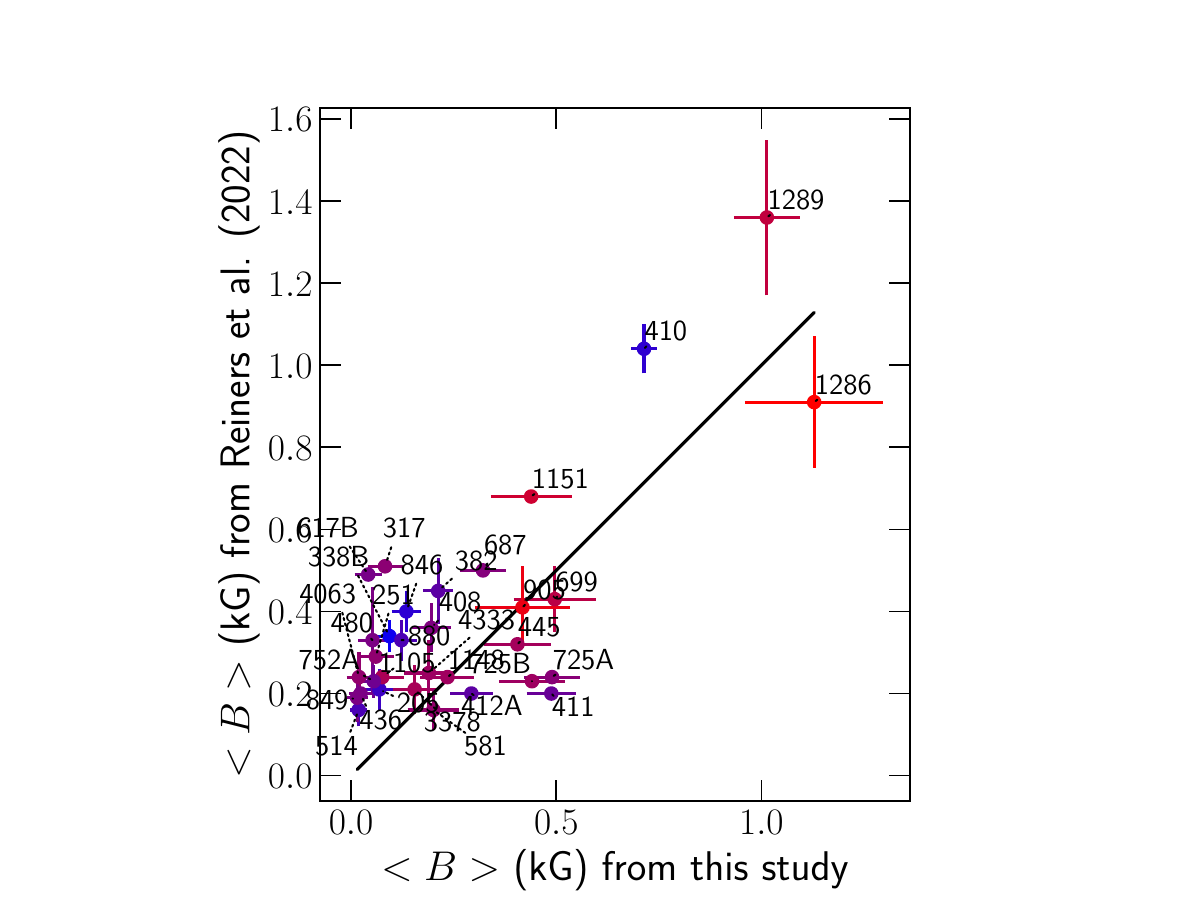}
	\caption{Same as Fig.~\ref{fig:bf-reiners} with labels indicating the names of the stars.}
	\label{fig:bf-reiners-label}
\end{figure}

\begin{figure}
	\includegraphics[width=\linewidth]{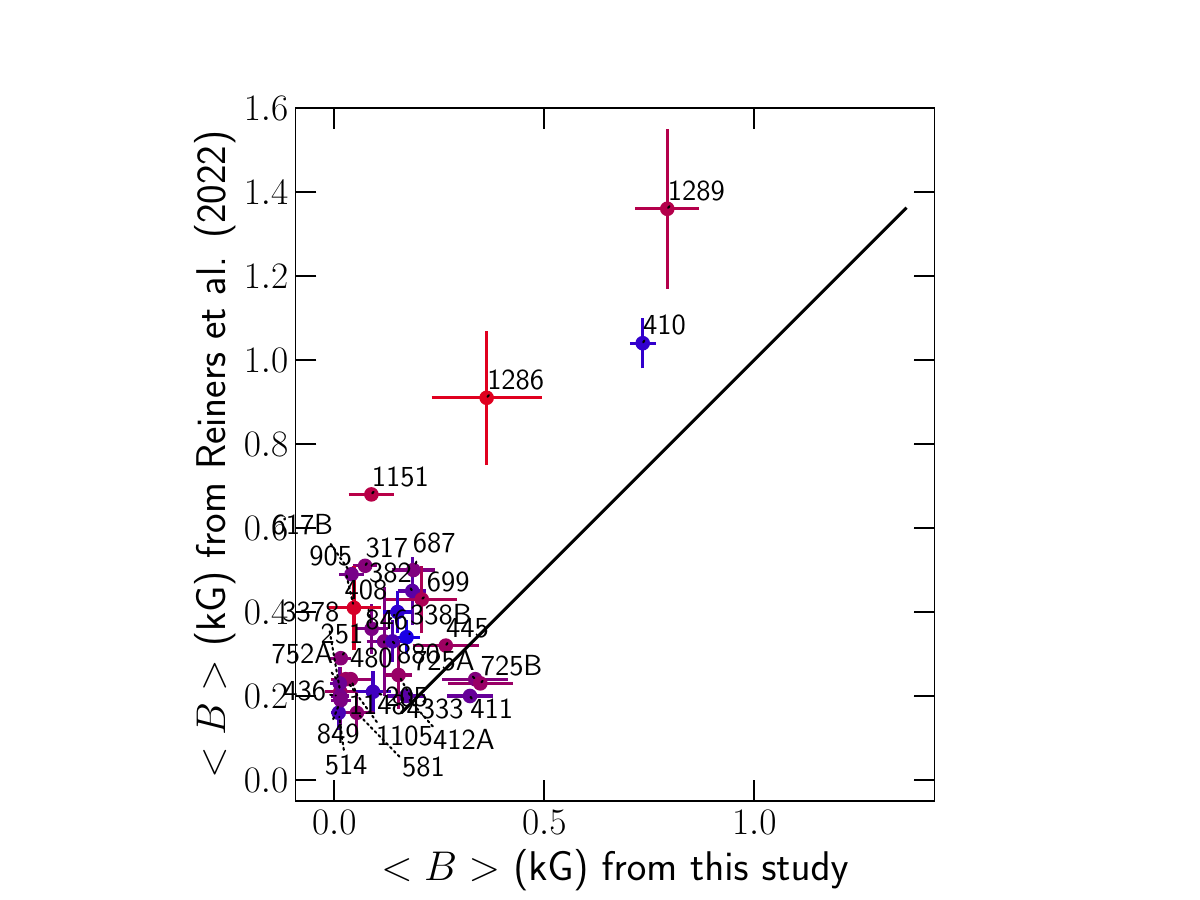}
	\caption{Same as Fig.~\ref{fig:bf-reiners-autologg} with labels indicating the names of the stars.}
	\label{fig:bf-reiners-autologg-label}
\end{figure}

\begin{figure}
	\includegraphics[width=\linewidth]{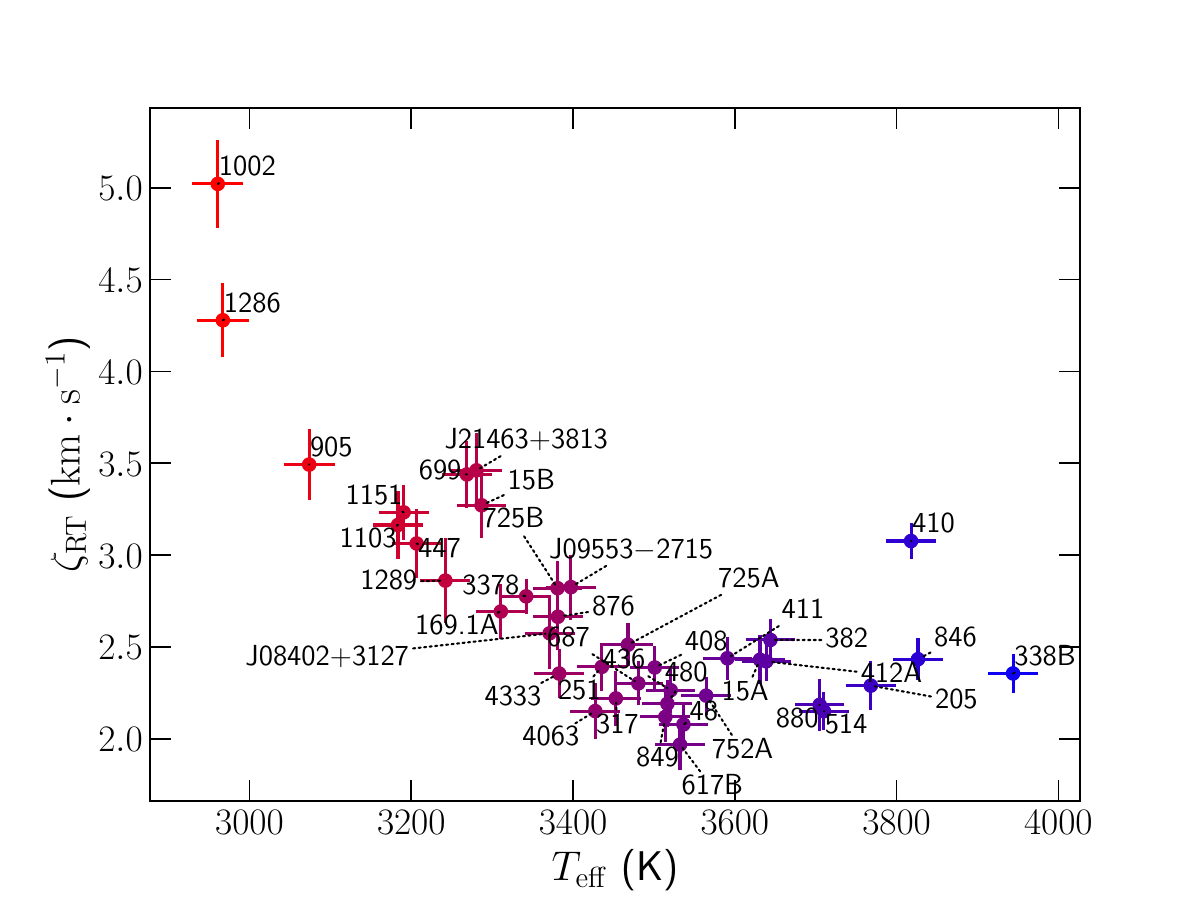}
	\caption{Same as Fig.~\ref{fig:macro-teff} with labels indicating the names of the stars.}
	\label{fig:macro-teff-label}
\end{figure}

\begin{figure}
	\includegraphics[width=\linewidth]{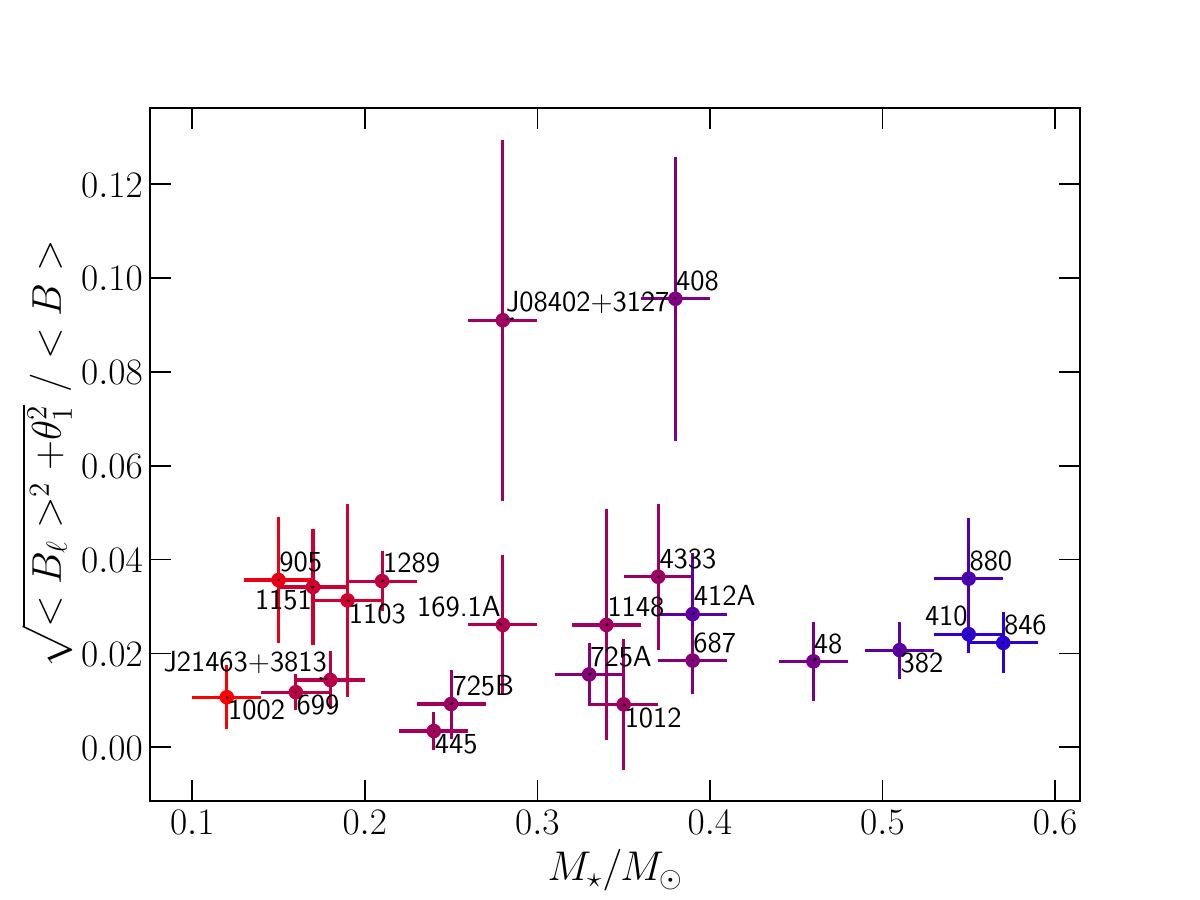}
	\caption{Same as Fig.~\ref{fig:bl-mass} with labels indicating the names of the stars.}
	\label{fig:bl-mass-label}
\end{figure}

Figures~\ref{fig:bf-reiners-label},
~\ref{fig:bf-reiners-autologg-label},
~\ref{fig:macro-teff-label} and ~\ref{fig:bl-mass-label}, present the same results as Figs~\ref{fig:bf-reiners},
~\ref{fig:bf-reiners-autologg},
~\ref{fig:macro-teff} and ~\ref{fig:bl-mass} with labels indicating the names of the targets.

\bsp	
\label{lastpage}
\end{document}